\documentclass{article}

\usepackage{PRIMEarxiv}

\usepackage[utf8]{inputenc} 
\usepackage[T1]{fontenc}    
\usepackage{hyperref}       
\usepackage{url}            
\usepackage{booktabs}       
\usepackage{amsfonts}       
\usepackage{nicefrac}       
\usepackage{microtype}      
\usepackage{lipsum}
\usepackage{fancyhdr}       
\usepackage{graphicx}       
\graphicspath{{media/}}     

\usepackage{amssymb}
\usepackage{csquotes}
\usepackage{textgreek}
\usepackage{float}
\usepackage{siunitx}
\usepackage{stmaryrd}

\usepackage{physics}
\usepackage{multirow}

\usepackage{caption}
\usepackage{subcaption}
\usepackage{xcolor}

\usepackage{booktabs}
\usepackage{array} 
\usepackage{colortbl}
\usepackage{diagbox}
\usepackage{stmaryrd}
\usepackage{empheq, etoolbox}
\title{Efficient and accurate simulation of the Smith-Zener pinning mechanism during grain growth using a front-tracking numerical framework}

\author{ \href{https://orcid.org/0000-0002-5962-7700}{\includegraphics[scale=0.06]{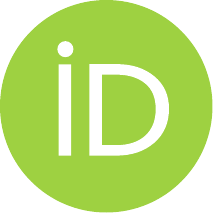}\hspace{1mm}Sebastian ~Florez}$^1$,
	\href{https://orcid.org/0000-0002-6677-2850}{\includegraphics[scale=0.06]{orcid.pdf}\hspace{1mm}Marc ~Bernacki}\thanks{Corresponding author: marc.bernacki@minesparis.psl.eu} $^1$ \\
	\\
	$^1$ Mines Paris, PSL University\\
 	Centre for material forming (CEMEF), UMR CNRS\\
 	 06904 Sophia Antipolis, France\\
}

\begin{document}
\maketitle

\begin{abstract}
This study proposes a new full-field approach for modeling grain boundary pinning by second phase particles in two-dimensional polycrystals.
These particles are of great importance during thermomechanical treatments, as they produce deviations from the microstructural evolution that the alloy produces in the absence of particles. This phenomenon, well-known as Smith-Zener pinning, is widely used by metallurgists to control the grain size during the metal forming process of many alloys. Predictive tools are then needed to accurately model this phenomenon. This article introduces a new methodology for the simulation of microstructural evolutions subjected to the presence of second phase particles. The methodology employs a Lagrangian 2D front-tracking methodology, while the particles are modeled using discretized circular shapes or pinning nodes. The evolution of the particles can be considered and modeled using a constant velocity of particle shrinking. This approach has the advantages of improving the limited description made of the phenomenon in vertex approaches, to be usable for a wide range of second-phase particle sizes and to improve calculation times compared to front-capturing type approaches.
\end{abstract}


\section{Introduction}\label{sec:introduction}
The in-service properties of metallic materials are intimately connected to their microstructures, which are inherited from hot forming processes. Thus, the design of high-value-added parts is nowadays systematically linked to precise control of the microstructure through optimization of processes and a detailed understanding of the underlying mechanisms \cite{rollett2017recrystallization} (recrystallization, grain growth, and solid-state phase transformations). It is with this perspective, for instance, that the concept of grain boundary engineering was developed, driven by the desire to improve various types of intergranular behavior (corrosion, segregation, crack propagation, creep, ductility, ...) through control and manipulation of grain boundaries (GB). Second-phase particles (SPP) represent one of the levers for potential microstructural control. \\

Particle pinning occurs when a GB meets a SPP, reducing the total surface occupied by the GB and consequently lowering the total grain boundary energy. This phenomenon, well-known as Smith-Zener pinning mechanism, was first rationalised by Smith \cite{Smith1948} and then detailed by Zener one year later \cite{Zener1949}. They
proposed a model to exhibit an equivalent $P_{SZ}$ pinning pressure and a possible stagnated grain size in the event that this pressure exceeds the capillary pressure $P_c$. Indeed, in practise, under certain conditions, SPP can strongly pin a microstructure, eventually leading to a limiting mean grain size during recrystallization and grain growth. Since these first developments to equate this phenomenon, many variants have been developed in order to dispel some of the initial hypotheses \cite{Louat1982, Nishizawa1997, Equation1998} but also to propose new point of views concerning this mechanism \cite{Bignon2024}. Predictive tools are then needed to accurately model this phenomenon and thus optimize the final grain size and in-use properties of the materials. 

Since thirty-five years, numerous full field modeling of the Smith-Zener phenomenon have been proposed, including Monte Carlo / Cellular Automata frameworks \cite{Srolovitz1984,Anderson1984,Anderson1989,Hassold1990,Gao1997,Kad1997,Phaneesh2012}, front-tracking or vertex \cite{Weygand1999, Couturier2003}, multi-phase fields \cite{Chang2009, Tonks2015, Moelans2006, Chang2014} and level-set approaches \cite{Agnoli2012, Agnoli2014a, Agnoli2015, Scholtes2016b, Villaret2020, Alvarado2021a, Alvarado2021b, Bernacki2024}.  Front-capturing approaches like level-set (LS) and multi-phase field (MPF) methods can reproduce more realistic contexts. Indeed, these approaches allow for easier and more accurate description of the actual shapes of second-phase particles, as well as the local interaction between precipitate/grain interfaces and GB. In the LS framework, the concept of incorporating inert SPP within a finite element (FE) framework was initially proposed for conducting 2D and 3D grain growth (GG) simulations \cite{Agnoli2012, Scholtes2016b, Villaret2020}, static recrystallization simulations \cite{Agnoli2014a} and also extend in order to take into account evolving SPP populations \cite{Alvarado2021a, Alvarado2021b}. This approach enables the consideration of SPPs without predefined assumptions about their size or morphology. It accommodates both isotropic and anisotropic particle/grain interface energies, regardless of whether the interfaces are coherent or incoherent. This approach will be considered here as a reference for discussing the obtained results with the new proposed front-tracking description of Smith-Zener pinning mechanism. The LS formulation developed by Alvarado et al. \cite{Alvarado2021a, Alvarado2021b} was used for all the LS simulations conducted in this article. The need to develop alternative approaches to LS or MPF methods indeed arises their generally prohibitive computational cost when large statistically representative 2D simulations or SPP with small sizes are aimed.  This last factor might restrict the mesoscopic study of Smith-Zener pinning mechanism, and thus it is important to explore alternative models with higher performances for an equivalent precision. \\

If vertex/front-tracking approaches are of prime interest to improve computational cost of representative full-field simulations, their use in context of Smith-Zener pinning modeling remains limited. Based on its own Vertex model \cite{Weygand1998First}, the first Vertex attempt was proposed by Weygand et al. in 1998 by considering the SPP as static vertices and a pinning position for the evolving GB \cite{Weygand1999}. An unpinning force was proposed in agreement with the Smith-Zener model in order to model unpinning events. Other simulations with the same model and heterogeneous SPP populations were proposed in \cite{Lepinoux2010}. While this approach allows for efficiency, it can obviously be questioned in terms of representativeness of the mechanism when the size of the particles is not negligible compared to the size of the grains, as often encountered in many materials and thermomechanical conditions. \\

Concerning existing front-tracking frameworks based on an explicit description of the SPP, one can cite the methodology developed in \cite{Couturier2003}, where a FE model, based on a variational formulation for grain boundary motion by viscous drag, is used to solve the equations governing GB motion of an arbitrary-shaped surface and its interaction with SPP in 3D. The model was later extended in \cite{Couturier2004, Couturier2005, Couturier2003} to take into account motion by curvature flow but in the context of a single grain boundary and led to impressive simulations of Smith-Zener pinning mechanisms compared to the Vertex state-of-the-art. A similar discussion was recently proposed in \cite{Mohles2020} with comparable simulations, but a different approach to describe the interaction between SPP and GB allowing to consider a richer description of SPP in terms of shape and energy relationship with the matrix. However, the existing methodologies remain then limited to the study of one or few grain interfaces in the context of static SPP and without considering intragranular properties.\\

This article presents the application of SPP-GB interaction in context of a front-tracking Lagrangian model. The basis of this Lagrangian model have been introduced in previous works \cite{Florez2020, Florez2020b, Florez2020c, Florez2020d, Florez2021a, Florez2021b}. Evolution of this modeling approach will be presented in the first part, while computational comparisons with the LS-FE model will be detailed in the second part. In the third part, discussions concerning the comparison and the complementarity with classical Vertex description of the Smith-Zener pinning phenomenon will be proposed. Limits and perspectives of this new framework will be detailed in the last part.

\section{Numerical Method in context of discretized second phase particles}\label{sec:numericalmethod_6}

\begin{figure}[!h]
\centering
\includegraphics[width=1.0\textwidth] {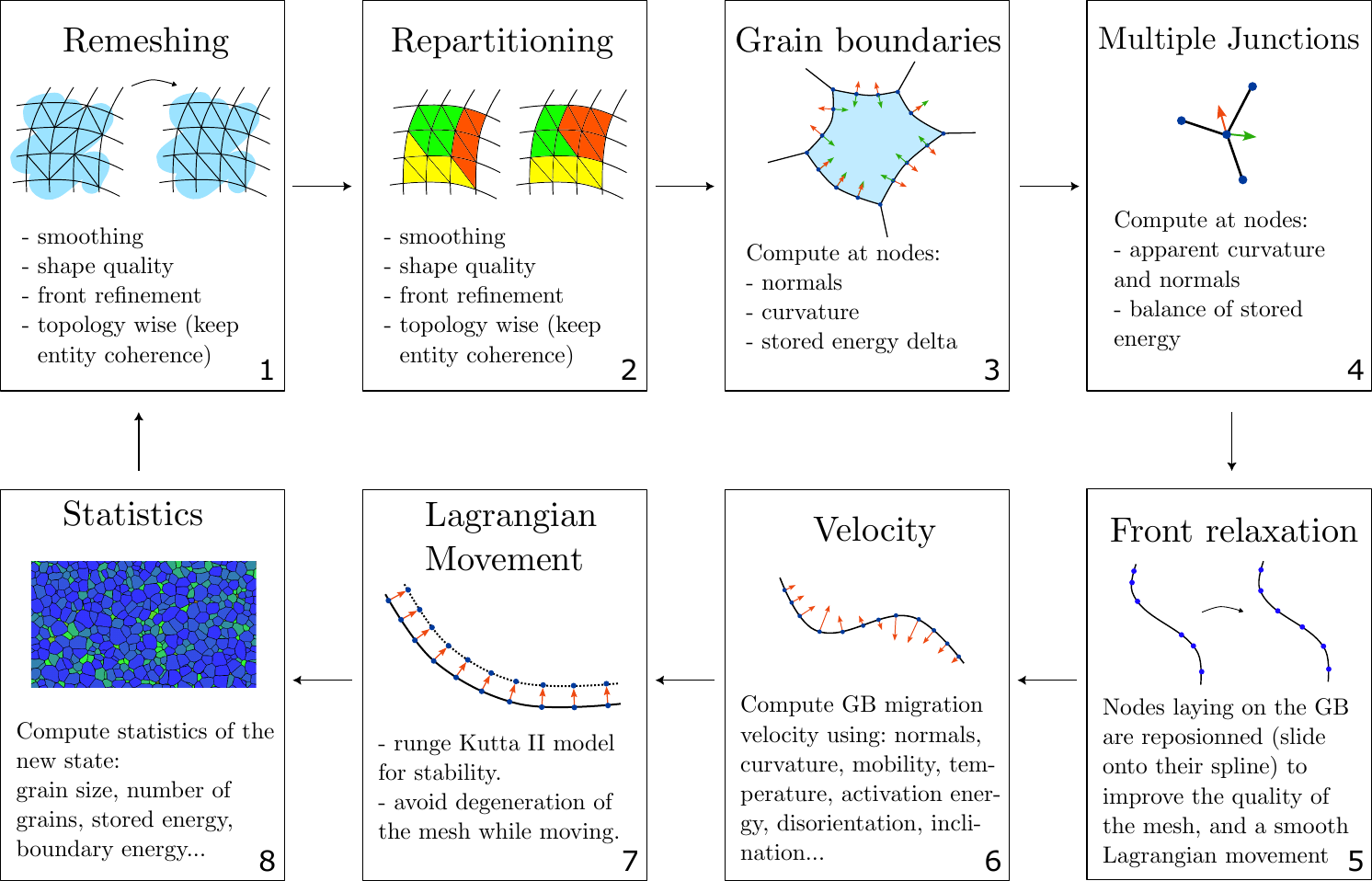}
\caption{Global loop for the TRM formulation \cite{Florez2020b}: 1. and 2. Remeshing procedure steps enabling to treat topological changes and maintains the mesh quality potentially with parallel computation \cite{Florez2020c}; 3. to 7. movement of interfaces using a Lagrangian scheme that updates the positions of the nodes of the mesh defining the GBs; and 8. output generation before the next time step.}
\label{fig:intro}
\end{figure}

In the context of 2D generic front-tracking models, the recent developments from Florez et al. concerning the ToRealMotion (TRM) code for "topological remeshing in Lagrangian framework for large interface motion" \cite{Florez2020b}, illustrated in Fig.\ref{fig:intro}, were validated for numerous metallurgical mechanisms. The TRM approach introduces the concept of using unstructured FE meshes for the detailed representation of grain interiors. The method consists of the movement of interfaces using a Lagrangian scheme that updates the positions of the nodes of the mesh defining the GBs (Figs.\ref{fig:intro}.3 to \ref{fig:intro}.7), while a remeshing procedure treats topological changes and maintains the mesh quality (Figs.\ref{fig:intro}.1 and \ref{fig:intro}.2). This strategy was adopted for several key reasons: first, to incorporate intragranular data such as stored energy to model nucleation and recrystallization; second, to accurately simulate significant domain deformations and discontinuous dynamic recrystallization mechanisms \cite{Florez2020d}; and third, to enhance the parallel processing capabilities of the method, surpassing those of conventional front-tracking models \cite{Florez2020c}. This method was also improved, considering the treatment of multiple junctions proposed by Barrales Mora \cite{BarralesMora2010}, to deal with anisotropic reduced mobility \cite{Florez2021a} and torque terms \cite{Florez2022} (Fig.\ref{fig:intro}.4). In this section, a methodology will be presented to apply the TRM model to SPP-GB interactions.\\

\subsection{Particles of ideal shape and initial discretization}

This 2D model uses circles to idealize and track the discretized SPP interfaces. During these simulations, on top of the data structure of the TRM model presented in \cite{Florez2020b}, a list of circles (defined by the position of their center and their radius {{$x_i,y_i$}, $r_i$}) is added. These circles are tracked until their radius $r_i=0$ (in case of evolutive SPP) or until the end of the simulation.\\

\begin{figure}[!h]
\centering
\includegraphics[width=1.0\textwidth] {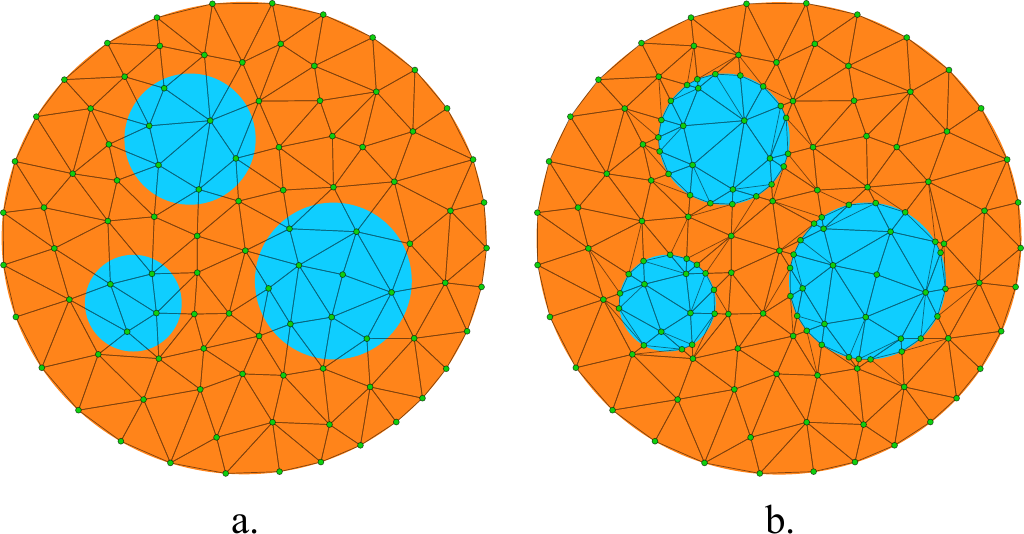}
\caption{Example of the immersion of circular particles (cyan) in a meshed domain (orange). a. Initial mesh, b. mesh after the remeshing defining the boundaries of particles with an explicit mesh.}
\label{Figure_Example_InitialRem}
\end{figure}

Of course, being a Lagrangian model, it is necessary to explicitly define the SPP-GB interfaces using a body-fitted mesh, for which, similarly to \cite{Florez2020d}, a remeshing procedure has been constructed. Initially, all segments crossed by the boundaries of a circle, are cut using the \emph{edge-split} operator (see Fig. \ref{Figure_Example_InitialRem}). Then, using a \emph{surface reconstruction} procedure, \cite{Florez2020b}, the elements contained within the circular domain are extracted from their original surface\footnote{In the TRM context, a \emph{surface} is defined as a set of elements and nodes. Surfaces typically represent a \emph{grain}, but in this scope, they represent grains and particles} and are attributed to a new surface representing the discrete particle.\\

\subsection{Kinetics at discretized SPP-GB interfaces}

\begin{figure}[!h]
\centering
\includegraphics[width=1.0\textwidth] {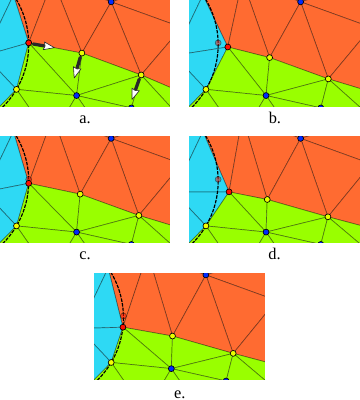}
\caption{Example of the cycle of the movement of a multiple junction in the boundary of a discretized particle (cyan). The initial position of the mutliple junction is displayed in red with an alpha component in all frames. a. Initial state and computation of the interfaces kinetics, b. displacement of the interfaces using the time-step and the velocity field, c. projection of the multiple junction of the meshed particle to the circle defining the particle domain, d. and e. new cycle, similar to b. and c.}
\label{Figure_Example_ProjectionMJ}
\end{figure}

The movement of multiple junctions is critical during the Smith-Zener pinning phenomenon, as they account for most of the SPP-GB interactions. In this model, we have used a free movement / projection approach (see Fig. \ref{Figure_Example_ProjectionMJ}): During a time step, nodes belonging to the discretized SPP-GBs interfaces are allowed to move freely according to their velocity $\vec{v}_i$. This velocity is computed using the methodologies presented in \cite{Florez2020b, Florez2020d, Florez2021a, Florez2021b}, then a projection of these nodes is made to the interfaces defined by the perfect circles. The projection is made to the nearest interface point (in this case, in the radial direction).\\

In this study, we calculate the velocity by assuming isotropic grain growth (homogeneous value in space of the mobility and grain boundary energy), leading to grain/grain/grain triple junctions achieving a Young's equilibrium at angles of $120^\circ$-$120^\circ$-$120^\circ$. Upon projection of these junctions onto the discretized SPP interface, if the SPP is large enough, the expected grain/SPP/grain equilibrium angles adjust to $90^\circ$-$180^\circ$-$90^\circ$. This presupposes that the chosen time step is sufficiently small to ensure gradual transitions.\\

Figure \ref{Figure_Example_ProjectionMJ} showcases the expected behavior of a multiple junction at a SPP interface, particularly near a curved GB. One can observed that, following a cycle of "migration-projection", the triple junction shifts from its initial location. This displacement is attributed to the angle made by the immediately adjacent edge of the GB being more obtuse towards the green grain compared to the orange grain, prompting the junction to move towards a more symmetrical stance. Repeating this cycle reveals a noticeable migration of the multiple junction, akin to sliding along the particle's surface.\\

\begin{figure}[!h]
\centering
\includegraphics[width=0.7\textwidth] {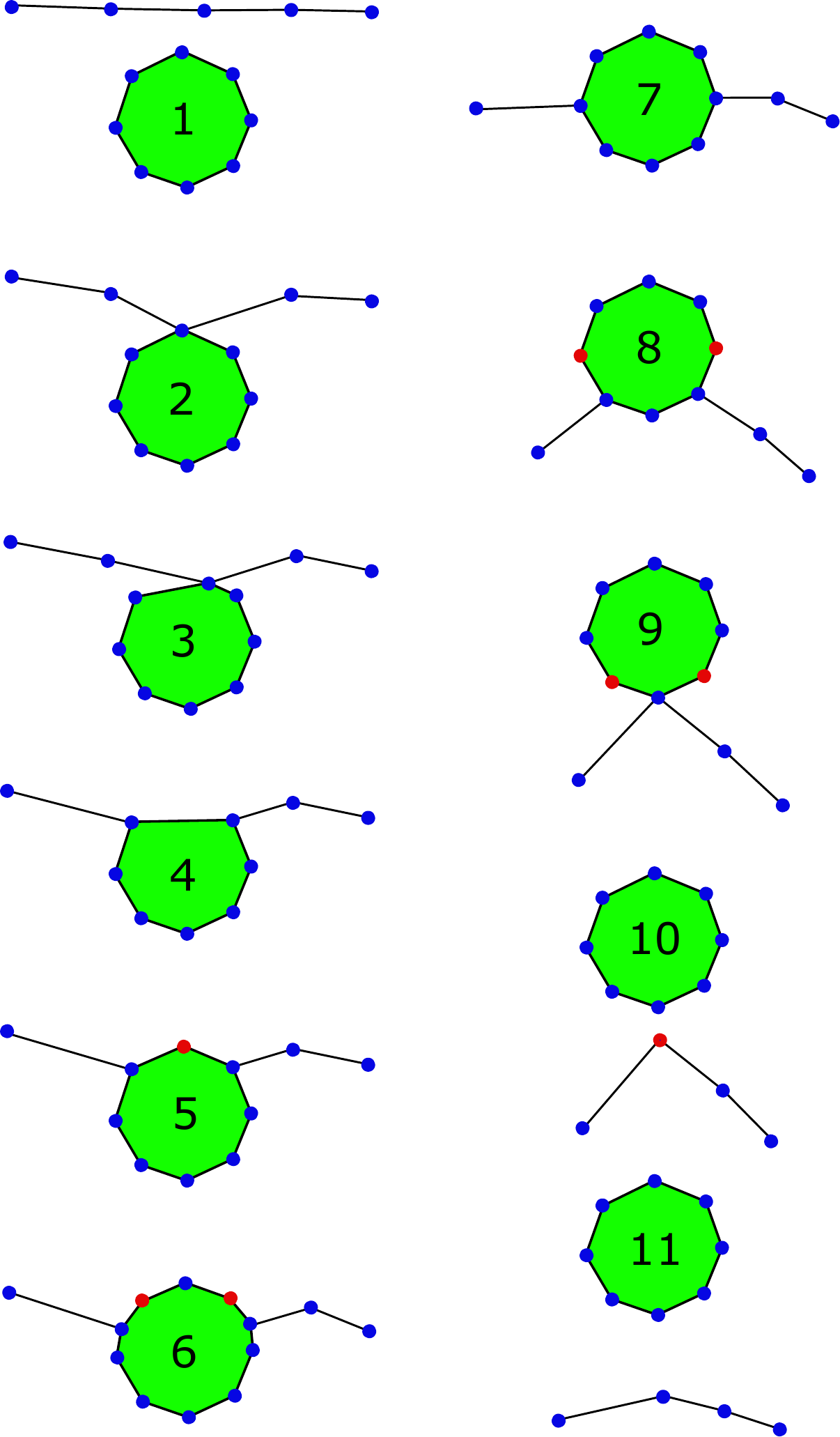}
\caption{Rough illustration of pinning and unpinning events of a GB in a discretized particle.}
\label{fig:unpinning}
\end{figure}

 More globally, the Figure \ref{fig:unpinning} illustrates in a rough way also the pinning and unpinning events of a GB in a discretized particle. For clarity, in this figure, only the nodal links between GB and particle boundaries are shown while the FE mesh is not. On every state, red nodes are new nodes (nodes not present in the previous state). Fig.\ref{fig:unpinning}-State 1 shows the initial state, the GB approaches the particle thanks curvature flow dynamics, Fig.\ref{fig:unpinning}-State 2 illustrates one of the nodes collapsing over its closest node on the particle to form a quadruple point. A higher curvature is now present at both sides of the GB. Fig.\ref{fig:unpinning}-State 3 depicts the GB advancing on its movement thanks to curvature flow and the dynamics of neighbouring nodes which are not shown in the picture, here the quadruple point moves to the right thanks to the balance of forces pulling in that direction, and the node slides over the imposed circular shape of the particle. Fig.\ref{fig:unpinning}-State 4 shows two nodes collapsing, one between the quadruple point and its neighbour, and the second, between other pair of GB node and particle node. Note how this collapse removes the quadruple point and transforms it into 2 triple junctions. Fig.\ref{fig:unpinning}-State 5 describes the appearance of a new node on the particle due to a segment in the particle (the one between the 2 triple junctions) being too long, the node splitting operation is used here. This node is positioned on the outline of the particle's circle. Fig.\ref{fig:unpinning}-State 6 shows two new splitting of the particle's outline, introducing 2 new nodes. Fig.\ref{fig:unpinning}-State 7 illustrates two collapsing operations between the triple junctions and the bottom neighbors. In fact, the GB pulls the triple junction to the bottom (due to the dynamics of neighboring nodes not shown in this figure) until they collapse with neighbors laying on the particle. The process in Fig.\ref{fig:unpinning}-States 6 and 7, continue (see states 8 and 9) until both sides of the GB collapse with the same node, to form, again, a quadruple point if the corresponding driving pressure is sufficient. The decomposition of this quadruple point leads to the unpinning of the GB from the particle in Fig.\ref{fig:unpinning}-State 10. Of course, since all these operations depend on the mesh size parameters imposed in the vicinity of the GB and particle, the unpinning events is linked to the local mesh size used on the GB. Once the GB is completely detached from the particle, the GB straightens up (Fig.\ref{fig:unpinning}-State 11). The process of unpinning is repeated for every GB-discretized particle pair in the simulation when it is possible (i.e. when the pinning is not stable). However, we have not observed this process to appear frequently as expected in 2D vertex simulations. Instead, the GBs in contact with particles move slower and become more stable than GBs not attached to any particle, producing that complete grains collapse over particles until a stable configuration is achieved in context of sufficient global Smith-Zener pinning pressure.\\

\begin{figure}[!h]
\centering
\includegraphics[width=0.8\textwidth] {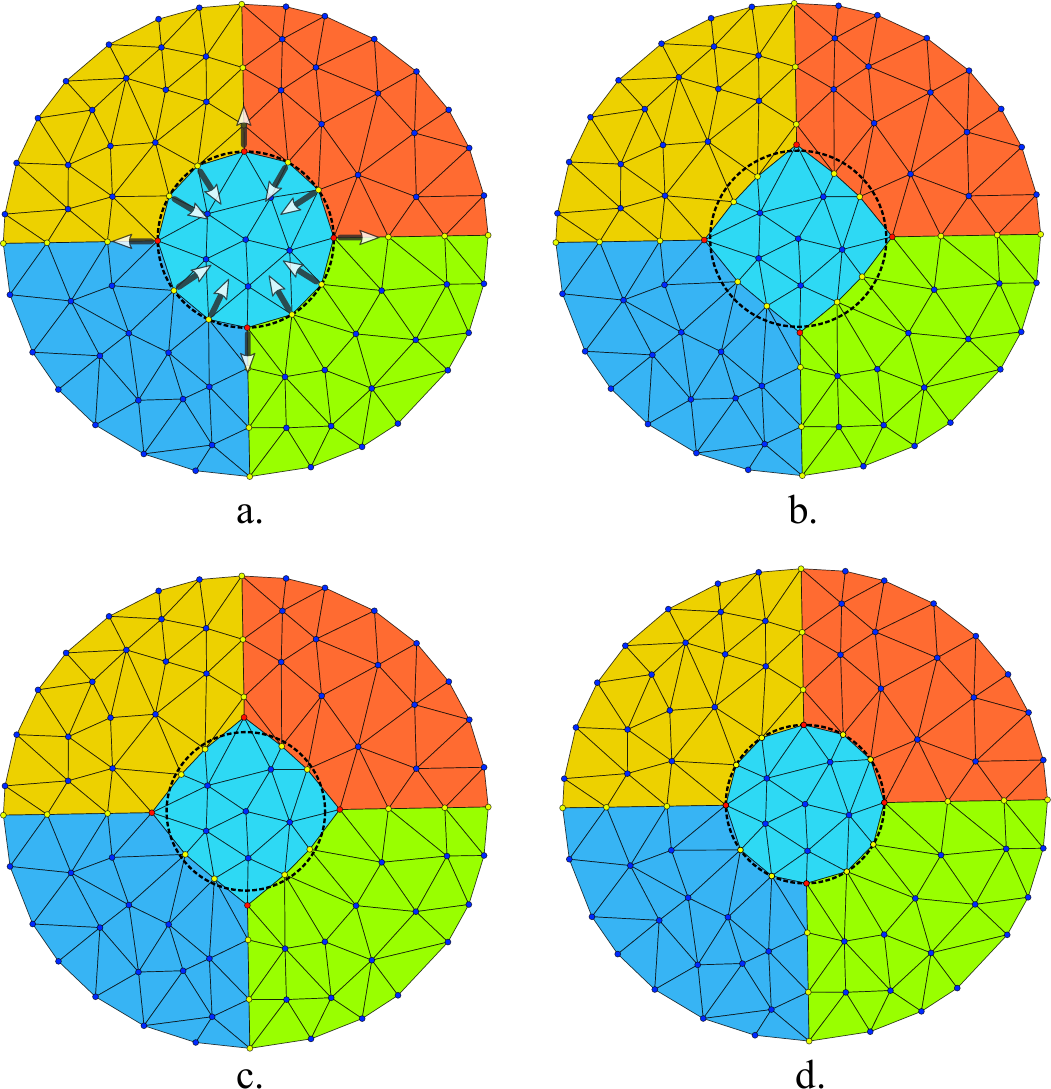}
\caption{Example of the life cycle of the evolution of a discretized particle. a. Computation of the dynamics of the interfaces, b. displacement of the interfaces using the time-step and velocity, c. transformation of the circle defining the particle, d. projection of the interfaces of the meshed particle to the circle defining the particle domain.}
\label{fig:Figure_Example_ParticuleEvolution}
\end{figure}

Finally, the TRM model also allows the evolution of particles by altering the radius $r_i$ of a circle that defines each discretized particle. Figure \ref{fig:Figure_Example_ParticuleEvolution} demonstrates this process over a single time step: initially, the model computes velocities and allows for the free movement of nodes, as depicted in Figures \ref{fig:Figure_Example_ParticuleEvolution}a and \ref{fig:Figure_Example_ParticuleEvolution}b. Subsequently, the circle that outlines the particle is updated (decreased in size in the example given by Figure \ref{fig:Figure_Example_ParticuleEvolution}c). The final step involves the projection procedure, illustrated in Figure \ref{fig:Figure_Example_ParticuleEvolution}d, which completes the particle evolution process within the model framework.

\section{Numerical results and validation in context of discretized second phase particles}\label{sec:numericalresults_6}
\subsection{Grain growth simulations}\label{testTRM1}

\begin{figure}[!h]
  \centering
  \begin{subfigure}{0.49\textwidth}
    \centering
    \includegraphics[scale=0.123]{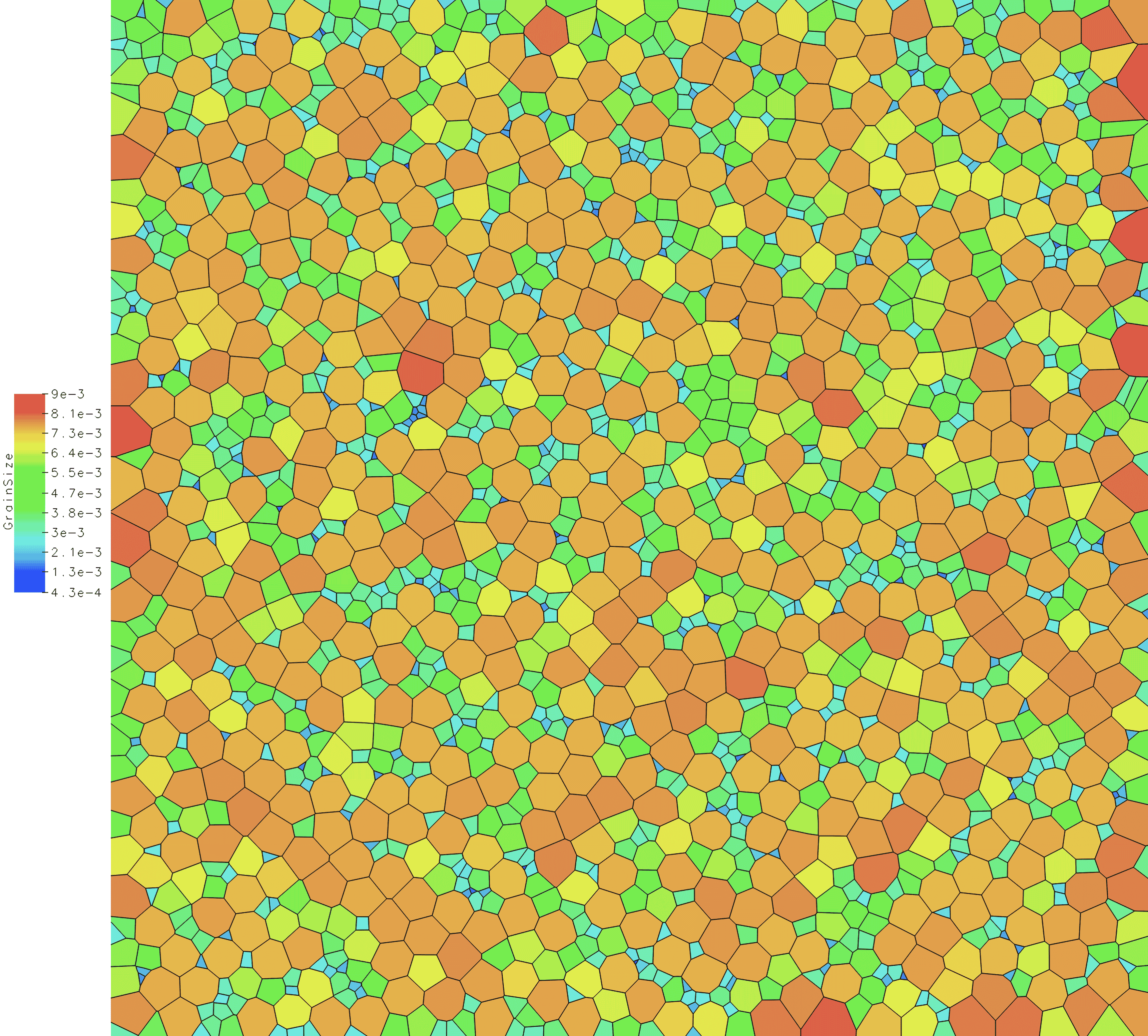}
    \label{fig:EBSDa}
  \end{subfigure}
  \begin{subfigure}{0.49\textwidth}
    \centering
    \includegraphics[scale=0.123]{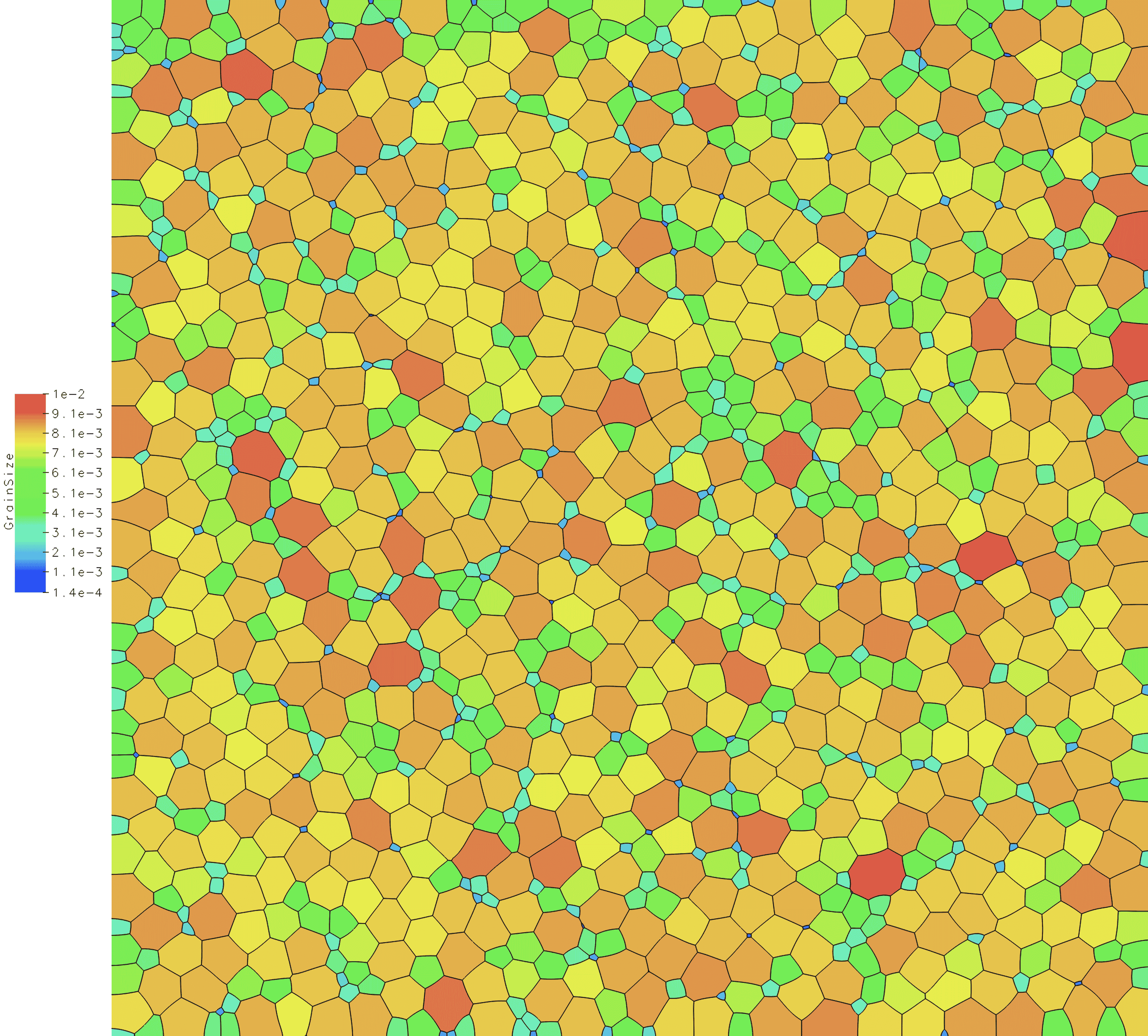}
    \label{fig:EBSDb}
  \end{subfigure}
   \begin{subfigure}{0.49\textwidth}
    \centering
    \includegraphics[scale=0.123]{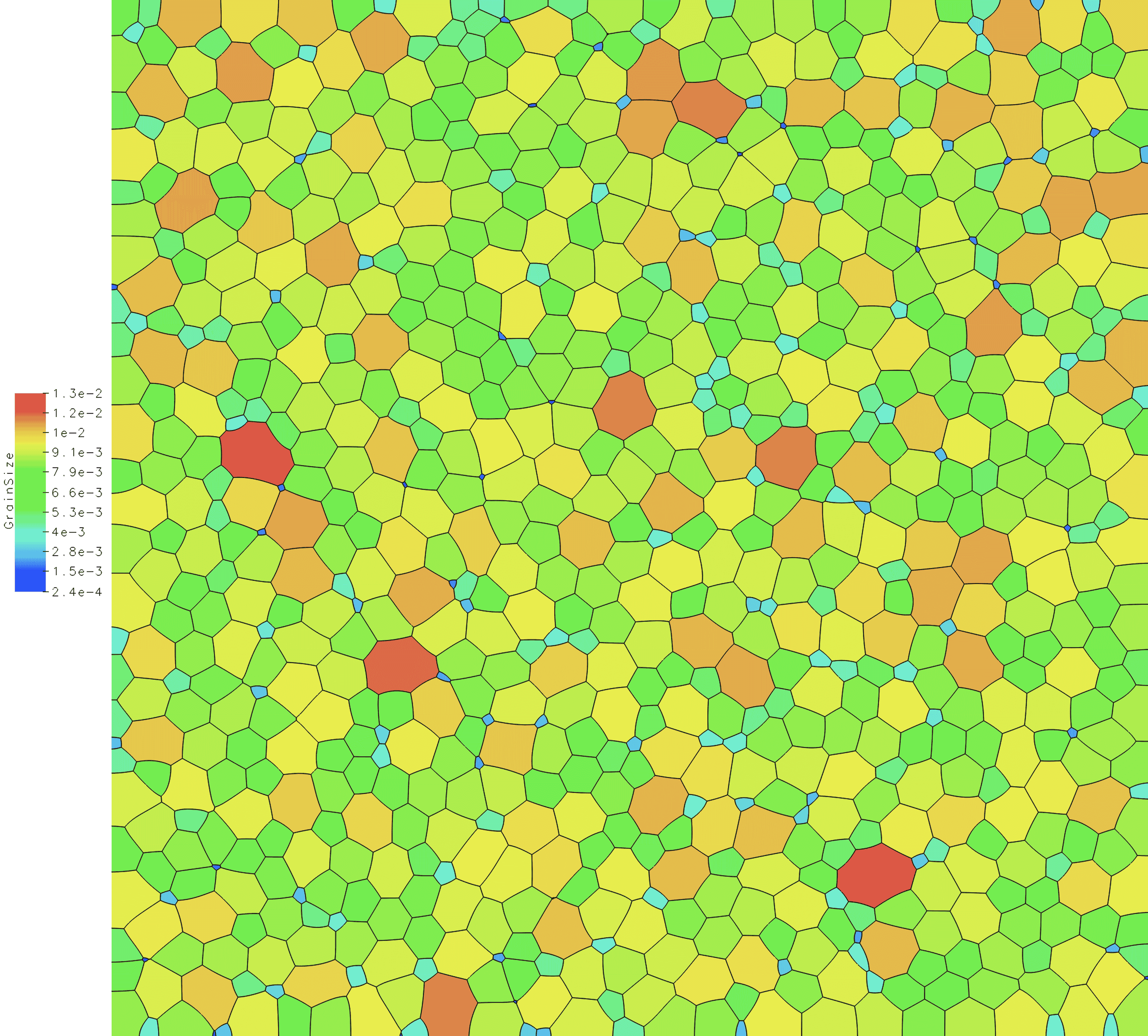}
    \label{fig:LVTMa}
  \end{subfigure}
  \begin{subfigure}{0.49\textwidth}
    \centering
    \includegraphics[scale=0.123]{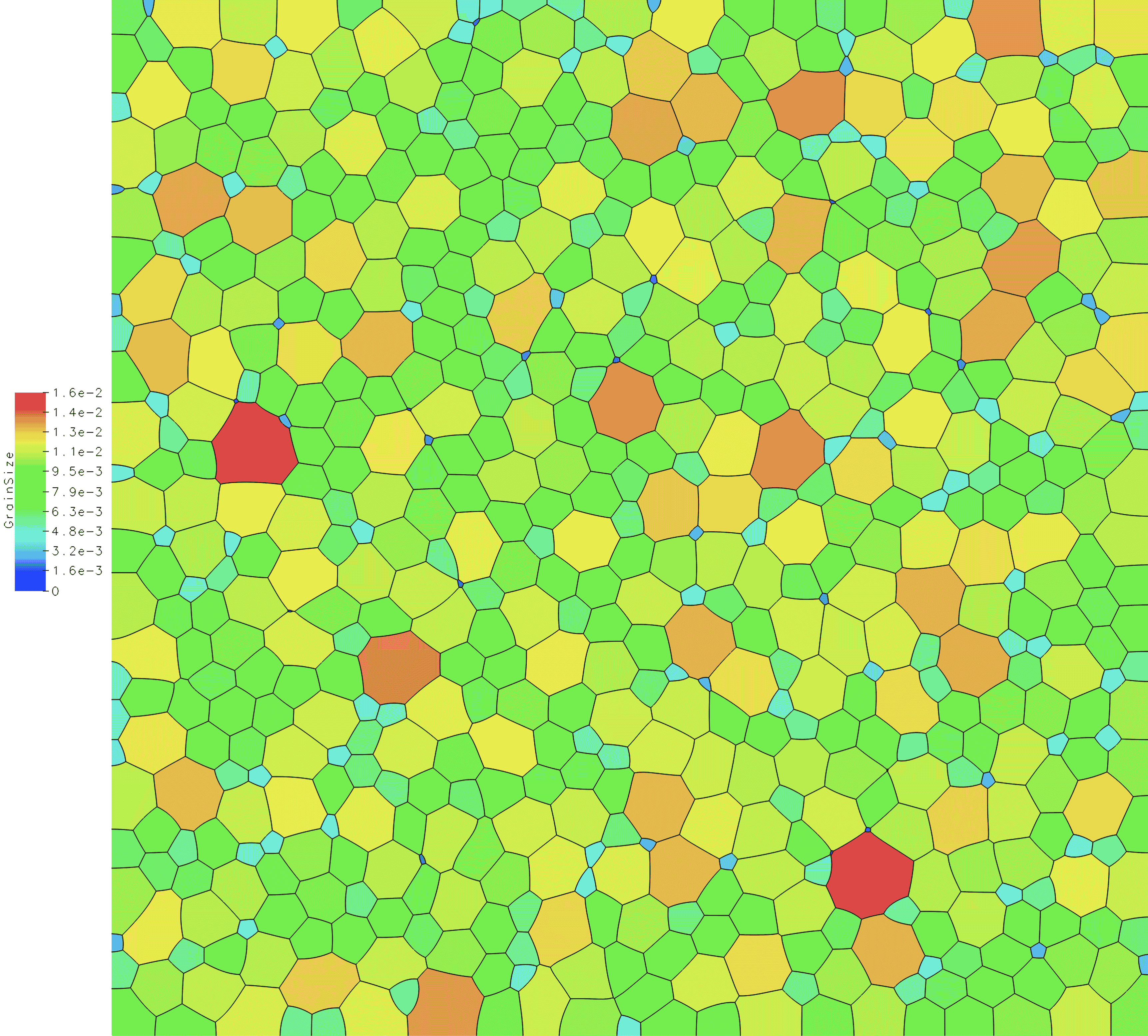}
    \label{fig:LVTMb}
  \end{subfigure}
\begin{subfigure}{0.49\textwidth}
    \centering
    \includegraphics[scale=0.123]{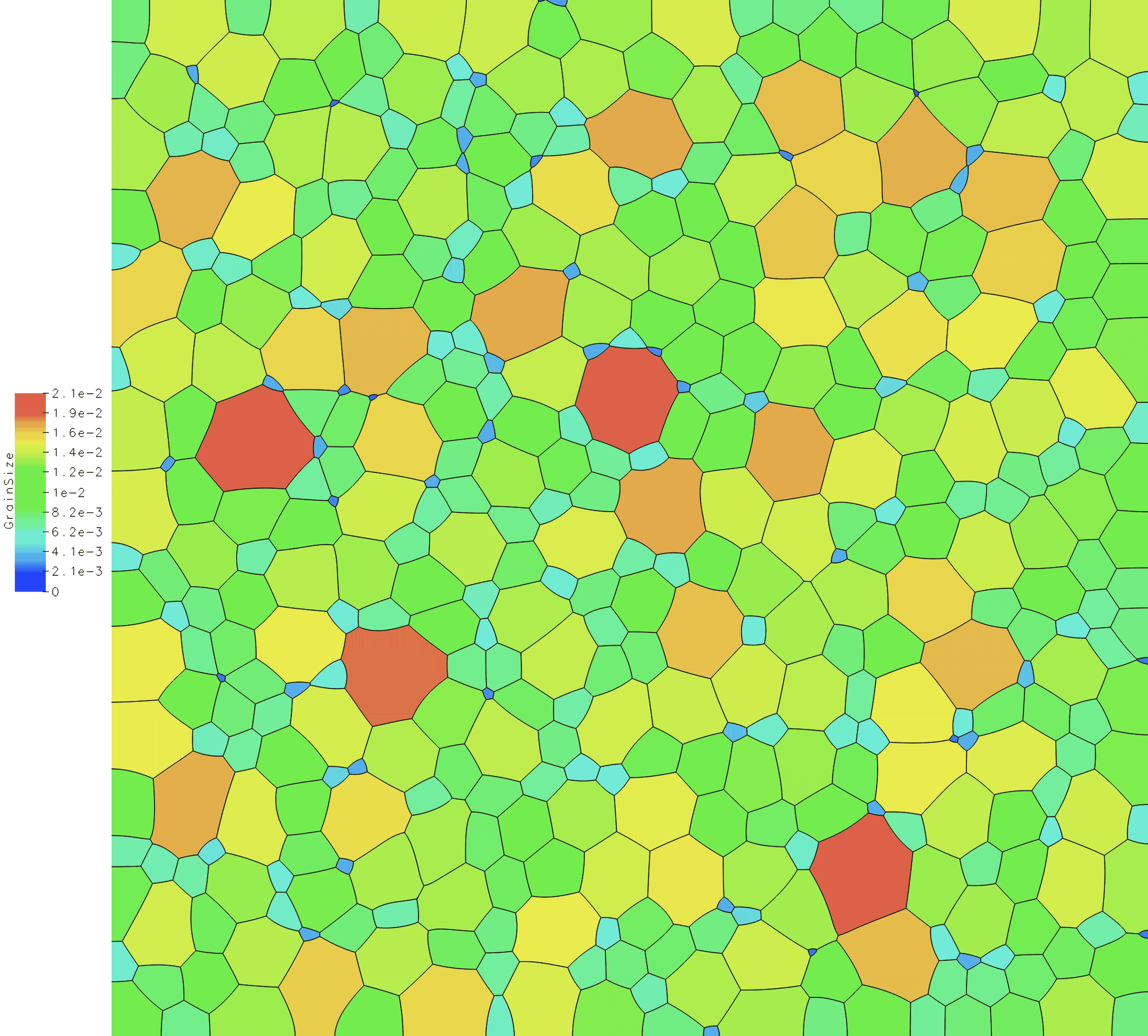}
    \label{fig:LVTMa}
  \end{subfigure}
  \begin{subfigure}{0.49\textwidth}
    \centering
    \includegraphics[scale=0.123]{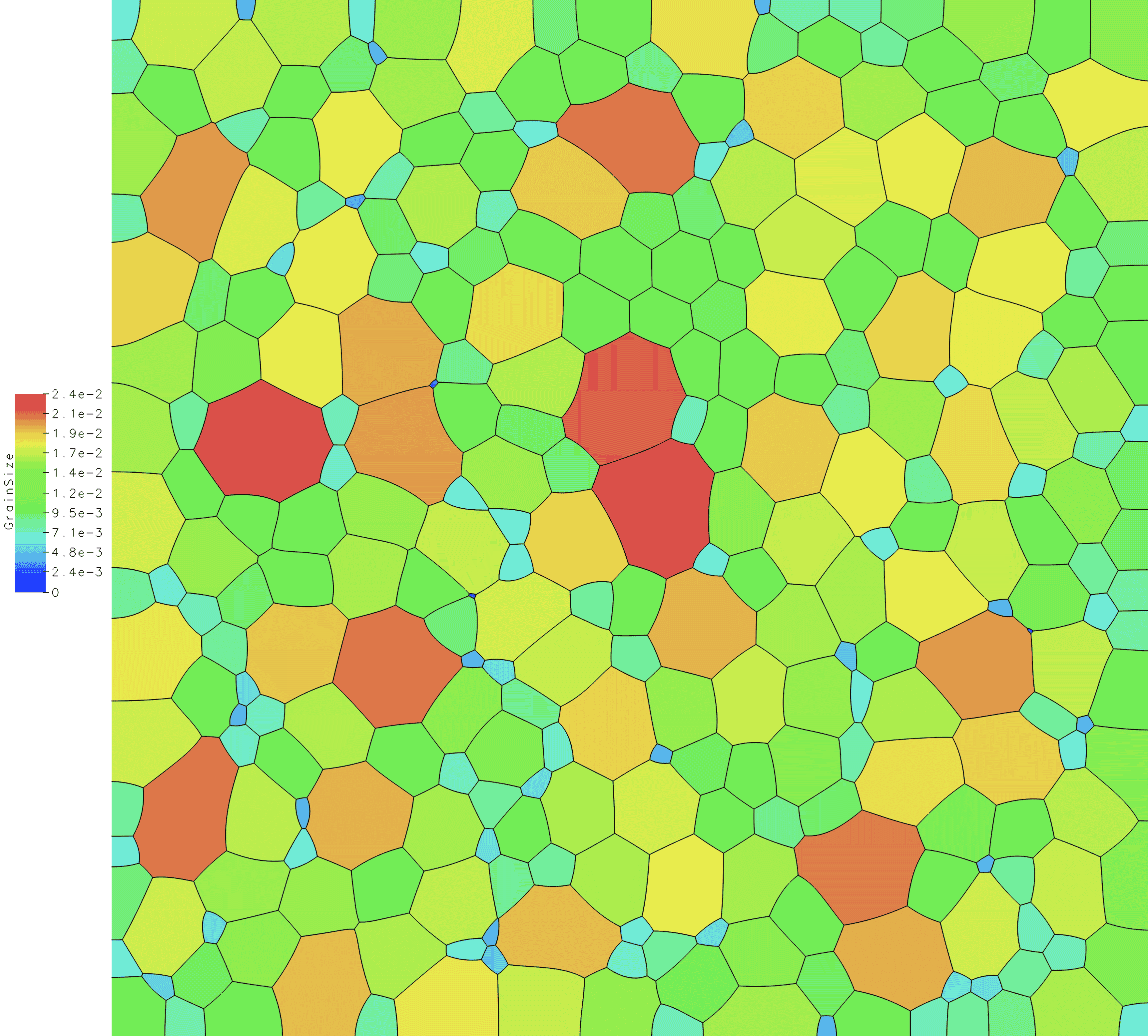}
    \label{fig:LVTMb}
  \end{subfigure}
  \caption{From left to right and top to bottom: microstructure evolution during an isothermal treatment ($T=$\SI{1060}{\celsius}) of 3h for AD730 alloys without considering the SPP; $t=\SI{0}{\second}$, $t=\SI{10}{\minute}$, $t=\SI{30}{\minute}$, $t=\SI{1}{\hour}$, $t=\SI{2}{\hour}$ and $t=\SI{3}{\hour}$ states are depicted. The field corresponds to the ECR in \SI{}{\milli\meter}.    }
  \label{fig:NoSPP}
\end{figure}

\begin{figure}[!h]
  \centering
  \begin{subfigure}{0.49\textwidth}
    \centering
    \includegraphics[scale=0.123]{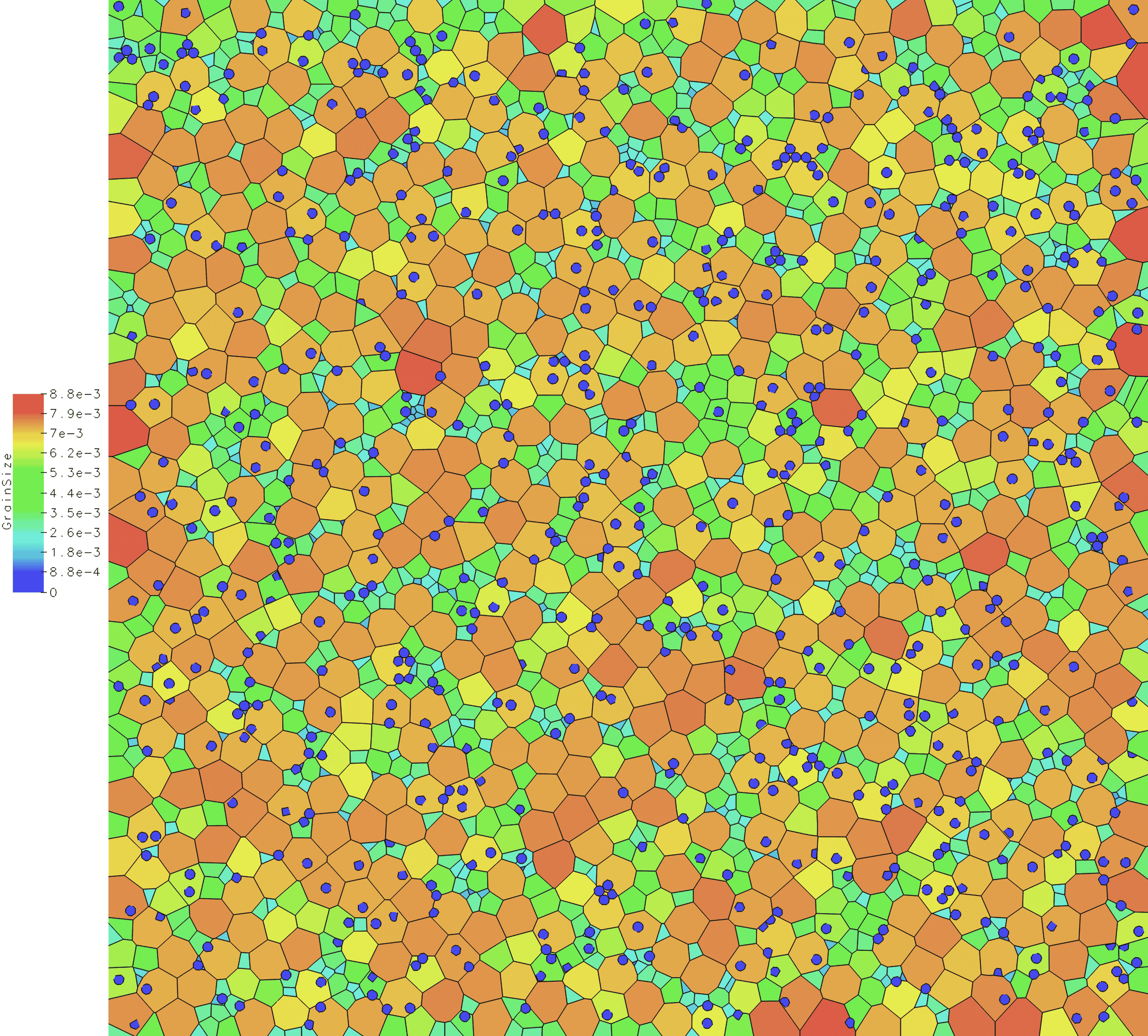}
    \label{fig:EBSDa}
  \end{subfigure}
  \begin{subfigure}{0.49\textwidth}
    \centering
    \includegraphics[scale=0.123]{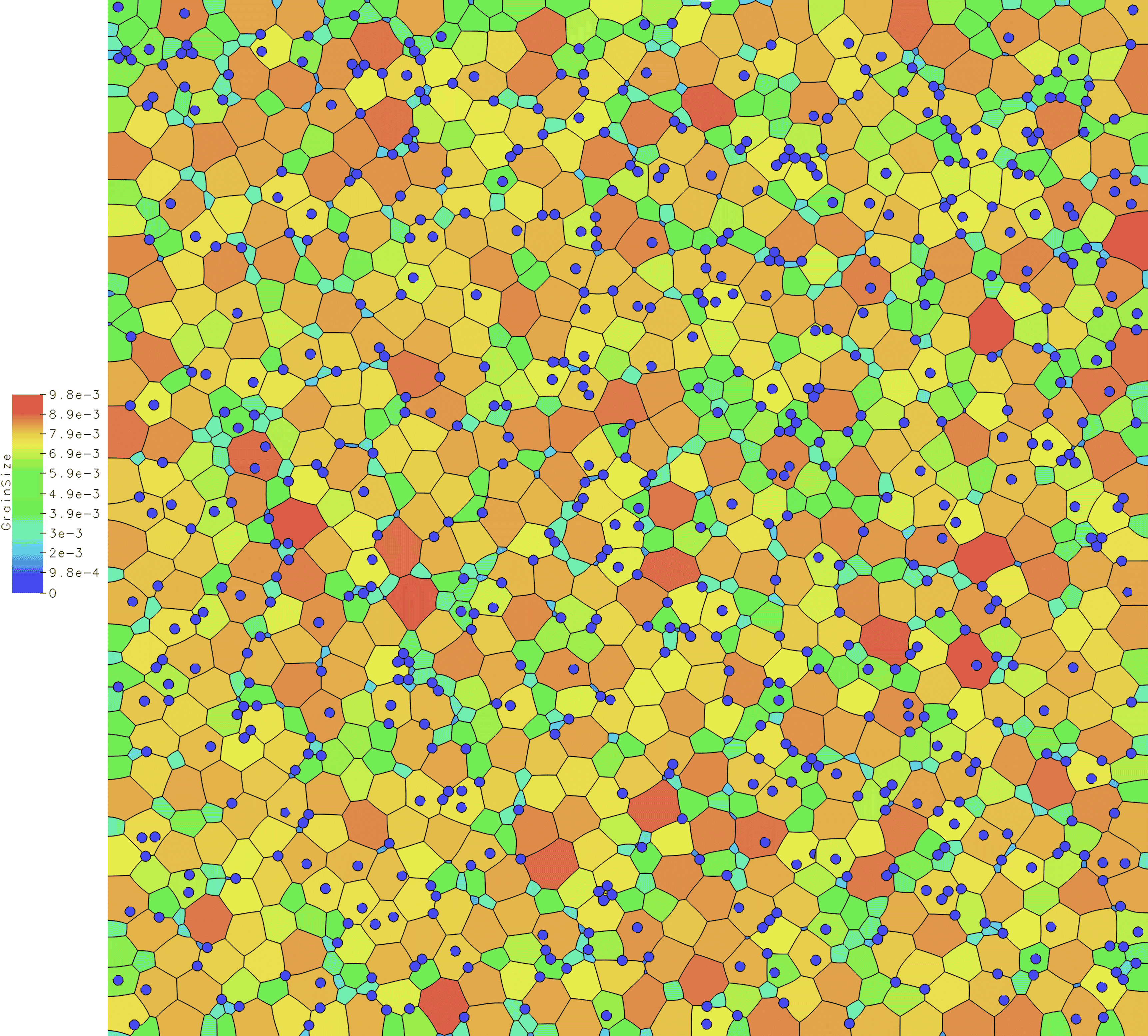}
    \label{fig:EBSDb}
  \end{subfigure}
   \begin{subfigure}{0.49\textwidth}
    \centering
    \includegraphics[scale=0.123]{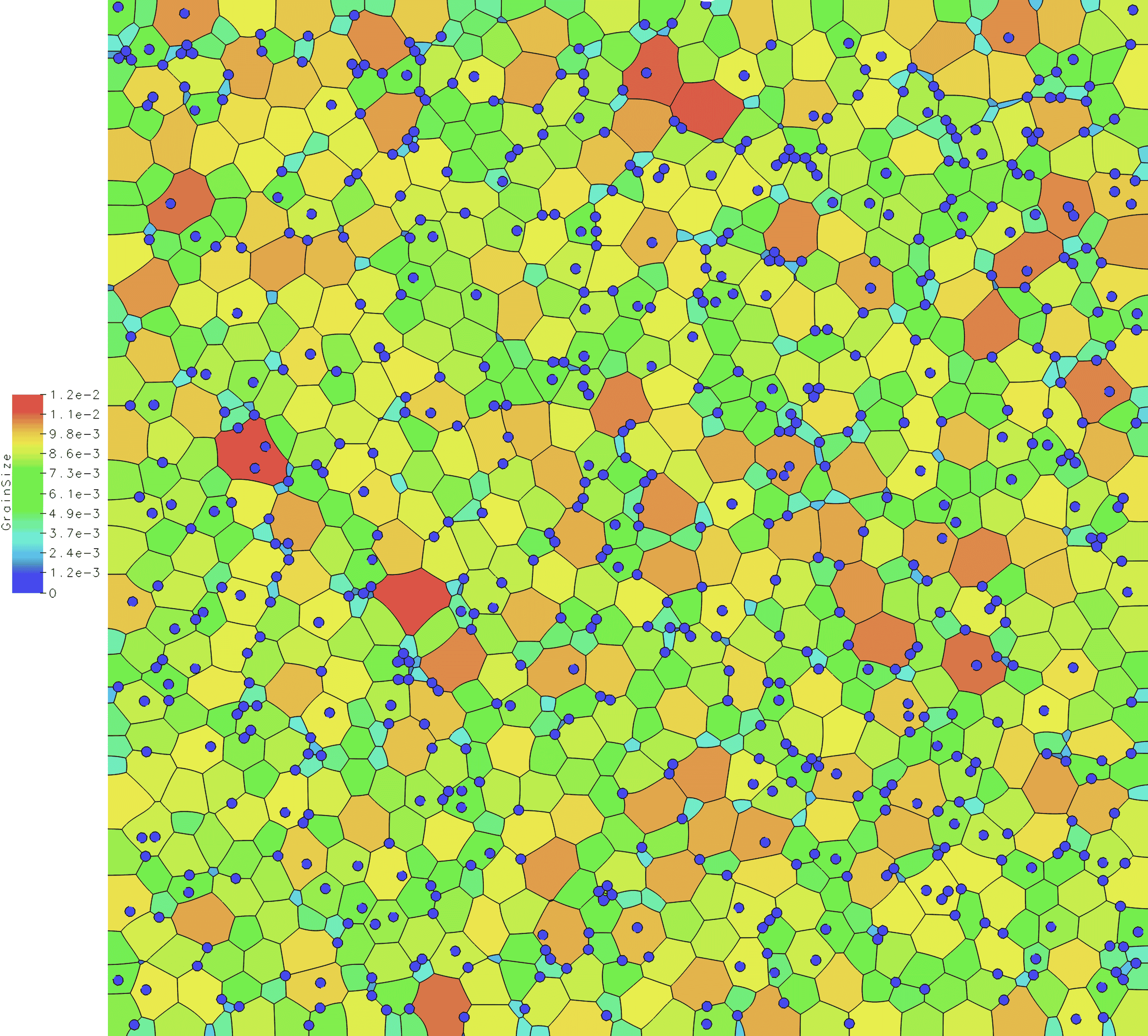}
    \label{fig:LVTMa}
  \end{subfigure}
  \begin{subfigure}{0.49\textwidth}
    \centering
    \includegraphics[scale=0.123]{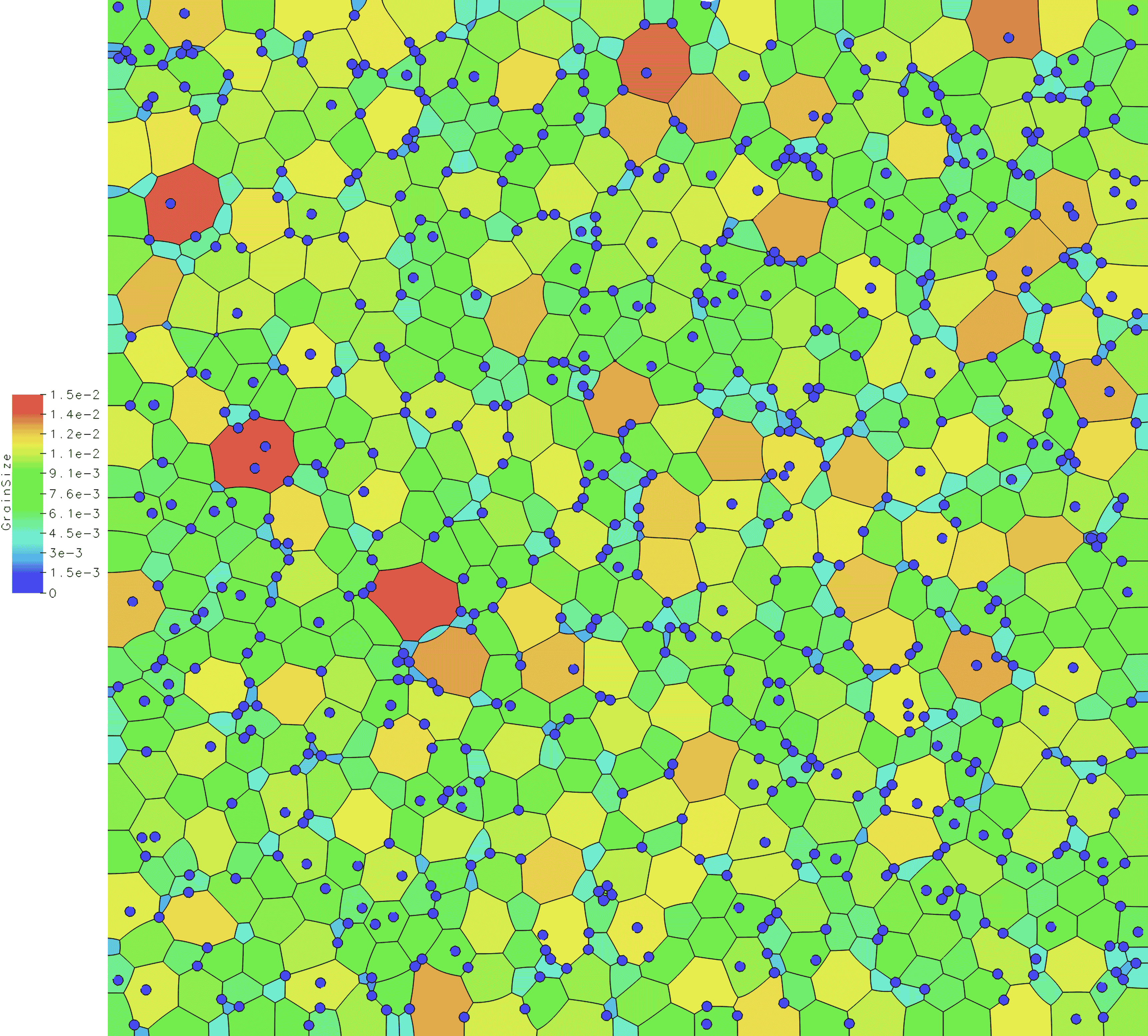}
    \label{fig:LVTMb}
  \end{subfigure}
\begin{subfigure}{0.49\textwidth}
    \centering
    \includegraphics[scale=0.123]{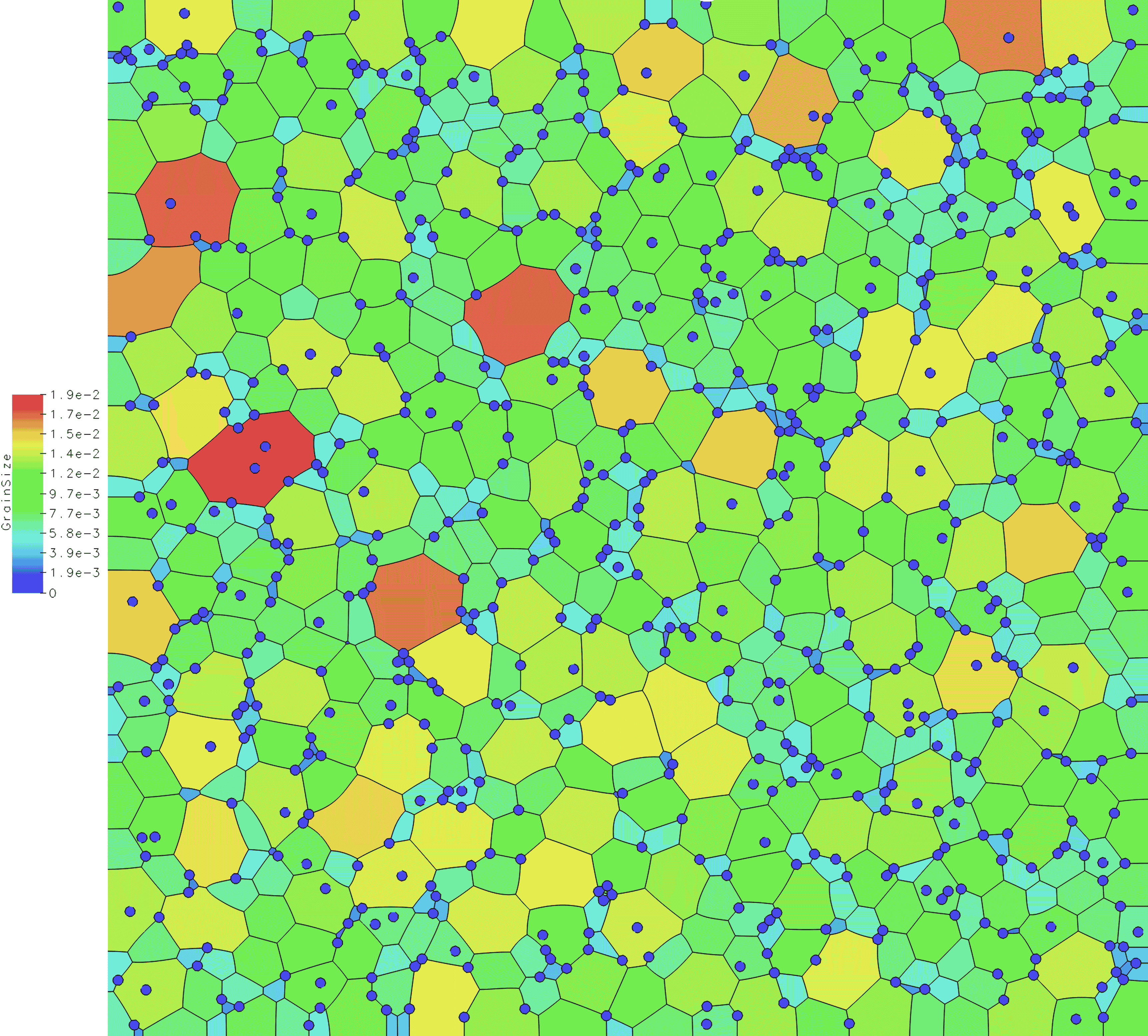}
    \label{fig:LVTMa}
  \end{subfigure}
  \begin{subfigure}{0.49\textwidth}
    \centering
    \includegraphics[scale=0.123]{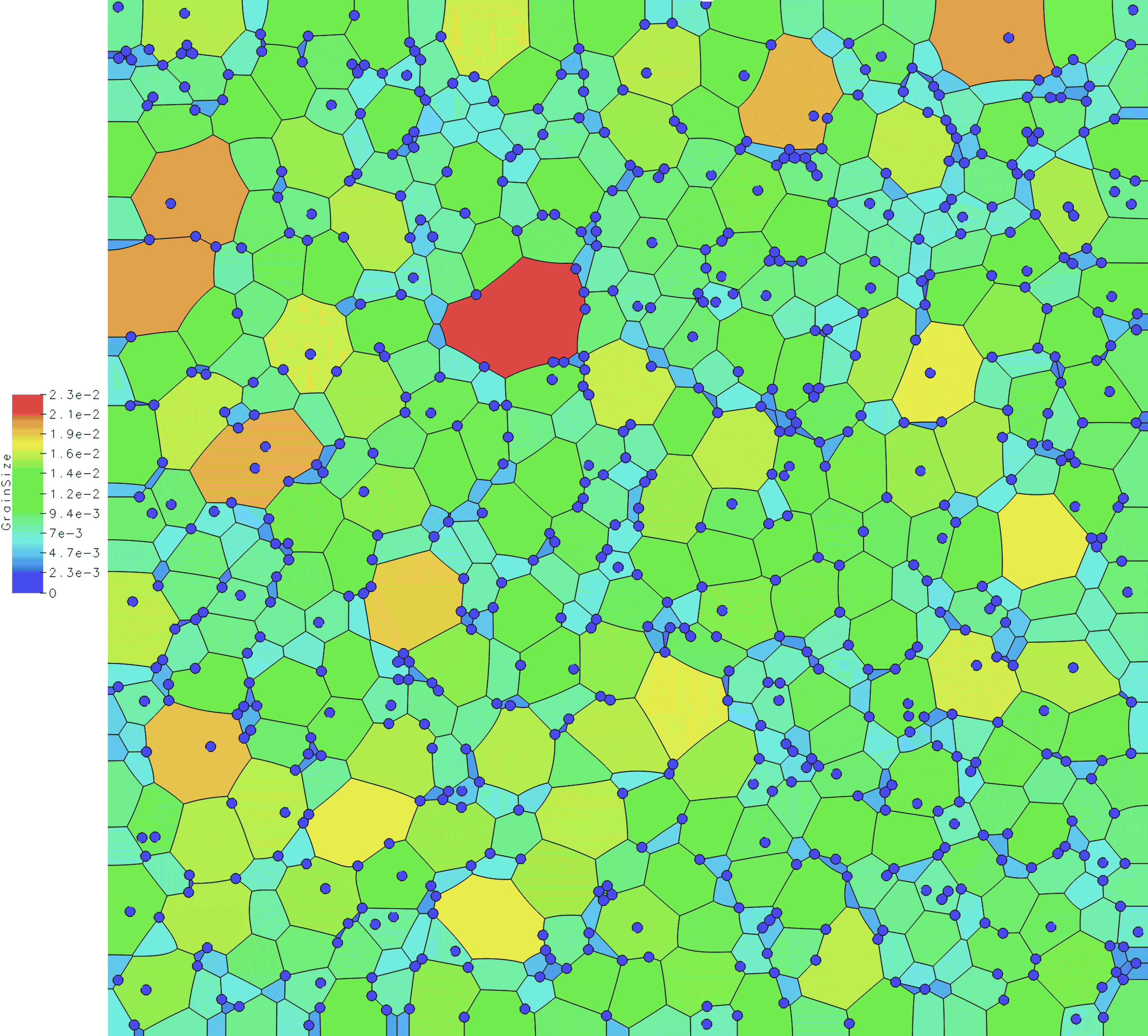}
    \label{fig:LVTMb}
  \end{subfigure}
  \caption{From left to right and top to bottom: microstructure evolution during an isothermal treatment ($T=$\SI{1060}{\celsius}) of 3h for AD730 alloys by considering static SPP; $t=\SI{0}{\second}$, $t=\SI{10}{\minute}$, $t=\SI{30}{\minute}$, $t=\SI{1}{\hour}$, $t=\SI{2}{\hour}$ and $t=\SI{3}{\hour}$ states are depicted. The field corresponds to the ECR in \SI{}{\milli\meter}.    }
  \label{fig:SSPP}
\end{figure}

\begin{figure}[!h]
  \centering
  \begin{subfigure}{0.49\textwidth}
    \centering
    \includegraphics[scale=0.123]{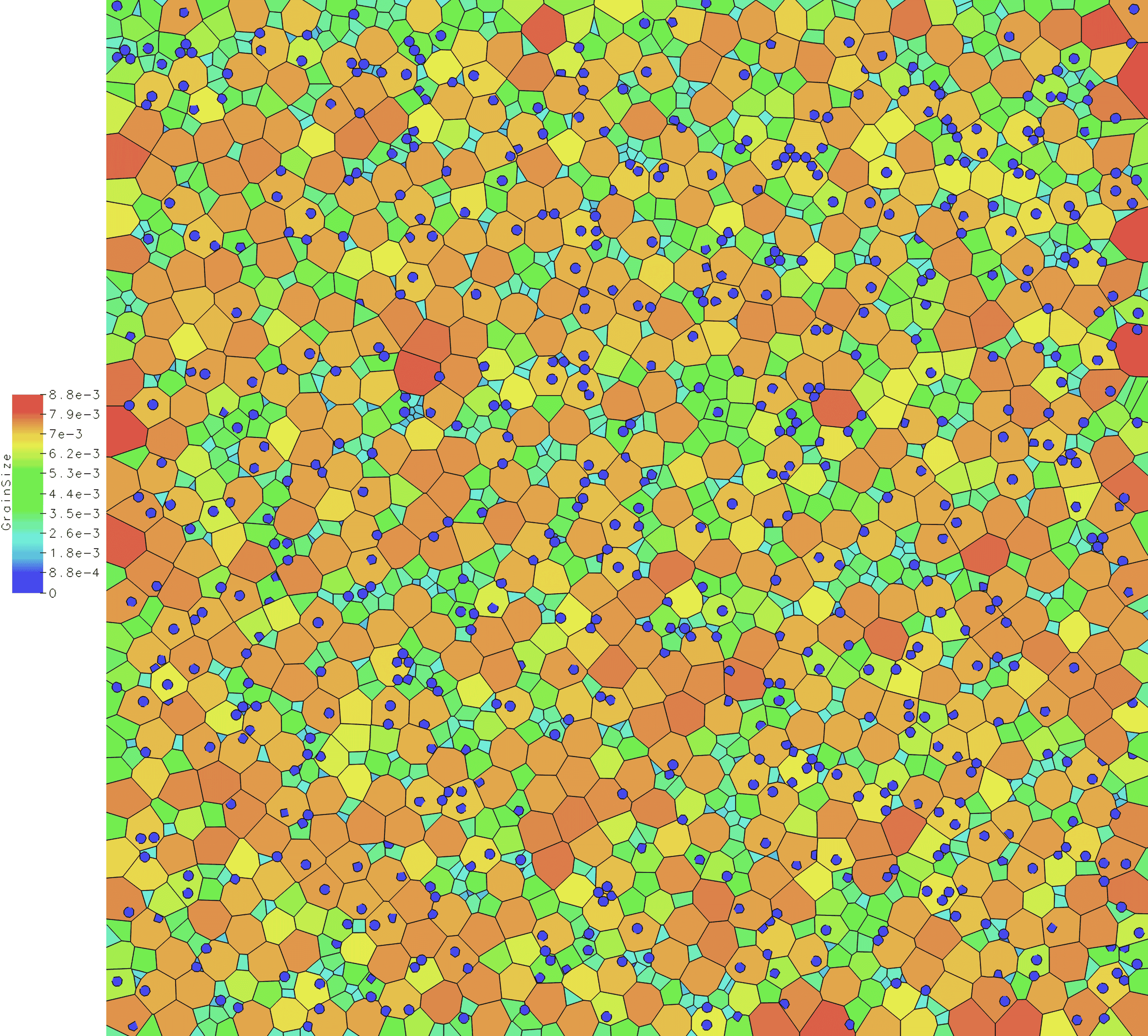}
    \label{fig:EBSDa}
  \end{subfigure}
  \begin{subfigure}{0.49\textwidth}
    \centering
    \includegraphics[scale=0.123]{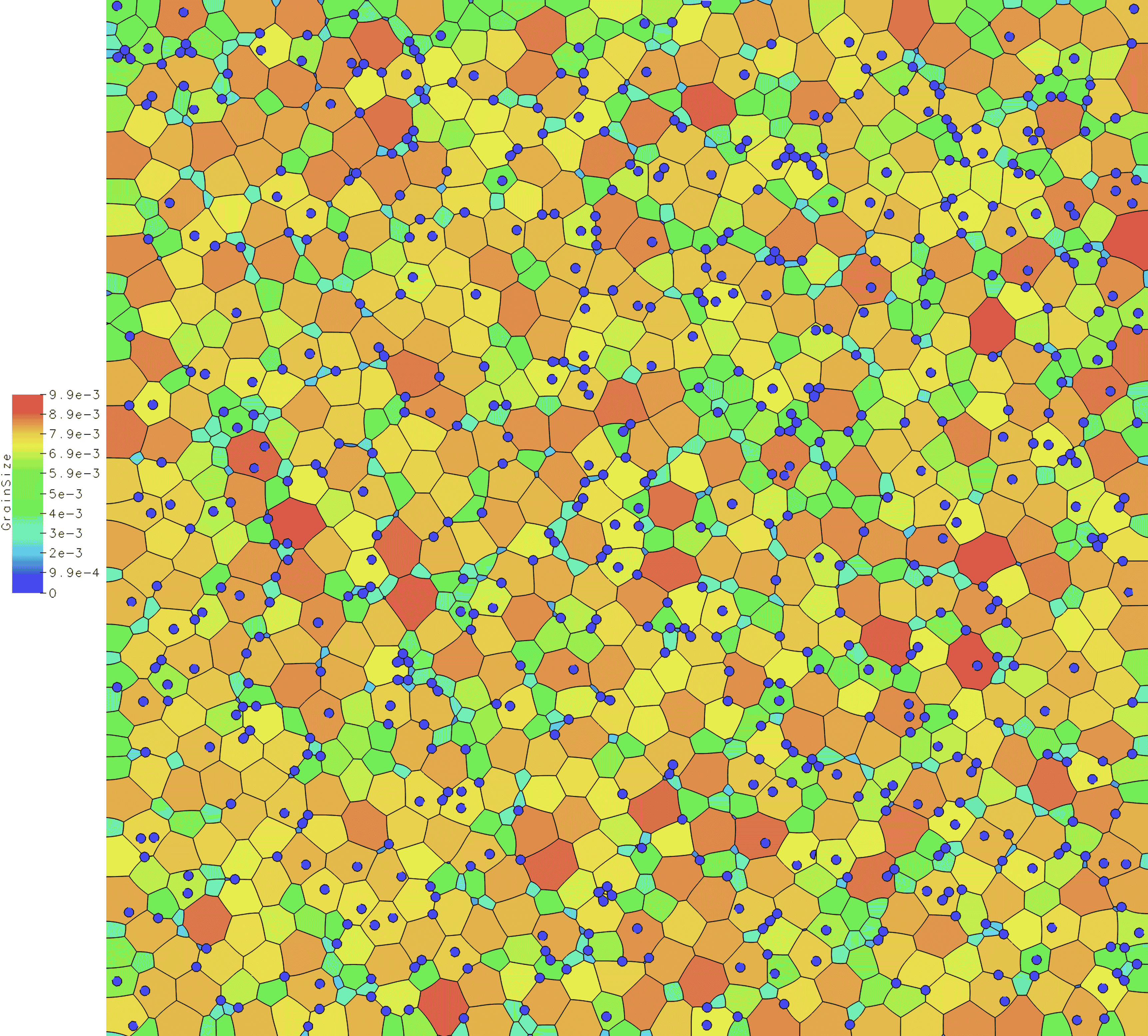}
    \label{fig:EBSDb}
  \end{subfigure}
   \begin{subfigure}{0.49\textwidth}
    \centering
    \includegraphics[scale=0.123]{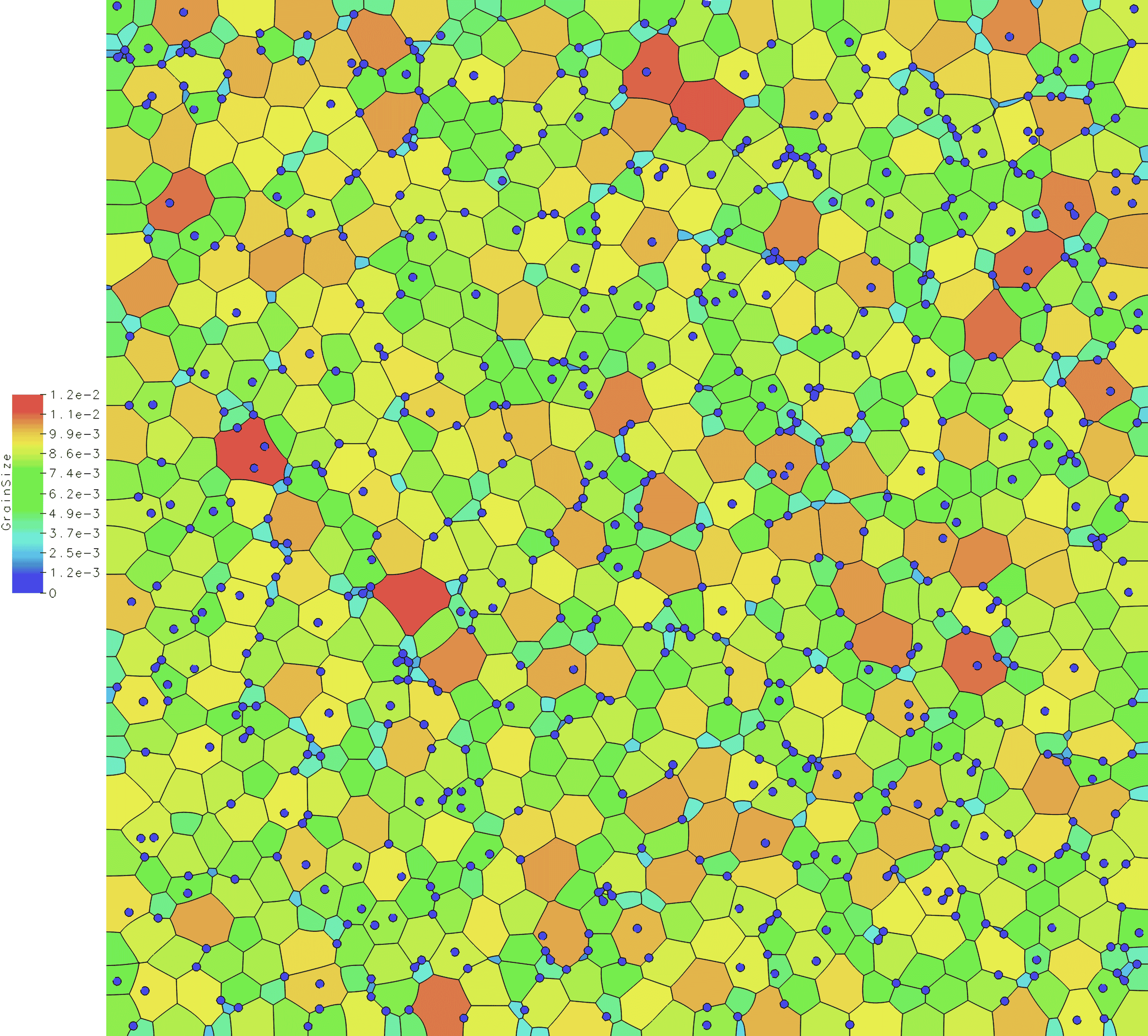}
    \label{fig:LVTMa}
  \end{subfigure}
  \begin{subfigure}{0.49\textwidth}
    \centering
    \includegraphics[scale=0.123]{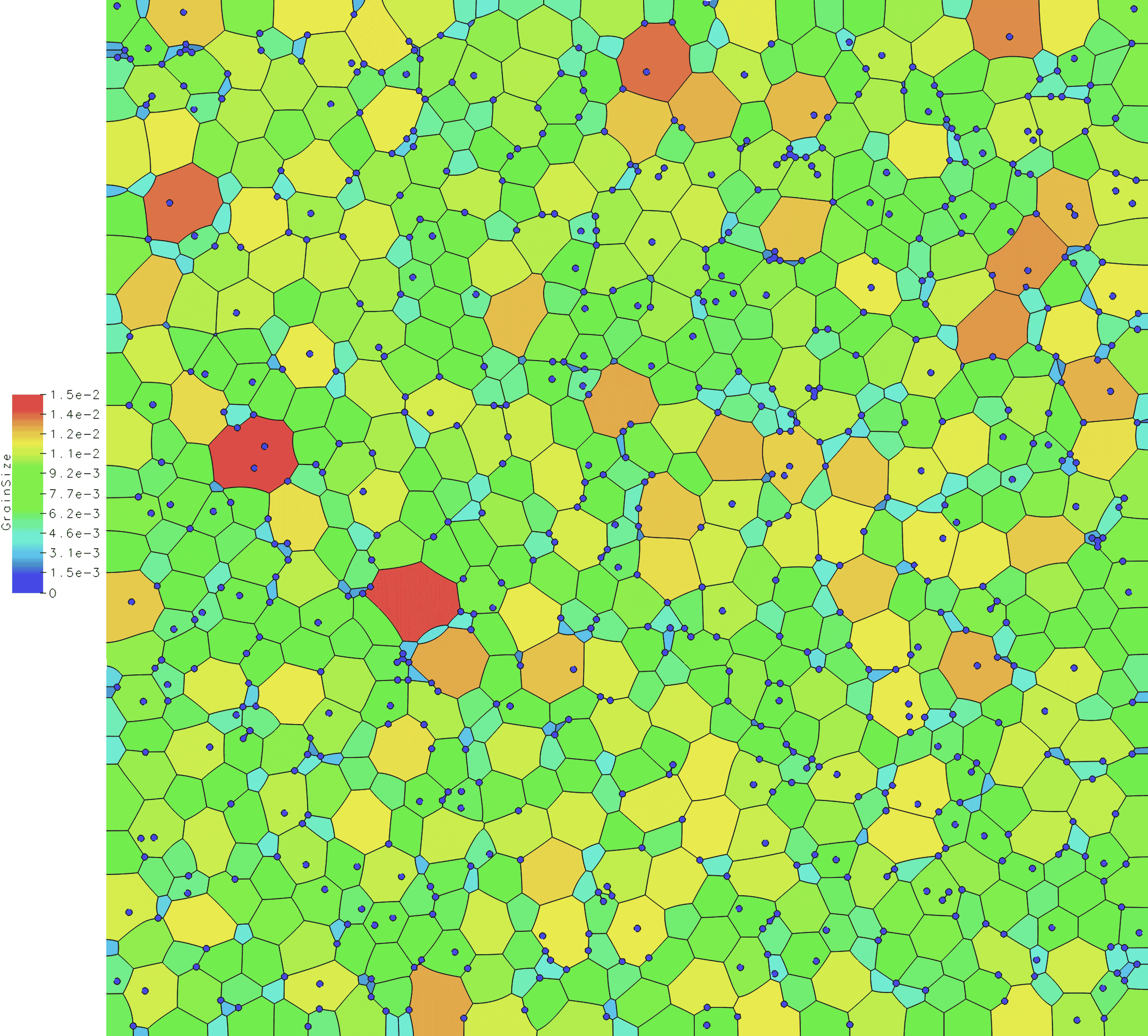}
    \label{fig:LVTMb}
  \end{subfigure}
\begin{subfigure}{0.49\textwidth}
    \centering
    \includegraphics[scale=0.123]{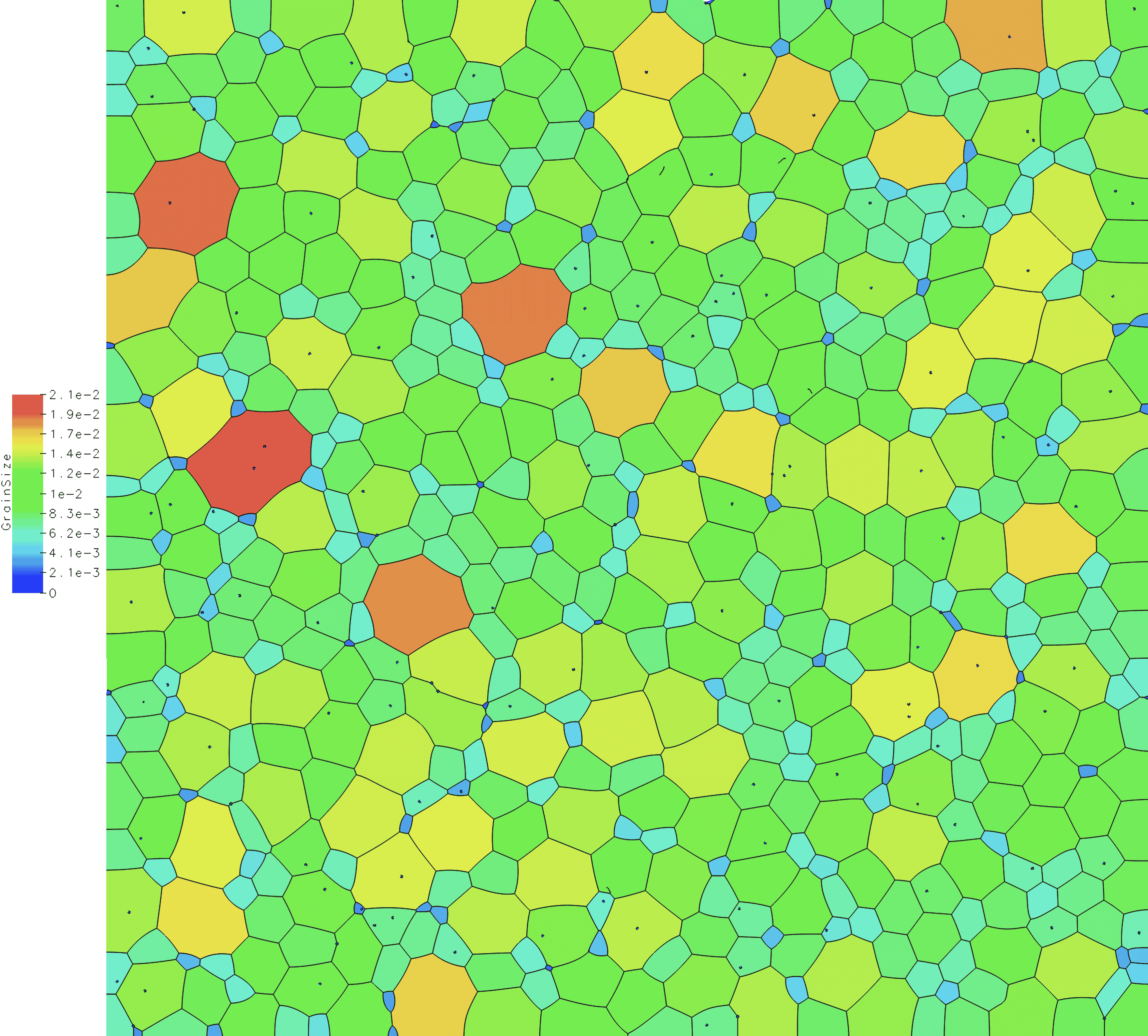}
    \label{fig:LVTMa}
  \end{subfigure}
  \begin{subfigure}{0.49\textwidth}
    \centering
    \includegraphics[scale=0.123]{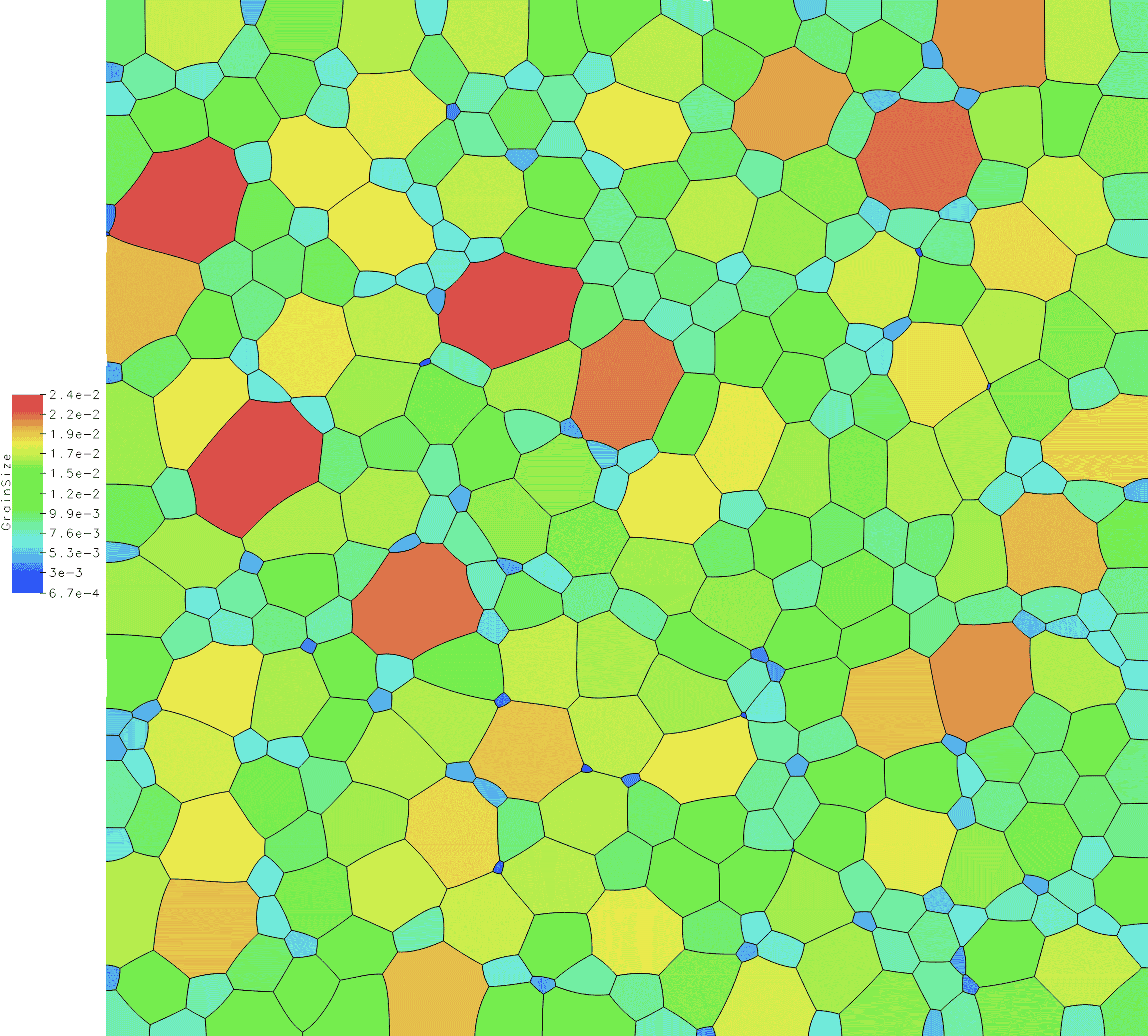}
    \label{fig:LVTMb}
  \end{subfigure}
  \caption{From left to right and top to bottom: microstructure evolution during an isothermal treatment ($T=$\SI{1060}{\celsius}) of 3h for AD730 alloys by considering static SPP; $t=\SI{0}{\second}$, $t=\SI{10}{\minute}$, $t=\SI{30}{\minute}$, $t=\SI{1}{\hour}$, $t=\SI{2}{\hour}$ and $t=\SI{3}{\hour}$ states are depicted. The field corresponds to the ECR in \SI{}{\milli\meter}.    }
  \label{fig:ESPP}
\end{figure}

The first set of simulations consists of four domains of 0.4 $mm$ x 0.4 $mm$ with different Laguerre-Voronoi tessellations \cite{Hitti2012} with around 1800 initial grain for each domain. The grain size is defined here at the equivalent circle radius or diameter (ECR or ECD), i.e. the radius or the diameter of the circle with the same area that the considered grain. The initial ECR distribution is imposed through a lognormal distribution ($\mu=2$\SI{7}{\micro\meter} and $\sigma=$\SI{7.6}{\micro\meter}). A monodisperse and spherical precipitate population, with an initial surface fraction $f_{spp}=5\%$ and an arithmetic mean radius $\overline{r}_{spp}=\SI{2}{\micro\meter}$ (around 768 SPP), is considered. All SPP are assumed incoherent.
As detailed in \cite{Florez2020b}, the classical curvature flow kinetic is verified for all GB:
\begin{equation}\label{eq:kinetic}
\vec{v}=-\mu\gamma\kappa\vec{n},
\end{equation}
with $\mu$ the GB mobility, $\gamma$ the GB energy, $\kappa$ the GB curvature, and $\vec{n}$ the outside unitary normal to the GB. As already highlighted, $\mu$ and $\gamma$ are assumed to be homogeneous in space. Concerning $\mu$, an Arrhenius law is assumed for its dependance to the temperature, when non-isothermal treatments are considered:
\begin{equation}\label{eq:arrhenius}
\mu\left(T\right)=\mu_0 e^{-Q/RT},
\end{equation}
with $\mu_0$ a constant pre-exponential term, $R$ the perfect gaz constant, $Q$ the activation energy for grain growth, and $T$ the absolute temperature. In the following of this section, representative values for the $AD730$ nickel base superalloy are used \cite{alvarado2021dissolution} ($\mu_0=\SI{2.9e37}{\milli\raiseto{4}\metre\per\second\per\joule}$, $Q=\SI{9.8e5}{\joule\per\mole}$, and $\gamma=\SI{0.6}{\joule\per\square\meter}$) and an isothermal heat treatment of 3 hours (10800s) at $T=$\SI{1060}{\celsius} is modeled. In context of evolving second phase particles, the precipitate interface velocity was set to $v_{spp}=\SI{2e-7}{\milli\meter\per\second}$ oriented towards the center of the SPP (dissolution context). It must be highlighted that these first simulations were designed to be comparable to the results proposed and validated in \cite{alvarado2021dissolution}. The figures \ref{fig:NoSPP}, \ref{fig:SSPP} and \ref{fig:ESPP} illustrate, respectively, for one of the generated digital polycrystal, the time evolution of the case without SPP, with static SPP and with evolutive SPP. 

\begin{figure}[!h]
\centering
\includegraphics[width=1\textwidth] {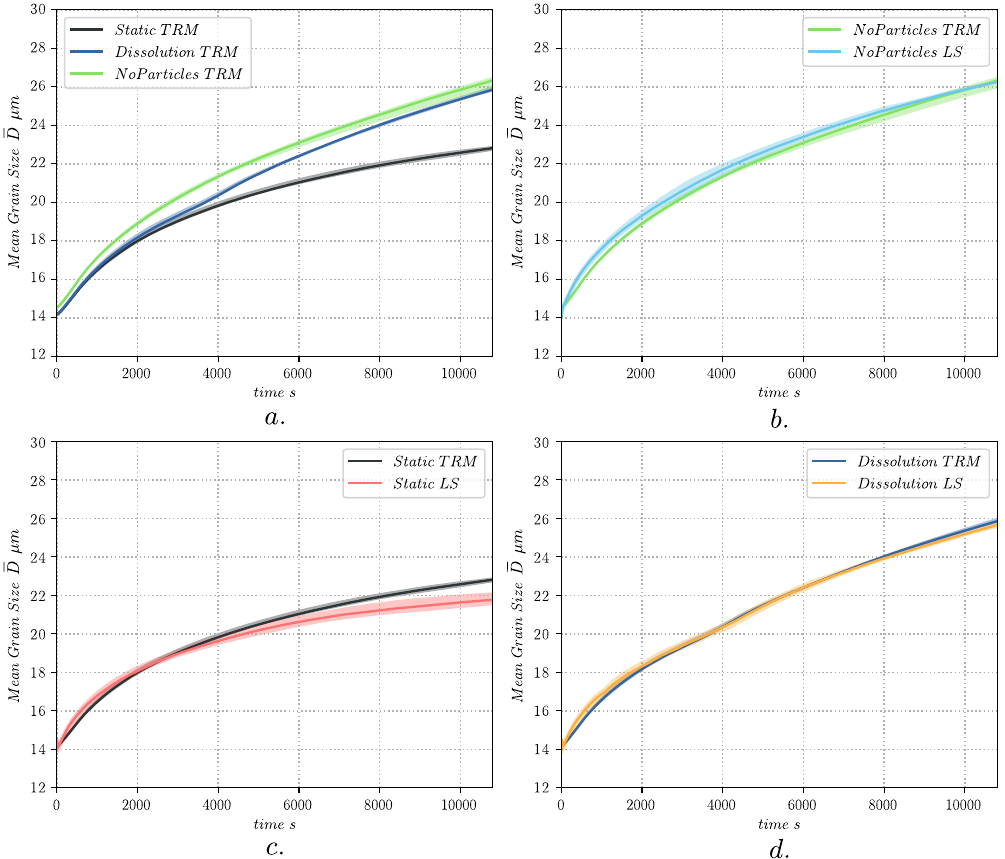}
\caption{GG evolution: a. with the TRM approach for the different scenarios, and comparison between the TRM and LS models concerning b. GG when no particles are considered in the microstructure, c. GG with a fixed SPP population, and d. GG under the effect of particles dissolution. Results for each case are averaged from the results of four different initial tessellations, with the range of results indicated by the semi-transparent shading of the same color as the corresponding curve.}
\label{1_SPP_TRM_04_04}
\end{figure}

Figure \ref{1_SPP_TRM_04_04} presents the comparison of the predicted grain size evolution between the TRM and LS model. First, the behavior obtained by the TRM method considering or not the presence and evolution of the SPP is illustrated in Figure \ref{1_SPP_TRM_04_04}.a. The results are coherent and well reproduce the Smith-Zener pinning mechanism. Moreover, they illustrate an excellent agreement between the predictions of the full-field models.

\subsection{Domain size influence and time calculation}

The TRM model was already discussed in terms of convergence study (number of grains) in context of pure grain growth without SPP \cite{Florez2020c}. The same discussion was operated here in context of static SPP. The  configuration considered in the previous section for static SPP, materials data, initial grain size distribution, SPP characteristics and temperature was considered by increasing the domain size.\\

\begin{figure}[!h]
\centering
\includegraphics[width=1\textwidth] {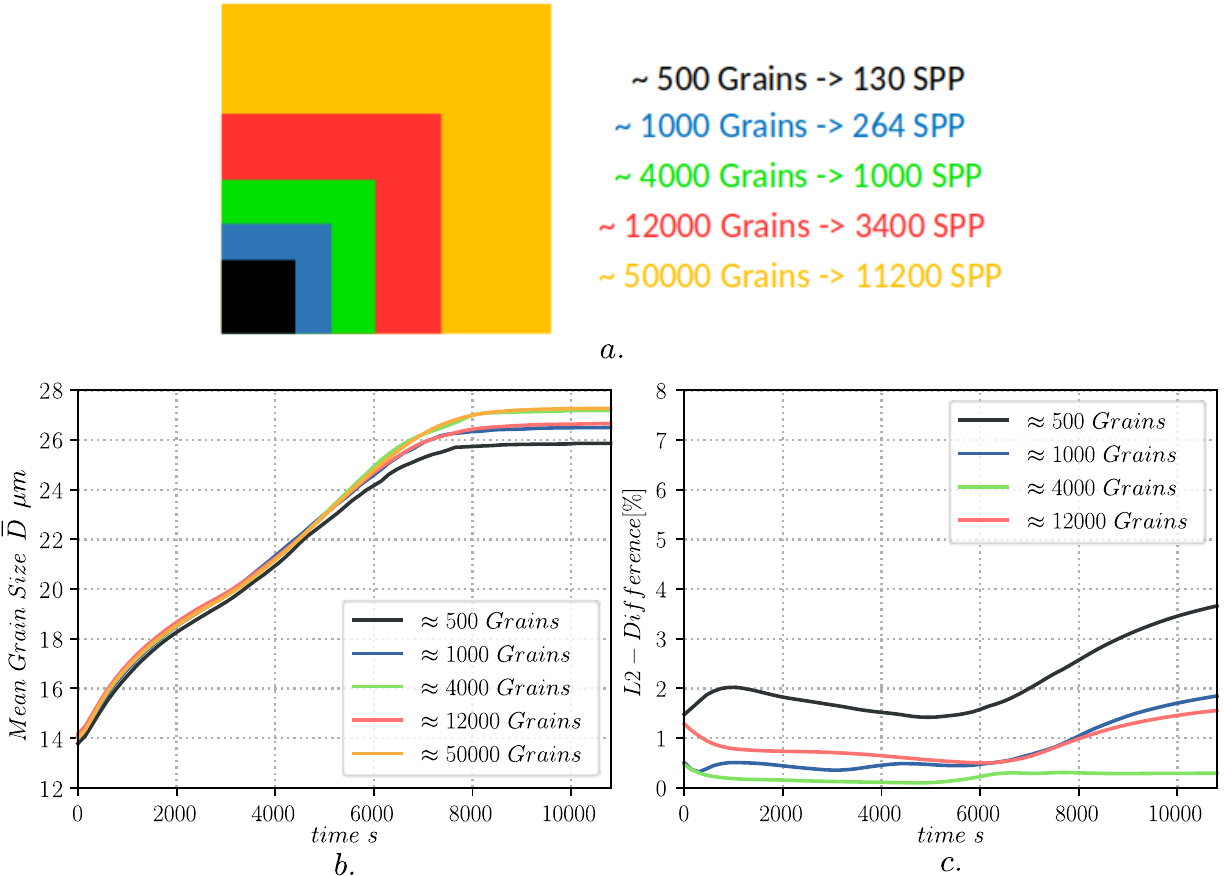}
\caption{Domain size evolution: a. different tested cases with the respective number of grains and particles. b. mean grain size evolution and c. the obtained $L2$ errors comparatively to the largest configuration.}
\label{Domain_size}
\end{figure}

The smallest domain contains 500 grains and 130 SPP, and the biggest one 50000 grains and 11200 SPP. Figure \ref{Domain_size}.a. presents the different domains with the respective number of grains and particles for each case. Figures \ref{Domain_size}.b and c. illustrate the mean grain size evolution and the $L2$ error considering the most significant case as the reference case.\\

First, there results illustrate the fast convergence of the proposed methodology in terms of number of grains.
Indeed, small domains exhibit small errors (around 3 \%) compared to the largest one. This means that it could be preferable in terms of computation resources to run several small test cases as proposed in section \ref{testTRM1}.a. than one large case as presented in section \ref{testTRM1}.b. Of course, comparatively to the optimal domain size discuss in \cite{Florez2020c}, the impact of Smith-Zener pinning pressure is here of the prime order to explain that the smallest configuration can already be predictive. \\

\begin{figure}[!h]
\centering
\includegraphics[width=1\textwidth] {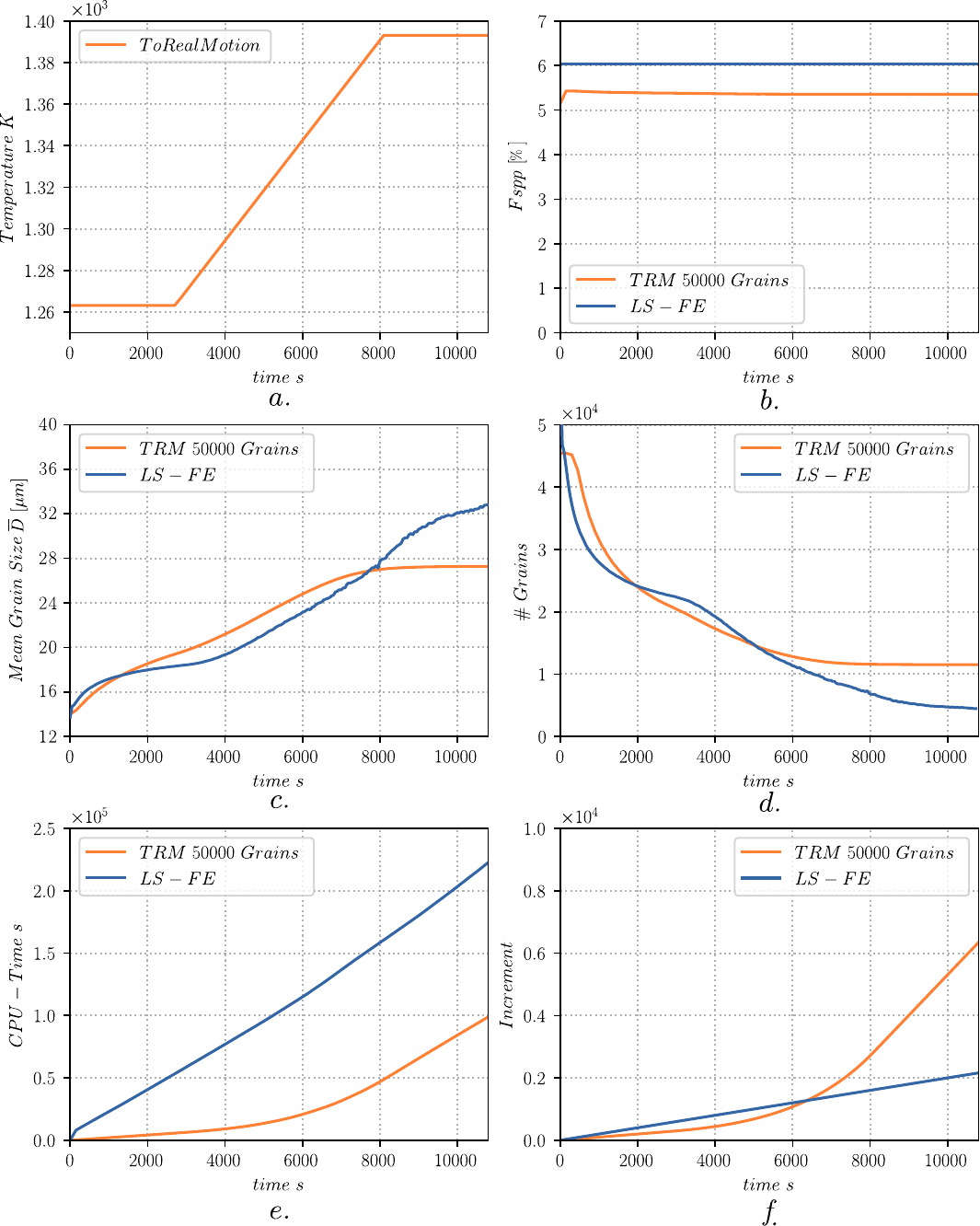}
\caption{Comparisons between the LS and TRM model of a polycrystal case with 50000 initial grains: a. Numerical heat treatment, b. Precipitate particle fraction, c. Mean grain size, d. Number of grains, e. Simulation time, f. Number of increments. }
\label{Domain_size2}
\end{figure}

Finally, a last test case is considered here to analyse the time calculation 
for a large simulation performed following the LS method and the new one proposed here. More precisely, for this test, the data presented in \cite{alvarado2021level} in context of a LS simulation are used. The initial microstructure consists of around 50000 grains generated with a Laguerre-Voronoi tessellation \cite{Hitti2012} for a domain size of $\SI{2}{\milli\meter}\times\SI{2}{\milli\meter}$. An arithmetic mean grain diameter $\overline{ECD}=\SI{9.5}{\micro\meter}$ and an initial monodisperse spherical particle population ($\overline{r}_{spp}=2$ $\mu m$) with a surface fraction $f_{spp}=6\%$ (around 20000 SPPs) are considered. The thermal path proposed in \cite{alvarado2021level} and reproduced in Fig.\ref{Domain_size2}.a is modeled. It corresponds to an isothermal treatment at \SI{990}{\celsius} for \SI{45}{\minute} followed by a linear increase of the temperature until \SI{1120}{\celsius} in
\SI{90}{minute} and the material is maintained at this temperature for others \SI{45}{\minute}. 
As in \cite{alvarado2021dissolution,alvarado2021level,nguyen2024pc}, grain boundary properties, $\mu$ and $\gamma$ are chosen as representative of the AD730 nickel base superalloy.\\

As a result of using a less refined mesh, the SPP fraction considered in the TRM approach appears to be slightly lower than the imposed fraction (see Fig.\ref{Domain_size2}b). This figure also illustrates the stability of the SPP fraction throughout the heat treatment. While one might therefore expect a slightly less pronounced pinning for the TRM approach, the opposite is ultimately observed at the end of the heat treatment. Indeed, if the predicted results (Figs.\ref{Domain_size2}c. and d.) are quite similar for both models, the $\overline{ECD}$ predicted by the LS method is slight bigger, then the number on grains at the end of the simulation is inferior than the one obtained with the TRM model. This result is very interesting because it illustrates a slightly different behavior between the two methods once the microstructure is globally fixed by a cloud of second-phase particles, as is the case here. The front-tracking approach proposed here makes any movement of a flat interface between two particles in 2D completely impossible, which is not entirely the case for an LS approach due to the approximate nature of curvature handling at particle joints. The error made by the LS approach, obviously dependent on the fineness of the finite element mesh used, can lead to some particle crossings that do not occur in the proposed approach. This explains the slightly different kinetics at the end of the heat treatment process. This discussion aligns with a similar analysis made in the mean-field approach proposed in \cite{Bignon2024}.\\

The comparison of computation times is also intriguing. Indeed, it has already been proven that the proposed method in 2D is much more efficient than front-capturing approaches (LS, MPF, etc.) in a pure grain growth and FE formulation context. This remains partially true here as illustrated in Fig.\ref{Domain_size2}e. Indeed, while the computation time is still twice as fast as the LS approach, the observed ratio is nowhere near that of a context without second-phase particles (up to 150 as observed in \cite{Florez2020b}). This result can be explained by two observations. First, the presence of SPP requires a FE mesh capable of describing them, and therefore an optimal mesh size adapted to both the grains and the particles, unlike cases without particles. Overall, the FE mesh size used is finer, which is already more demanding for the simulation as a whole. Secondly, handling interactions at the particle outlines requires additional operations, which become increasingly numerous as the number of contacts between particles and grain boundaries increases during the heat treatment, as illustrated in Fig.\ref{Domain_size2}e. Moreover, the time step is also adjusted to manage these interactions while ensuring the stability of the solution \cite{Florez2020b}; thus, the number of time iterations also increases as a function of the length of grain boundaries pinned by the particles, as shown in Fig.\ref{Domain_size2}f. These various elements highlight the advantages of the method, as it remains overall faster than front-capturing approaches, but also its limitations. Indeed, one could imagine that, as with LS and MPF approaches, highly dense populations of very fine particles relative to the grain size could diminish the benefits of the proposed method according to the criteria outlined in this paragraph. A mixed formulation is proposed in the next part to overcome this limitation.

\section{Discretized second phase particles and Z-Nodes, a new combination}\label{sec:znode}
\subsection{Z-Nodes, interest and principle}\label{ZNodeIntro}
As discussed in the introduction, Vertex/ front-tracking approaches traditionally treat second-phase particles as specific nodes within the discretization \cite{Weygand1999,Lepinoux2010}, incorporating pinning laws that more or less accurately account for the Smith-Zener pinning phenomenon, without truly being able to critically assess the resulting outcomes. We therefore questioned whether the TRM method could be enhanced to measure the actual benefit of discretizing second-phase particles compared to traditional Vertex approaches. Additionally, we aimed to develop a hybrid method to describe very small SPP in relation to grain sizes, a range inaccessible to front-capturing approaches within acceptable computation times.\medbreak

Thus, the TRM algorithm was modified to accommodate a new class of nodes called 'Z-Nodes'. Z-Nodes are modeled thanks to a few rules imposed over the remeshing and GB migration procedure (see Fig.\ref{fig:intro}). Z-Nodes are allowed to be collapsed with other nodes, however, the remaining node will always be transformed into a Z-Node and will be positioned at the same coordinates as the initial Z-Node. Two Z-Nodes can not be collapsed. Node gliding is removed from Z-Nodes, however edge swapping can still occur on their vicinity. Readers interested in the algorithmic details of these operators are invited to consult \cite{Florez2020d}. The movement of Z-Nodes due to GB migration is also not allowed. Following these rules, classical pinning node of the Vertex approach, can also be considered in the TRM framework.\medbreak

Once again, the AD730 alloy is considered in this section with the same initial GSD and following now a $\SI{5}{\hour}$ isothermal at $\SI{1060}{\celsius}$. If the reduced mobility remains representative of this alloy at this temperature, various distributions of second-phase particles will be considered. These populations do not necessarily have sizes and fractions representative of this alloy at this temperature but serve solely as a numerical playground to compare the two approaches to handling second-phase particles in terms of accuracy, prediction, and computational cost.

\subsection{First illustration concerning Z-Nodes}
\begin{figure}[!h]
  \centering
  \begin{subfigure}{0.32\textwidth}
    \centering
    \includegraphics[scale=0.19]{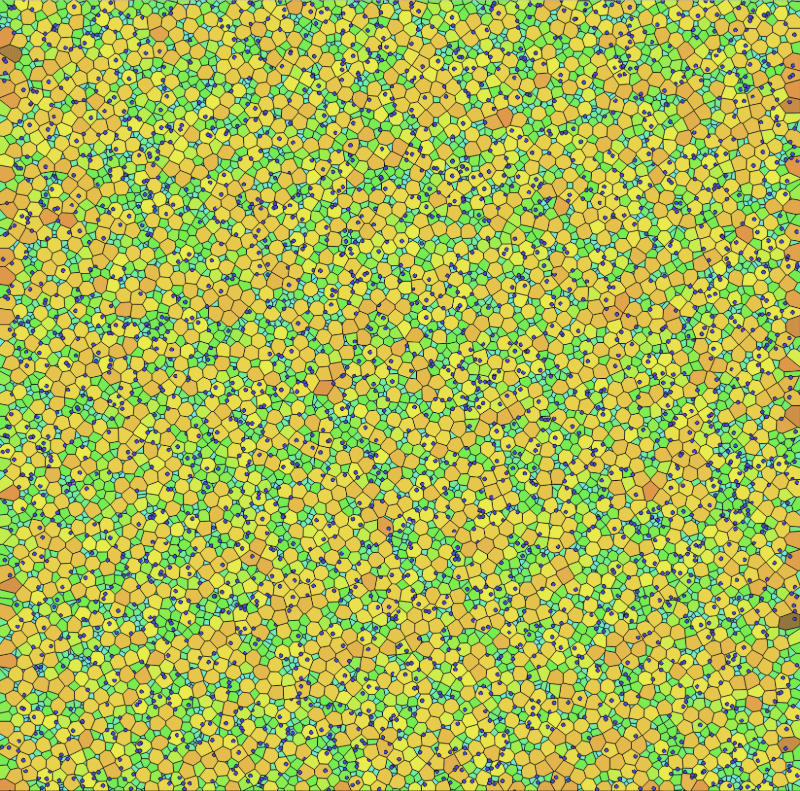}
    \caption{}
    \label{subfig:ZNode11}
  \end{subfigure}
  \begin{subfigure}{0.32\textwidth}
    \centering
    \includegraphics[scale=0.19]{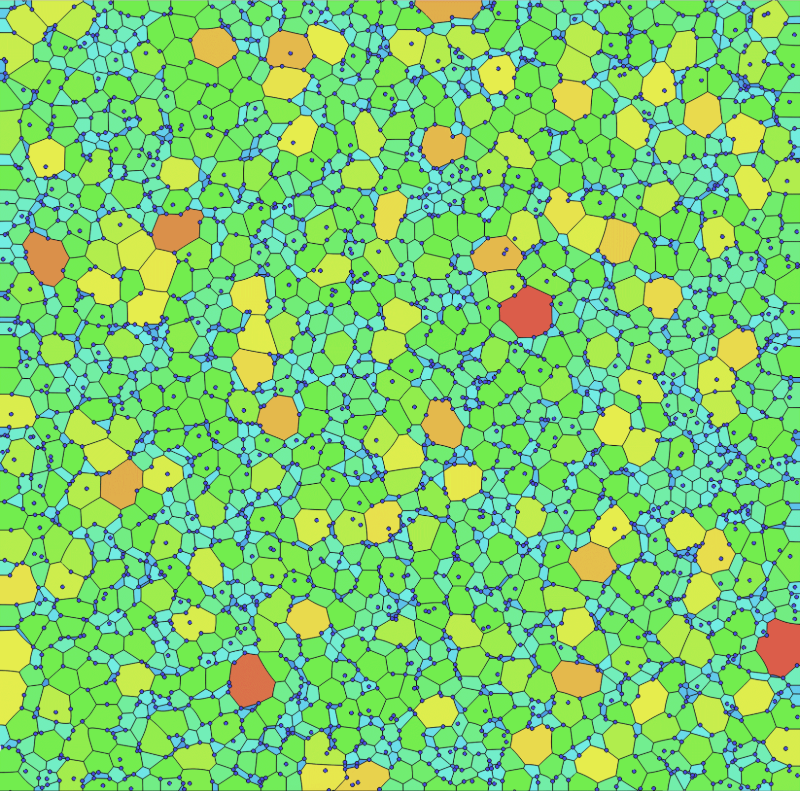}
    \caption{}
    \label{subfig:ZNode12}
  \end{subfigure}
   \begin{subfigure}{0.32\textwidth}
    \centering
    \includegraphics[scale=0.19]{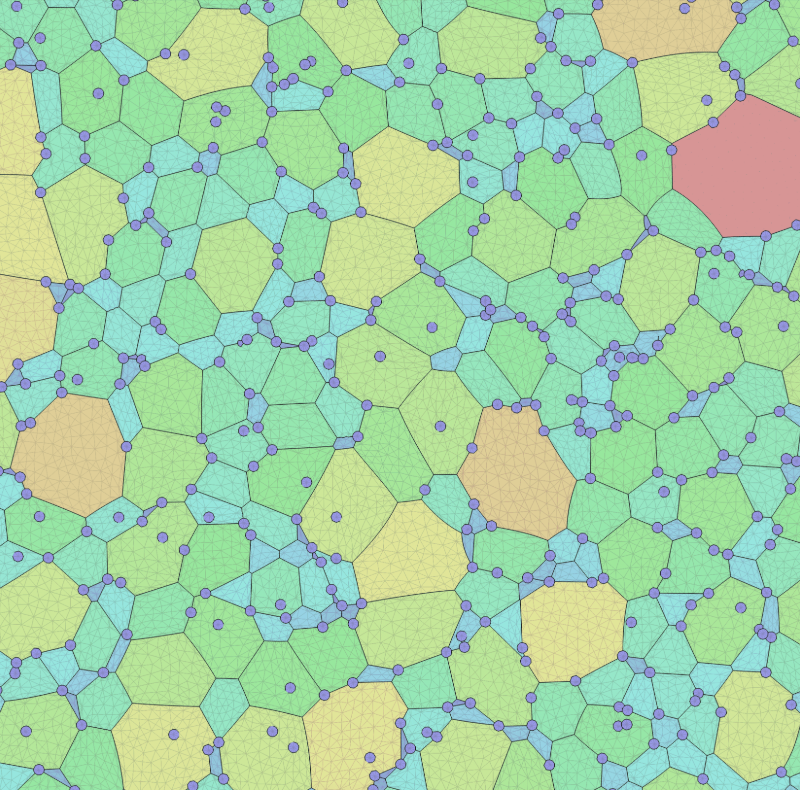}
    \caption{}
    \label{subfig:ZNode13}
  \end{subfigure}
  \begin{subfigure}{0.32\textwidth}
    \centering
    \includegraphics[scale=0.19]{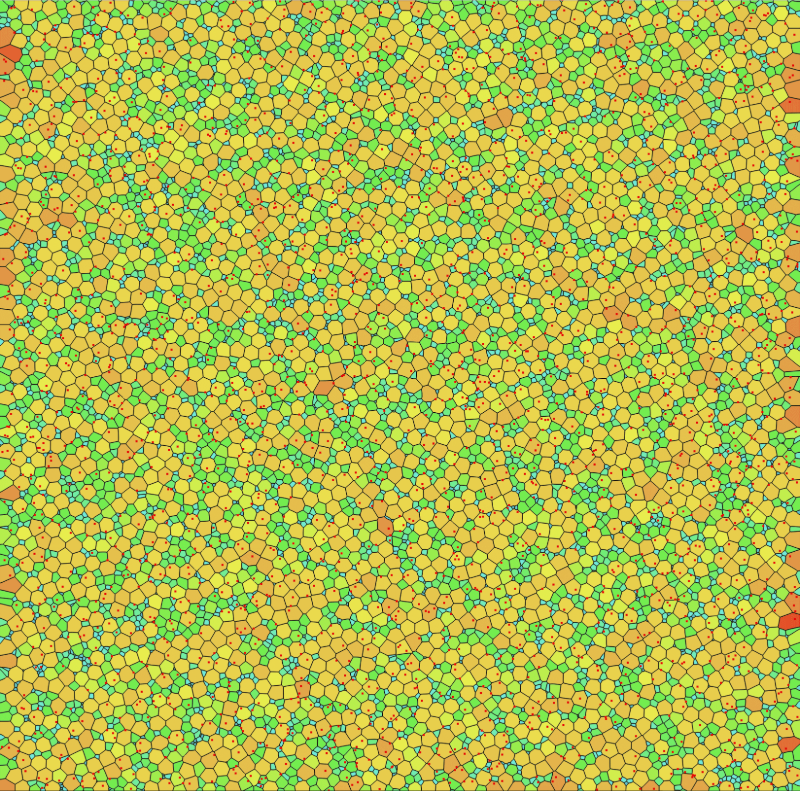}
    \caption{}
    \label{subfig:ZNode14}
  \end{subfigure}
\begin{subfigure}{0.32\textwidth}
    \centering
    \includegraphics[scale=0.19]{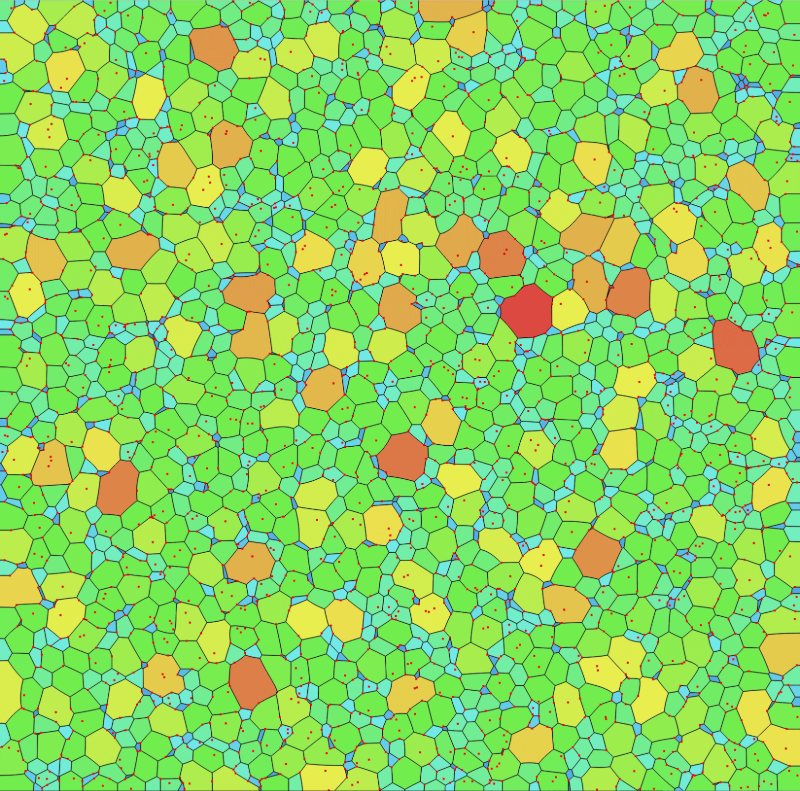}
    \caption{}
    \label{subfig:ZNode15}
  \end{subfigure}
  \begin{subfigure}{0.32\textwidth}
    \centering
    \includegraphics[scale=0.19]{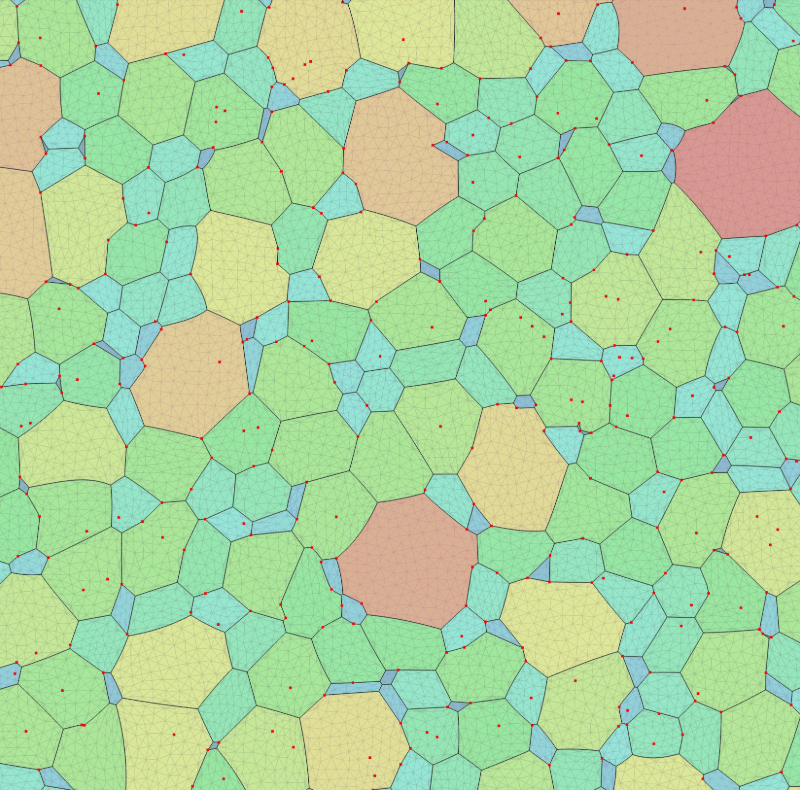}
    \caption{}
    \label{subfig:ZNode16}
  \end{subfigure}
  \begin{subfigure}{0.4\textwidth}
    \centering
    \includegraphics[scale=0.4]{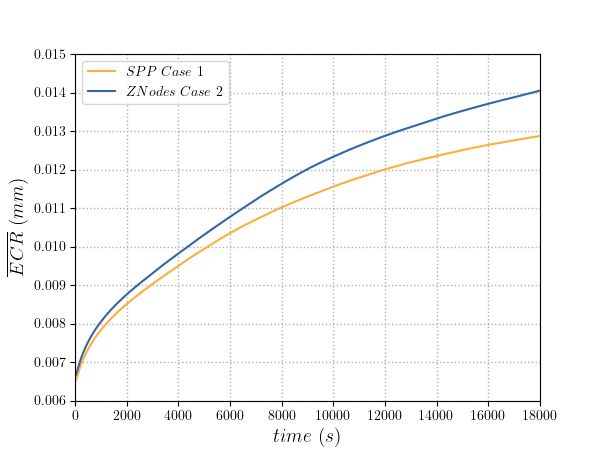}
    \caption{}
    \label{subfig:ZNode17}
  \end{subfigure}
  \begin{subfigure}{0.29\textwidth}
    \centering
    \includegraphics[scale=0.4]{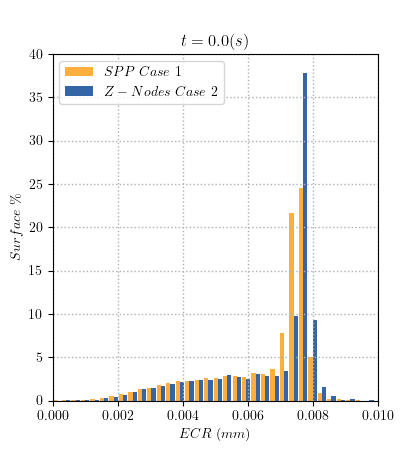}
    \caption{}
    \label{subfig:ZNode18}
  \end{subfigure}
  \begin{subfigure}{0.29\textwidth}
    \centering
    \includegraphics[scale=0.4]{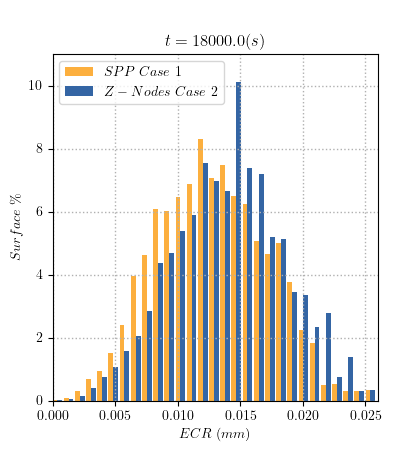}
    \caption{}
    \label{subfig:ZNode19}
  \end{subfigure}
  \caption{First illustration for the comparative test cases between the modeling of Smith-Zener pinning mechanism by considering the discretization of the SPPs and the Z-Nodes strategy: (top) with discretized SPP in dark blue, from (a) to (c): Initial microstructure, after the 5h thermal treatment, and a zoom (in the 0.3mm$\times$0.3mm central zone) at the final time with the FE mesh in black; (middle) with Z-Nodes in red, from (d) to (f): Initial microstructure, after the 5h thermal treatment, and a zoom (in the 0.3mm$\times$0.3mm central zone) at the final time; (bottom) from (g) to (i): evolution of the arithmetic mean of the ECR for both strategies, the initial grain size distributions weighted by the surface, and the final grain size distributions weighted by the surface.}
  \label{fig:ZNode1}
\end{figure}

The figure \ref{fig:ZNode1}  provides an initial illustration of the methodological difference proposed in the case of second-phase particles that are relatively large compared to grain size. Specifically, this illustration is identical to the case study in the previous section with a static second phase particle population: a monodisperse population with a size of $r=$\SI{2}{\micro\meter} and a surface fraction of 5\% is considered. The ratio between the initial average grain size and the particle size is therefore $\overline{ECR}/{\bar{r}}=3.5$. The figure \ref{subfig:ZNode11} illustrates the initial microstructure where the second-phase particles are discretized, while the figure \ref{subfig:ZNode14} shows the same initial microstructure where the second-phase particles are not discretized (Z-Nodes at the same positions are represented by red points). 
It is important to note that the Z-Nodes are merely points and that the thickness of the red points has no intrinsic meaning other than to aid in visualizing the positions of the second-phase particles. Although a much coarser mesh could be used for the Z-Nodes type approach, the same simulation characteristics were applied for this initial case: a time step of $\Delta t=$\SI{10}{s}, a mesh size of $h=\SI{300}{\nano\meter}$ and an output at every minute of the $\SI{5}{\hour}$ heat treatment. Figures \ref{subfig:ZNode12} and \ref{subfig:ZNode15} describe the final microstructures for both strategies in the global calculation domain whereas the Figures \ref{subfig:ZNode13} and \ref{subfig:ZNode16} correspond to a zoom on the $\SI{0.3}{\milli\meter}\times\SI{0.3}{\milli\meter}$ central domain for both methodologies at the final time with the corresponding FE meshes superimposed in black. The figures \ref{subfig:ZNode17}, \ref{subfig:ZNode18}, and \ref{subfig:ZNode19} shows, respectively, for both methodologies the time evolution of the arithmetic mean grain size ($\overline{ECR}$), the initial grain size distributions ($ECR$) weighted by the surface, and the final ones.

This first comparative case allows for several observations. As expected, the Z-Nodes type approach tends to underestimate the Smith-Zener pinning pressure compared to the more precise case where the SPPs are discretized, especially as the SPPs here have a relatively large size. This is evident in the evolution of the $\overline{ECR}$ (see Fig.\ref{subfig:ZNode17}) and the final distributions (see Fig.\ref{subfig:ZNode19}). However, the difference remains reasonable (a shift of $\SI{1}{\micro\meter}$ concerning the average grain size after $\SI{5}{\hour}$ of heat treatment at $\SI{1060}{\celsius}$). Based on Figures \ref{subfig:ZNode12} versus \ref{subfig:ZNode15} and \ref{subfig:ZNode13} versus \ref{subfig:ZNode16}, where the final grain morphologies are depicted, the error appears to mainly concern an accelerated disappearance of small grains when the particles are not discretized. It is worth noting that the difference in the initial grain size distributions, although the microstructures are supposed to be identical, stems from the surface occupied by the SPPs in the first simulation which is not present in the Z-Nodes approach. As a result, the initial grain size distribution (see Fig.\ref{subfig:ZNode18}) appears slightly shifted to the right for the Z-Nodes case. As the FE mesh size used is similar here in both strategies (see Fig.\ref{subfig:ZNode13} and Fig.\ref{subfig:ZNode16}), the CPU time for both simulations are very similar and are summarized in the Table \ref{tab:simulation_settings} of the next section dedicated to a more global comparison between both methodologies.

\subsection{Comparative study between both approaches}

The purpose of this section is to illustrate the capability of the proposed front-tracking method to handle 2D polycrystal simulations that have never been investigated by other approaches of the same type and are inaccessible in terms of reasonable computation times for LS and MPF methods. Accordingly, a computational campaign was conducted with increasingly fine particle sizes, reaching up to about $\SI{125}{\nano\meter}$. Table \ref{tab:simulation_settings} summarizes the different parameters considered in each simulation and also some final information after the \SI{5}{\hour} isothermal treatment. In this strategy, the ratio $f/\bar{r}$, appearing in classical Smith-Zener pinning pressure and limiting grain size predictions \cite{Humphreys2017,manohar1998,bernacki2024digital,bernacki2024kinetic}, was assumed constant and will be discussed in the following. The cases 1 and 2 correspond to the simulations describe in Fig.\ref{fig:ZNode1} and introduced in the previous section.\medbreak

\begin{table}[ht]
\centering
\begin{tabular}{|c||c|c|c|c|c|c|c|c|}
\hline
\rowcolor{lightgray} \small$\raisebox{-1ex}{Id}$\textbackslash $\raisebox{1ex}{Par.}$ & \small $\Omega (mm^2$) &\small  \raisebox{1ex}{$\# G_0$}/\raisebox{-1ex}{$\# G_f$} &\small $h$(nm) & \small $\bar{r}$(nm) & \small $f$\% & \small\# SPP &\small $\overline{ECR}_f (\mu m)$&\small CPU Time\\
\hline\hline
\small Case 1 & \small$0.8\times 0.8$ &\small $\raisebox{1ex}{6807}$/$\raisebox{-1ex}{2083}$  &\small 300 &\small 2000 &\small 5 &\small 2546 &\small 12.87 &\small $\sim$ 8h \\
\hline
\small Case 2 &\small $0.8\times 0.8$ &\small $\raisebox{1ex}{6807}$/$\raisebox{-1ex}{1723}$ &\small 300 &\small $\times$ &\small $\times$ &\small 2546 &\small 14.05&\small $\sim$ 7h30min\\
\hline
\small Case 3 &\small $0.8\times 0.8$ &\small $\raisebox{1ex}{6807}$/$\raisebox{-1ex}{2581}$ &\small 250 &\small 1000 &\small 2.5 &\small 5092  &\small 11.03 &\small $\sim$ 8h30min\\
\hline 
\small Case 4 &\small $0.8\times 0.8$ &\small $\raisebox{1ex}{6807}$/$\raisebox{-1ex}{2459}$&\small 300 &\small $\times$ &\small $\times$ &\small 5092 &\small 11.56&\small $\sim$ 9h\\
\hline
\small Case 5 &\small $0.8\times 0.8$ &\small $\raisebox{1ex}{6807}$/$\raisebox{-1ex}{3610}$&\small 150 &\small 500 &\small 1.25 &\small 10185 & \small 8.59 &\small $\sim$ 19h\\
\hline
\small Case 6 &\small $0.8\times 0.8$ &\small$\raisebox{1ex}{6807}$/$\raisebox{-1ex}{3271}$ &\small 300 &\small $\times$ &\small $\times$ &\small 10185 &\small 9.58 &\small $\sim$8h30 min\\
\hline
\small Case 7 &\small $0.5\times 0.5$ &\small $\raisebox{1ex}{2655}$/$\raisebox{-1ex}{1720}$ &\small 75 &\small 250 &\small 0.625 &\small 9174 &\small 7.95 &\small $\sim$ 23h\\
\hline
\small Case 8 &\small $0.5\times 0.5$ &\small $\raisebox{1ex}{2655}$/$\raisebox{-1ex}{1738}$ &\small 300 &\small $\times$ &\small $\times$ &\small 9174 &\small 8.82&\small $\sim$ 3h\\
\hline
\small Case 9 &\small $0.25\times 0.25$ &\small $\raisebox{1ex}{667}$/$\raisebox{-1ex}{531}$ &\small 32.5 &\small 125 &\small 0.3125 &\small 3826  &\small 7.26&\small $\sim$ 38h30min\\
\hline
\small Case 10 &\small $0.25\times 0.25$ &\small $\raisebox{1ex}{667}$/$\raisebox{-1ex}{532}$ &\small 300 &\small $\times$ &\small $\times$ &\small 3826 &\small 7.18&\small $\sim$ 35min \\
\hline
\end{tabular}
\caption{Simulation settings: $\Omega$ corresponds to the domain dimensions, $\# G_0 / \# G_f$ to the initial number of grains and the final one at $t=$\SI{5}{\hour}. $h$ to the FE mesh size used, $\bar{r}$ to the size of static circular SPP, $f\%$ to the surface fraction of SPP, $\# SPP$ to the initial number of SPP or Z-Nodes, $\overline{ECR}_f$ to the arithmetic mean grain size at $t=$\SI{5}{\hour}, and CPU Time to the total calculation time necessary to simulate the test cases until $t=$\SI{5}{\hour} with output data at each minute.}
\label{tab:simulation_settings}
\end{table}

\begin{figure}[!h]
  \centering
  \begin{subfigure}{0.49\textwidth}
    \centering
    \includegraphics[scale=0.14]{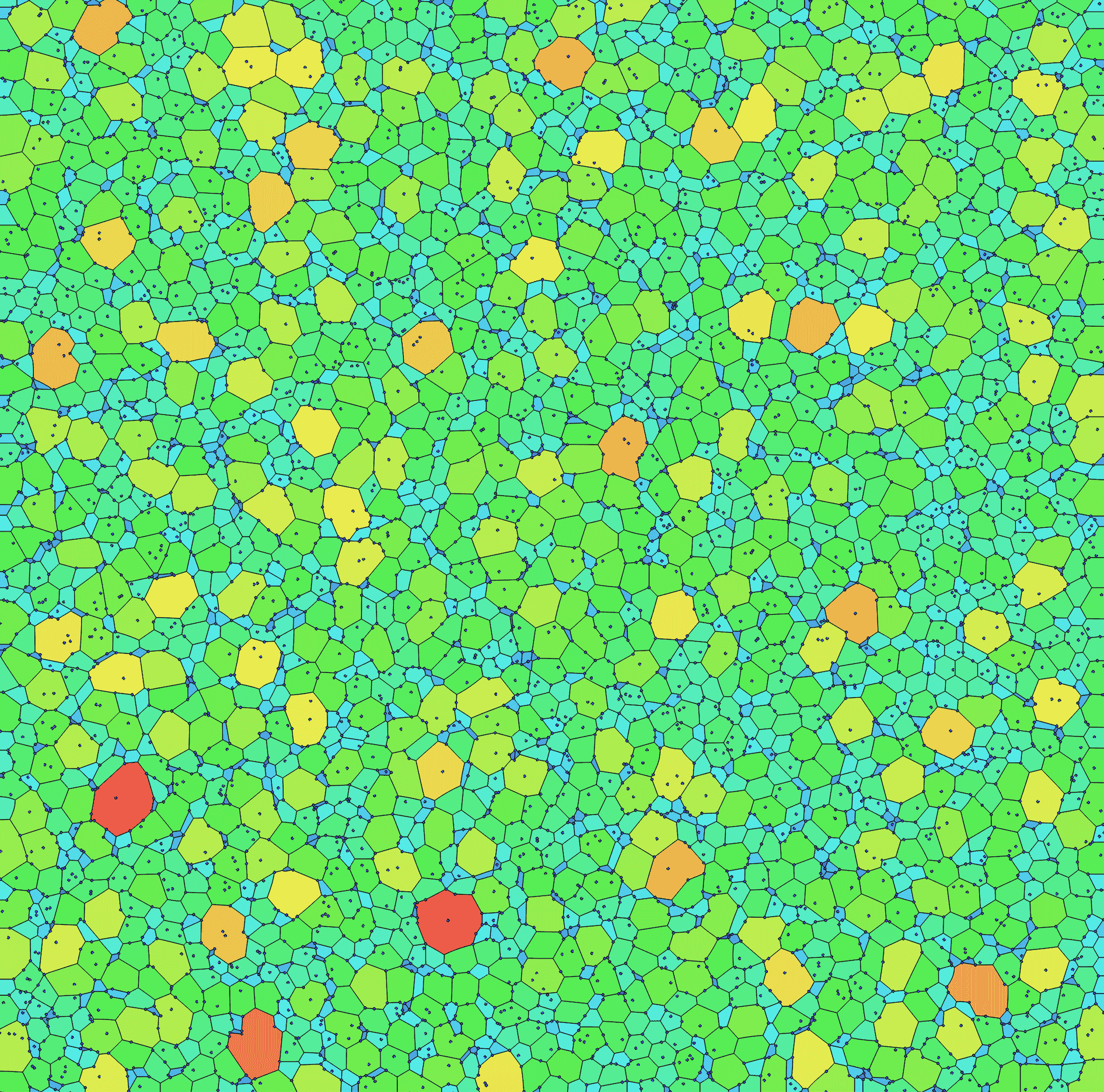}
    \label{fig:C3}
  \end{subfigure}
  \begin{subfigure}{0.49\textwidth}
    \centering
    \includegraphics[scale=0.14]{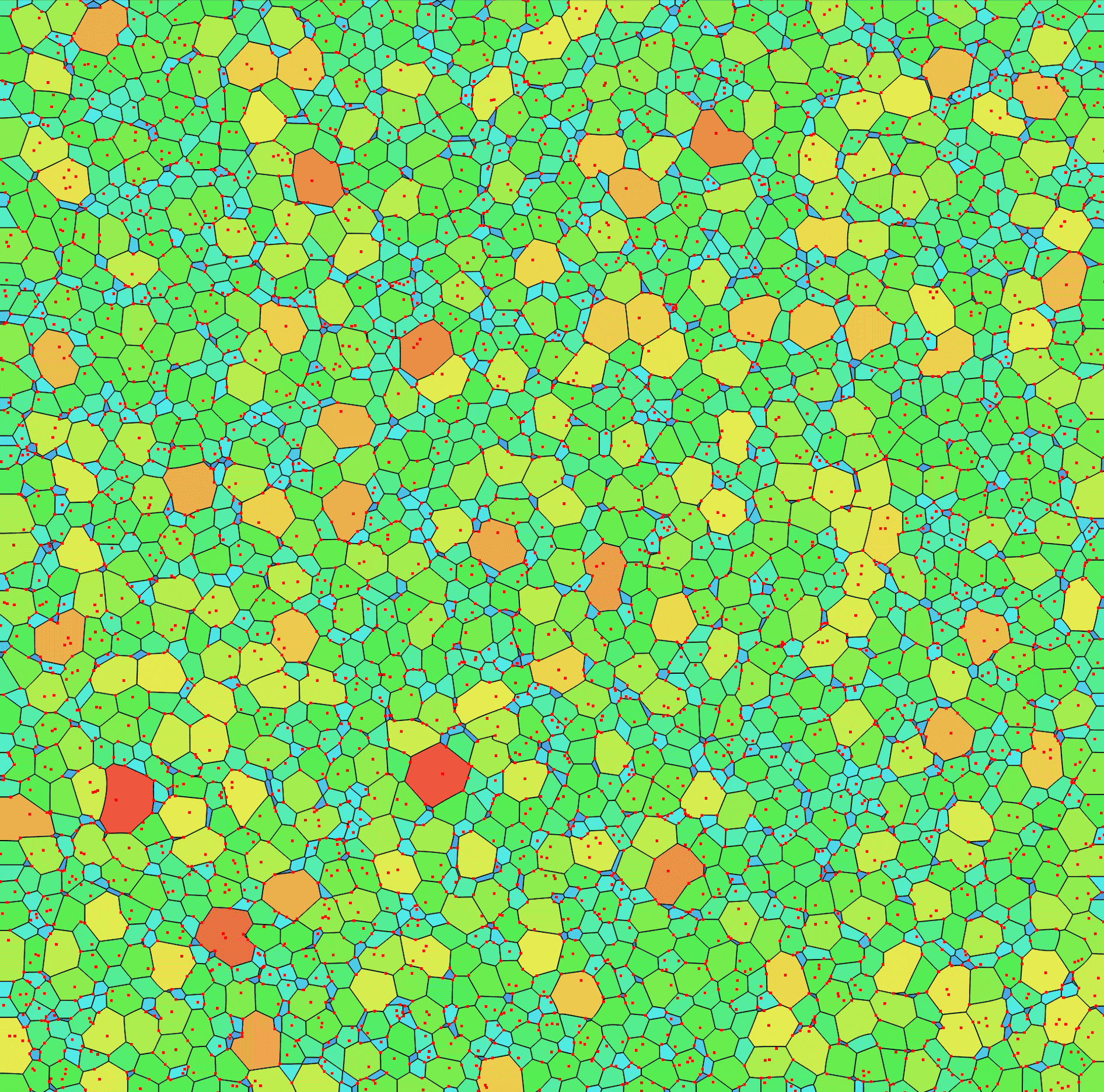}
    \label{fig:C4}
  \end{subfigure}
   \begin{subfigure}{0.49\textwidth}
    \centering
    \includegraphics[scale=0.14]{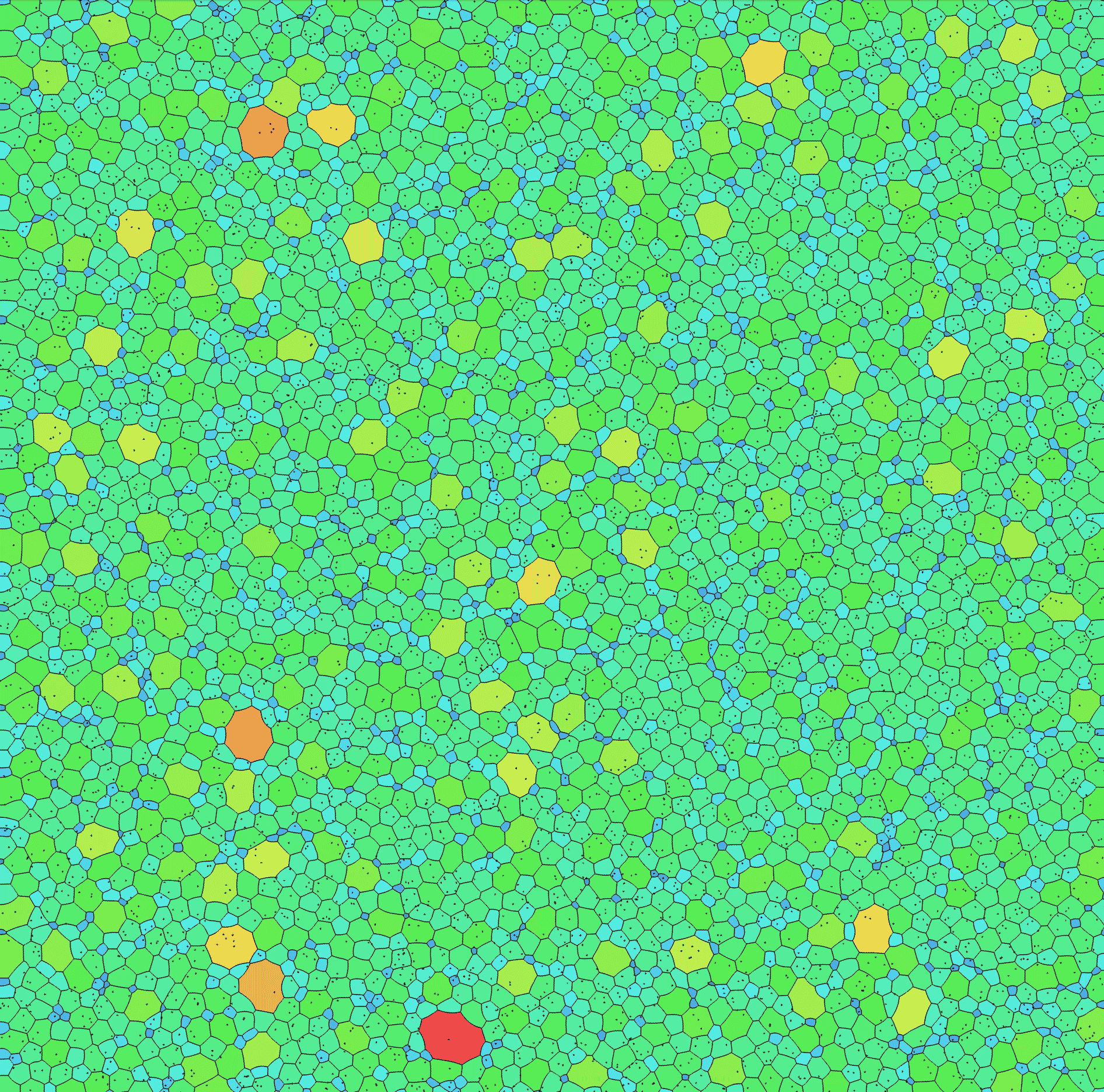}
    \label{fig:C5}
  \end{subfigure}
  \begin{subfigure}{0.49\textwidth}
    \centering
    \includegraphics[scale=0.14]{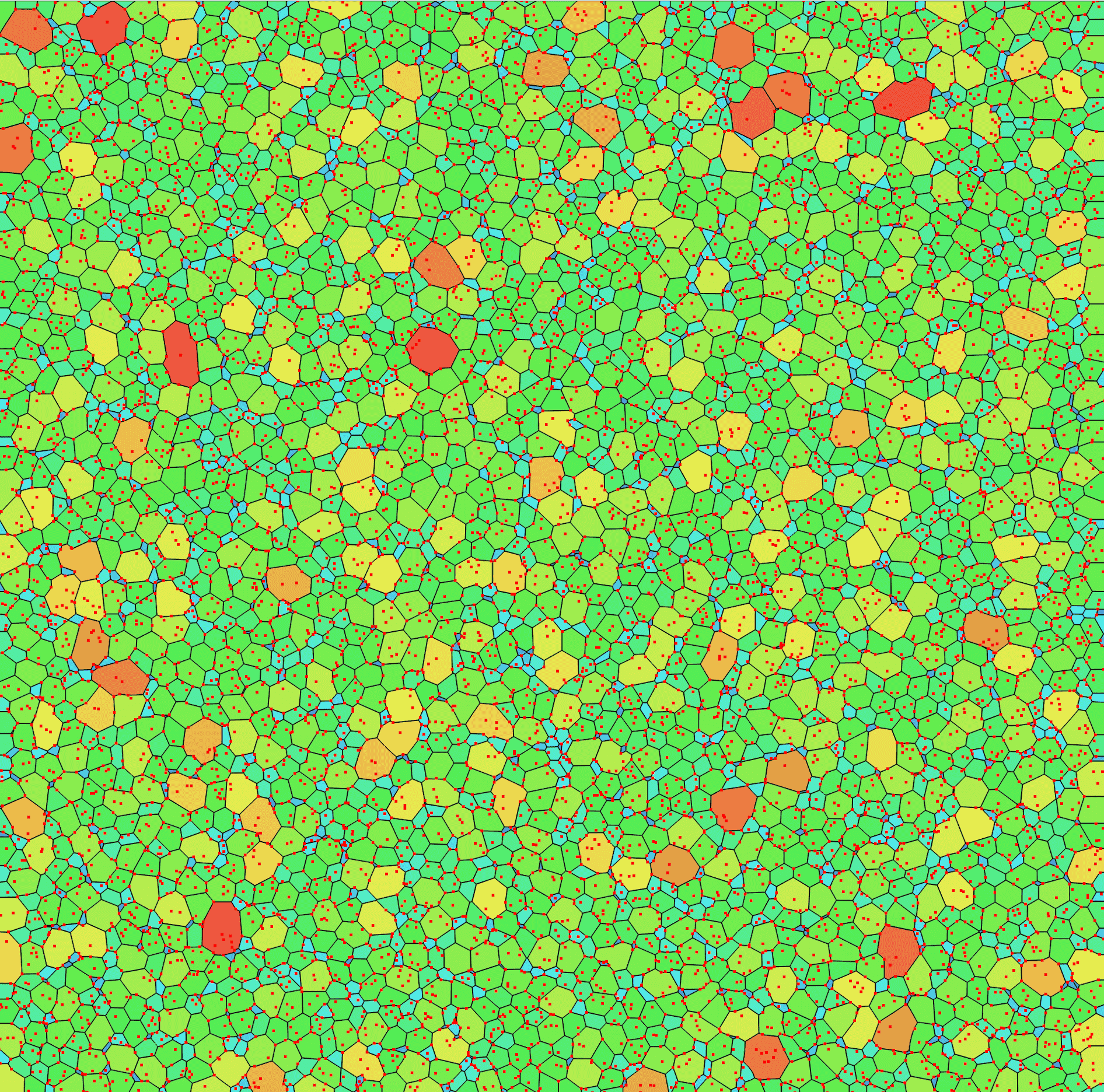}
    \label{fig:C6}
  \end{subfigure}
\begin{subfigure}{0.49\textwidth}
    \centering
    \includegraphics[scale=0.14]{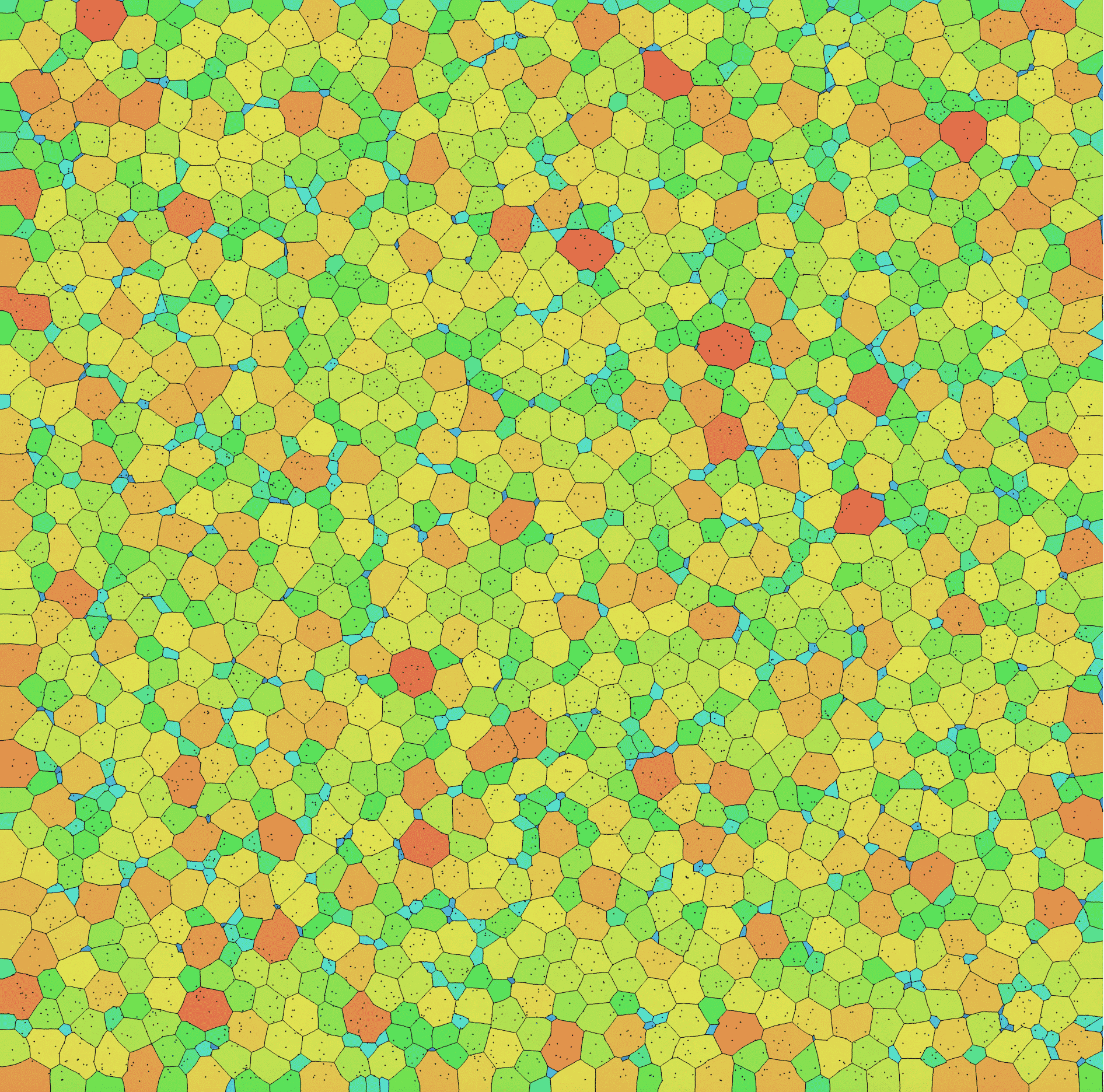}
    \label{fig:C7}
  \end{subfigure}
  \begin{subfigure}{0.49\textwidth}
    \centering
    \includegraphics[scale=0.14]{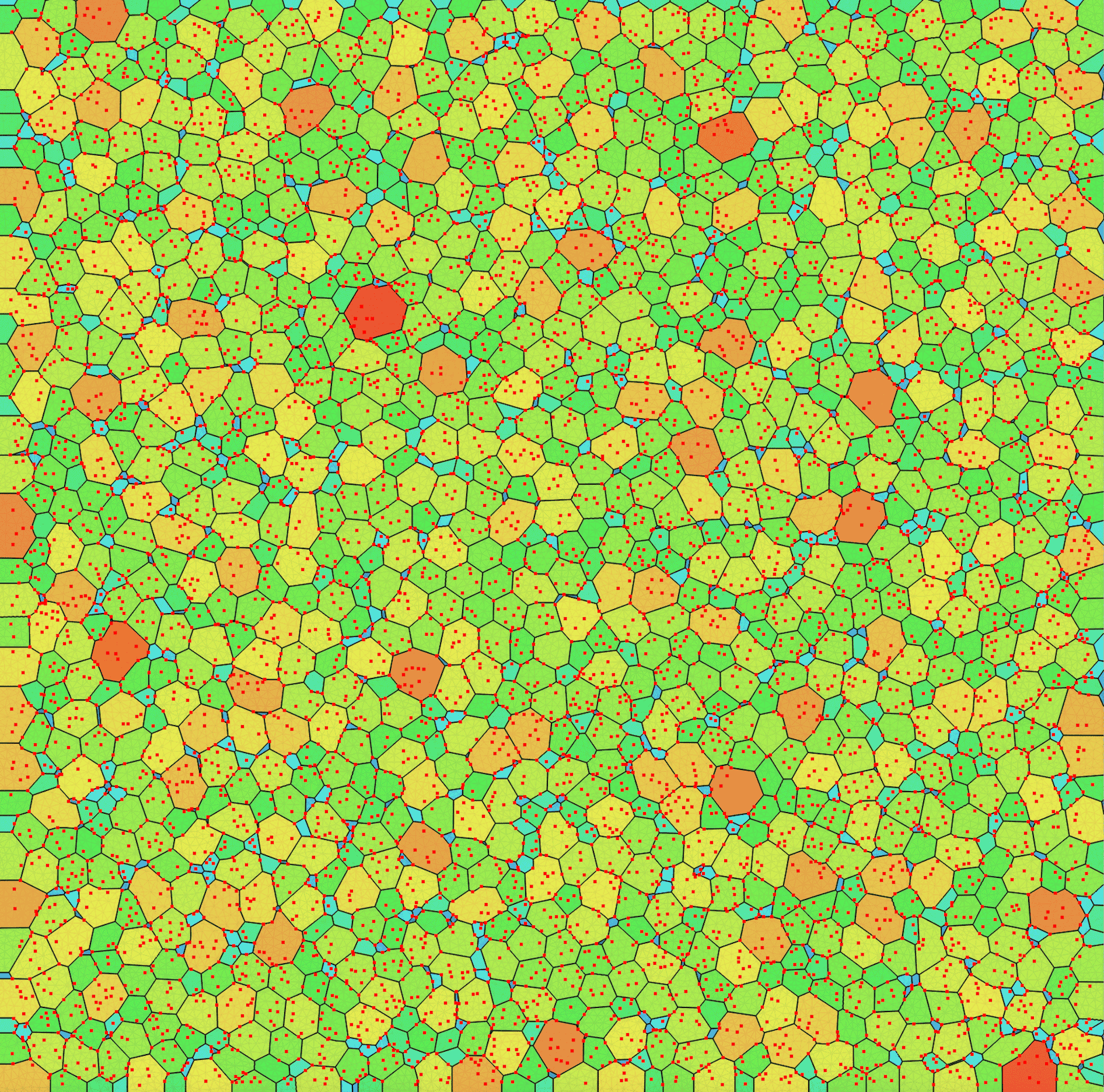}
    \label{fig:C8}
  \end{subfigure}
  \begin{subfigure}{0.49\textwidth}
    \centering
    \includegraphics[scale=0.14]{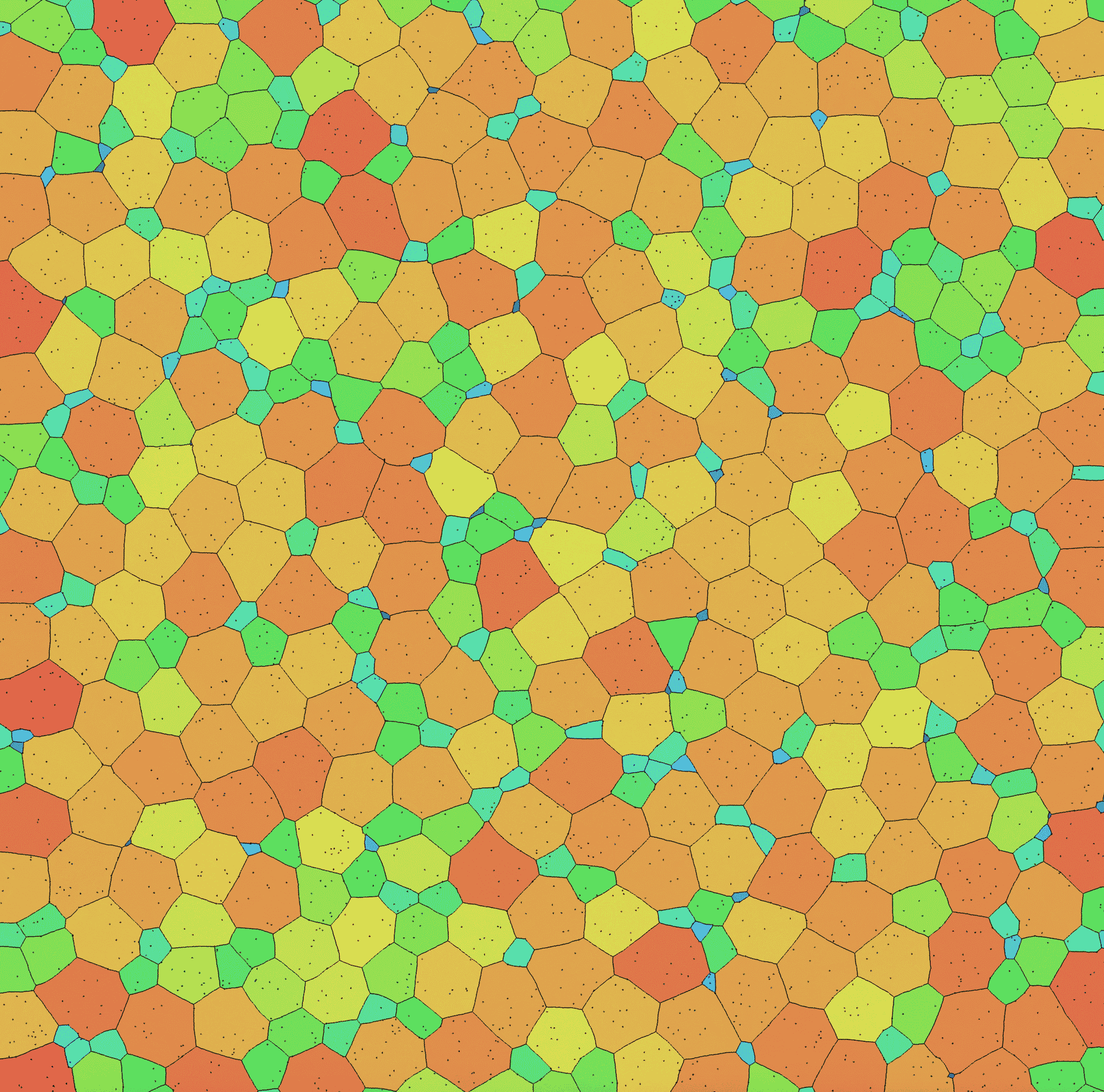}
    \label{fig:C9}
  \end{subfigure}
  \begin{subfigure}{0.49\textwidth}
    \centering
    \includegraphics[scale=0.14]{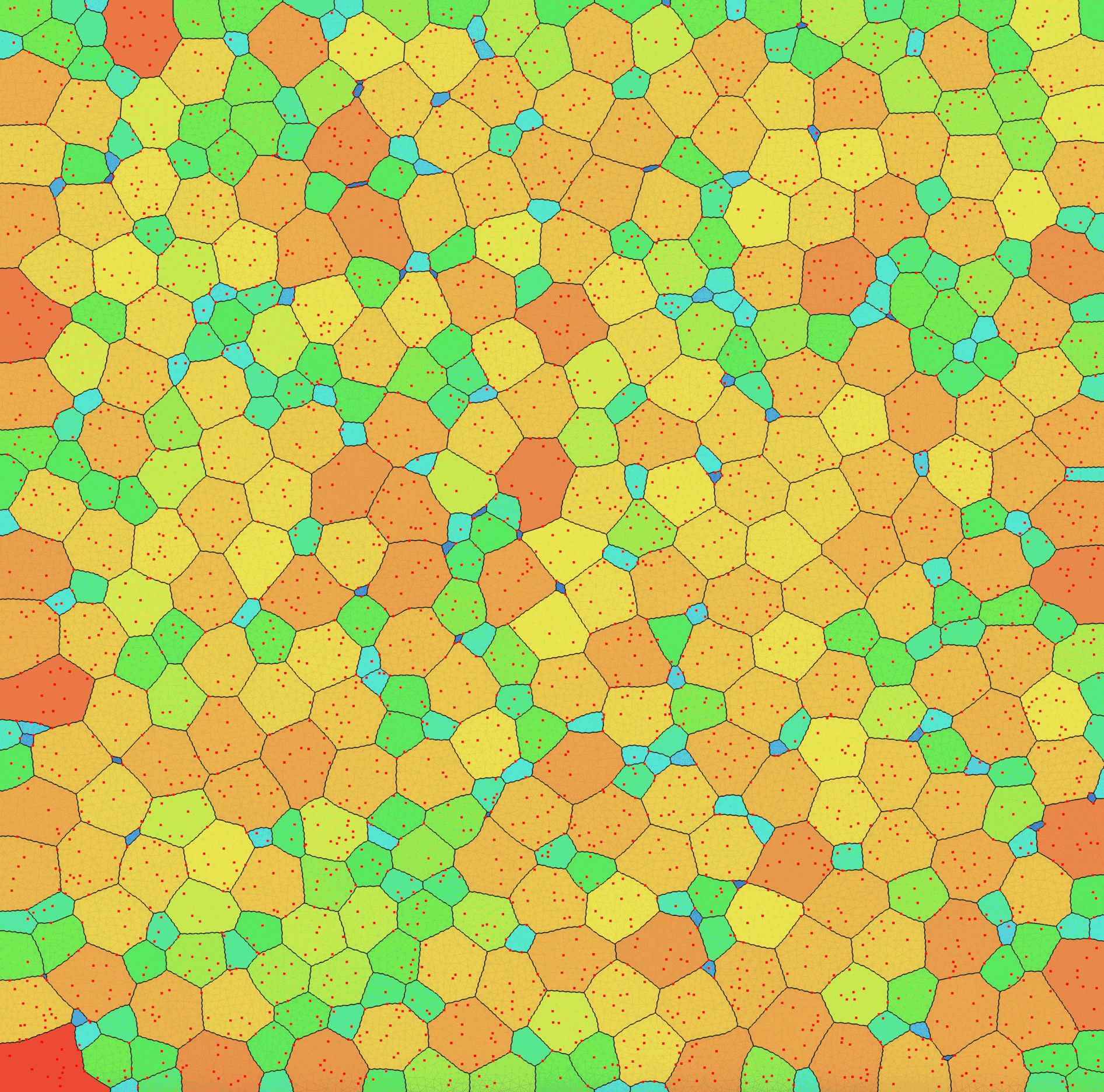}
    \label{fig:C10}
  \end{subfigure}
  \caption{For $p\in\llbracket 1,4\rrbracket$, final state of the cases $2p+1$ in the left side and the cases $2p+2$ in the right side.}
  \label{fig:cases3to10evolution}
\end{figure}

\begin{figure}[!h]
  \centering
  \begin{subfigure}{0.49\textwidth}
    \centering
    \includegraphics[scale=0.1]{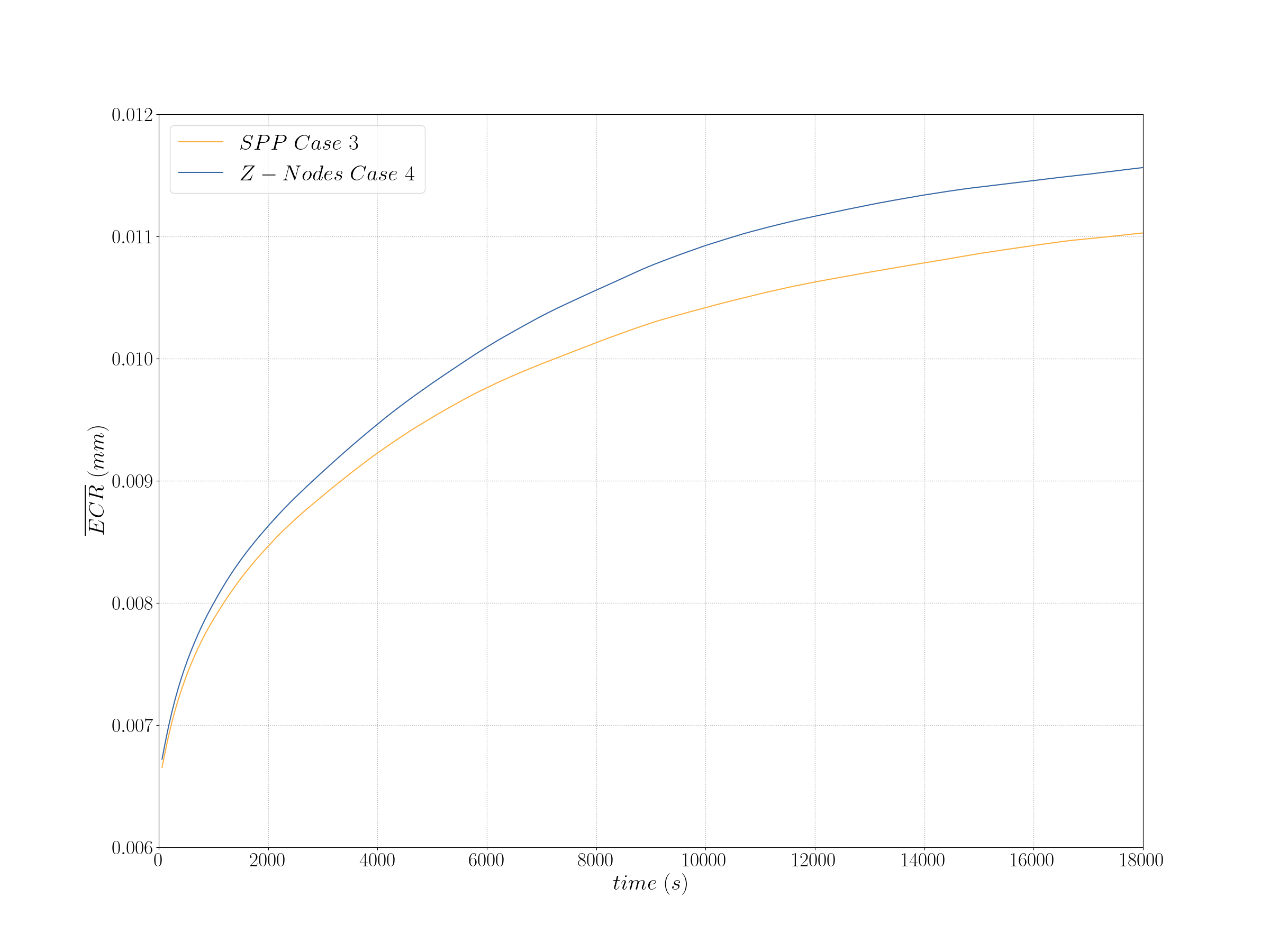}
    \label{fig:ECR2}
  \end{subfigure}
  \begin{subfigure}{0.49\textwidth}
    \centering
    \includegraphics[scale=0.1]{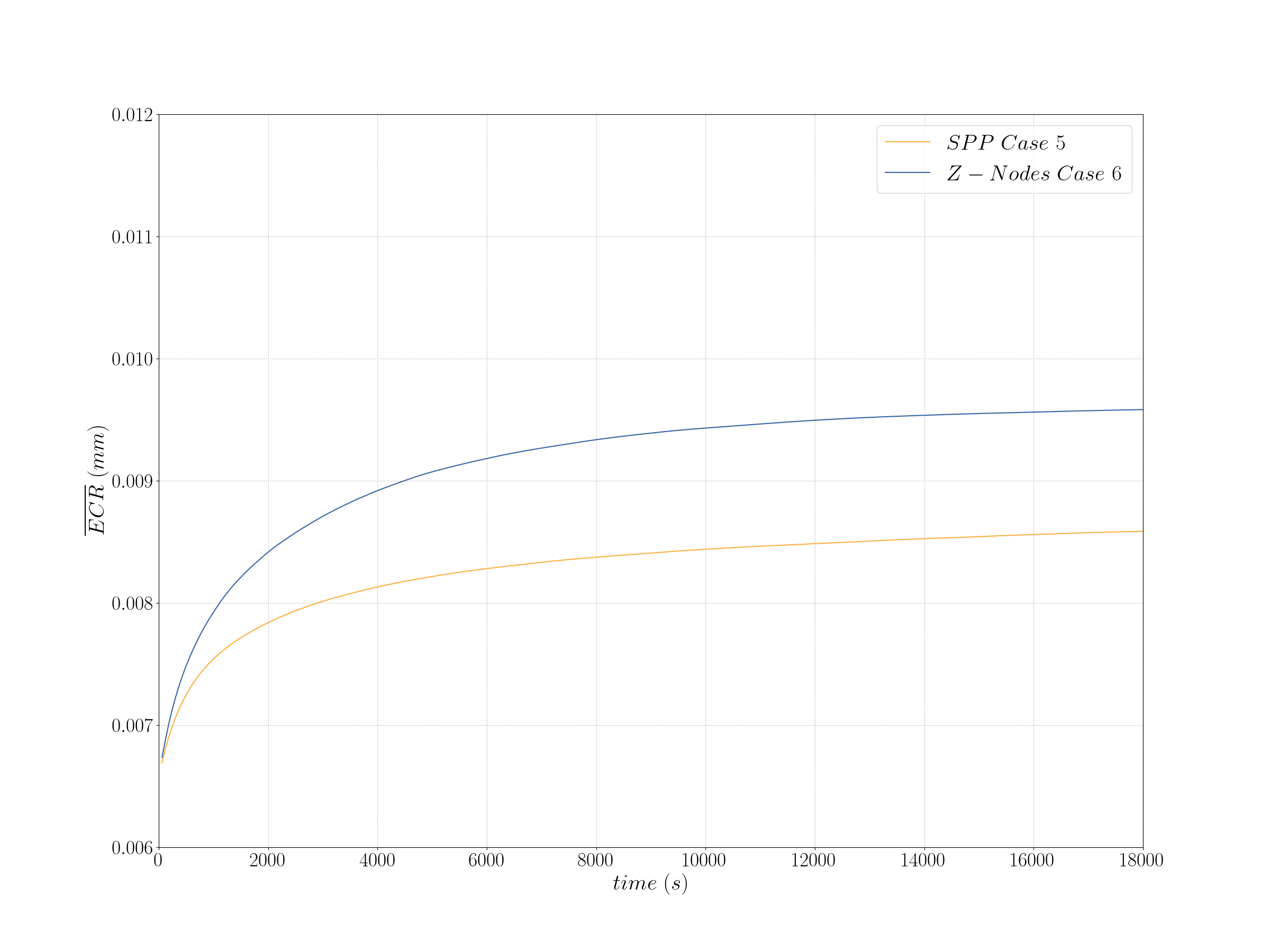}
    \label{fig:ECR3}
  \end{subfigure}
   \begin{subfigure}{0.49\textwidth}
    \centering
    \includegraphics[scale=0.1]{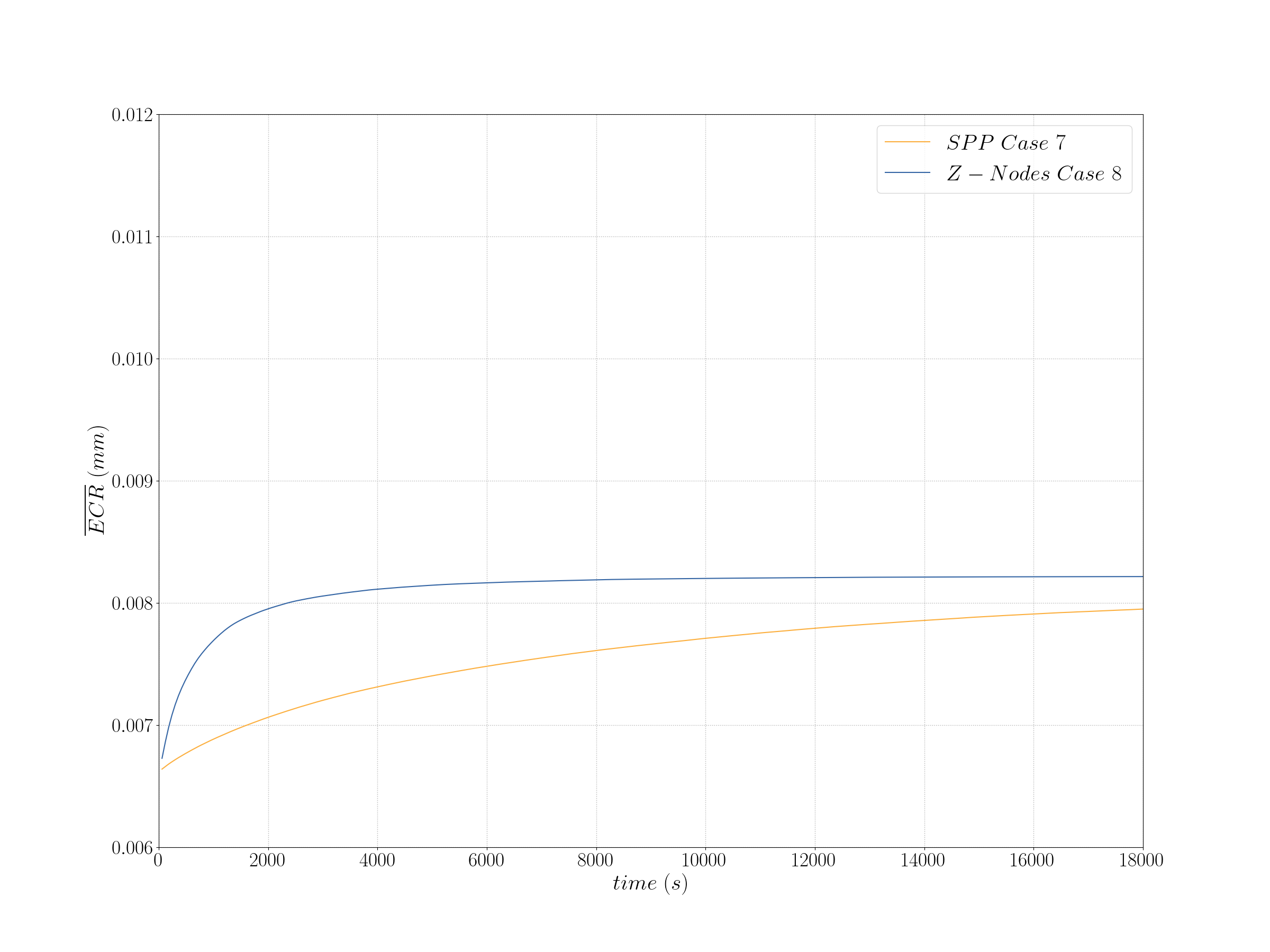}
    \label{fig:ECR4}
  \end{subfigure}
  \begin{subfigure}{0.49\textwidth}
    \centering
    \includegraphics[scale=0.1]{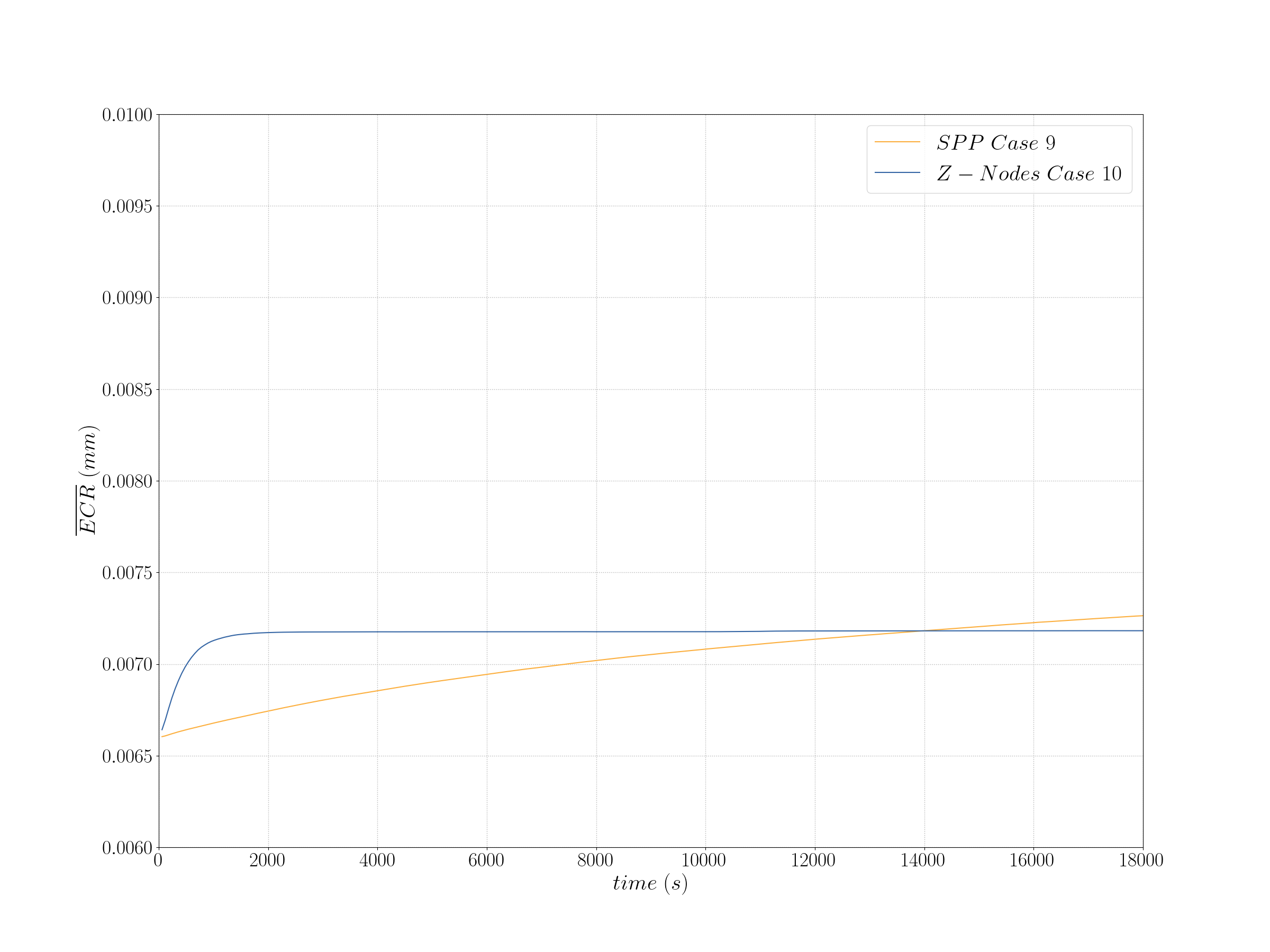}
    \label{fig:ECR5}
  \end{subfigure}
  \caption{For $p\in\llbracket 1,4\rrbracket$, comparisons of $\overline{ECR}$ evolutions between configurations with SPP discretization (cases $2p+1$) and with Z-Nodes (cases $2p+2$).}
  \label{fig:ecrevolution}
\end{figure}

\begin{figure}[!h]
  \centering
  \begin{subfigure}{0.49\textwidth}
    \centering
    \includegraphics[scale=0.1]{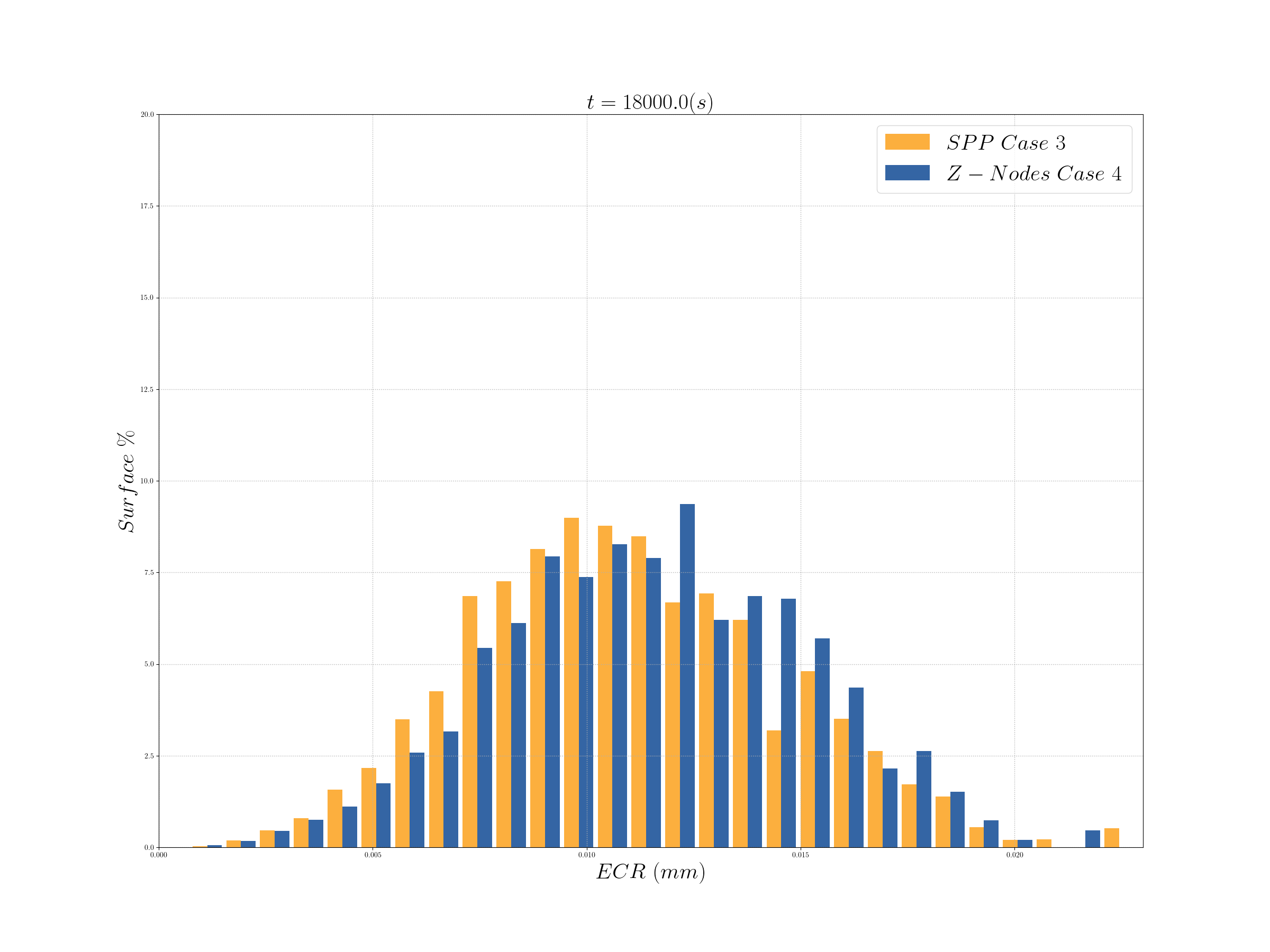}
    \label{fig:D2}
  \end{subfigure}
  \begin{subfigure}{0.49\textwidth}
    \centering
    \includegraphics[scale=0.1]{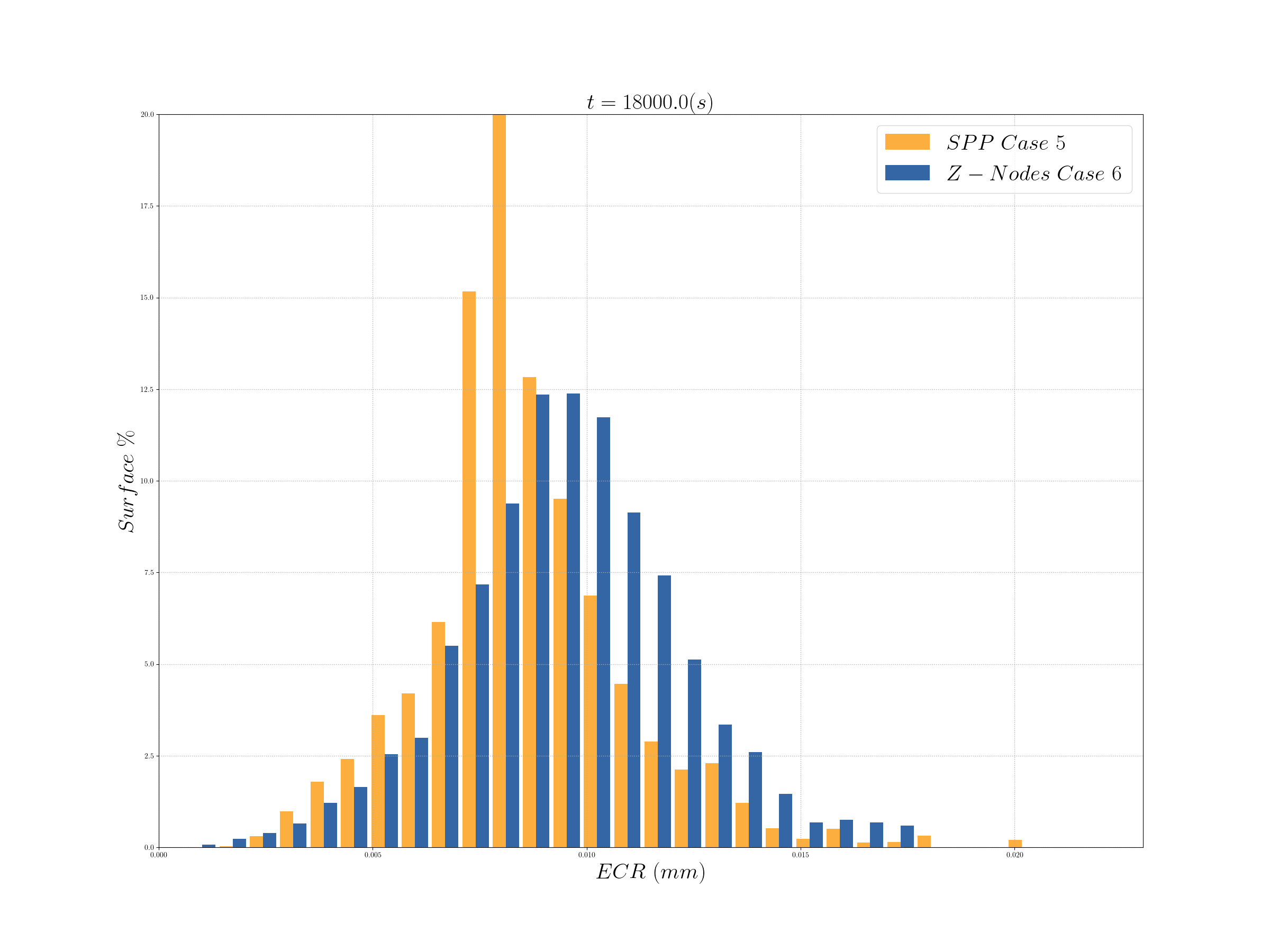}
    \label{fig:D3}
  \end{subfigure}
   \begin{subfigure}{0.49\textwidth}
    \centering
    \includegraphics[scale=0.1]{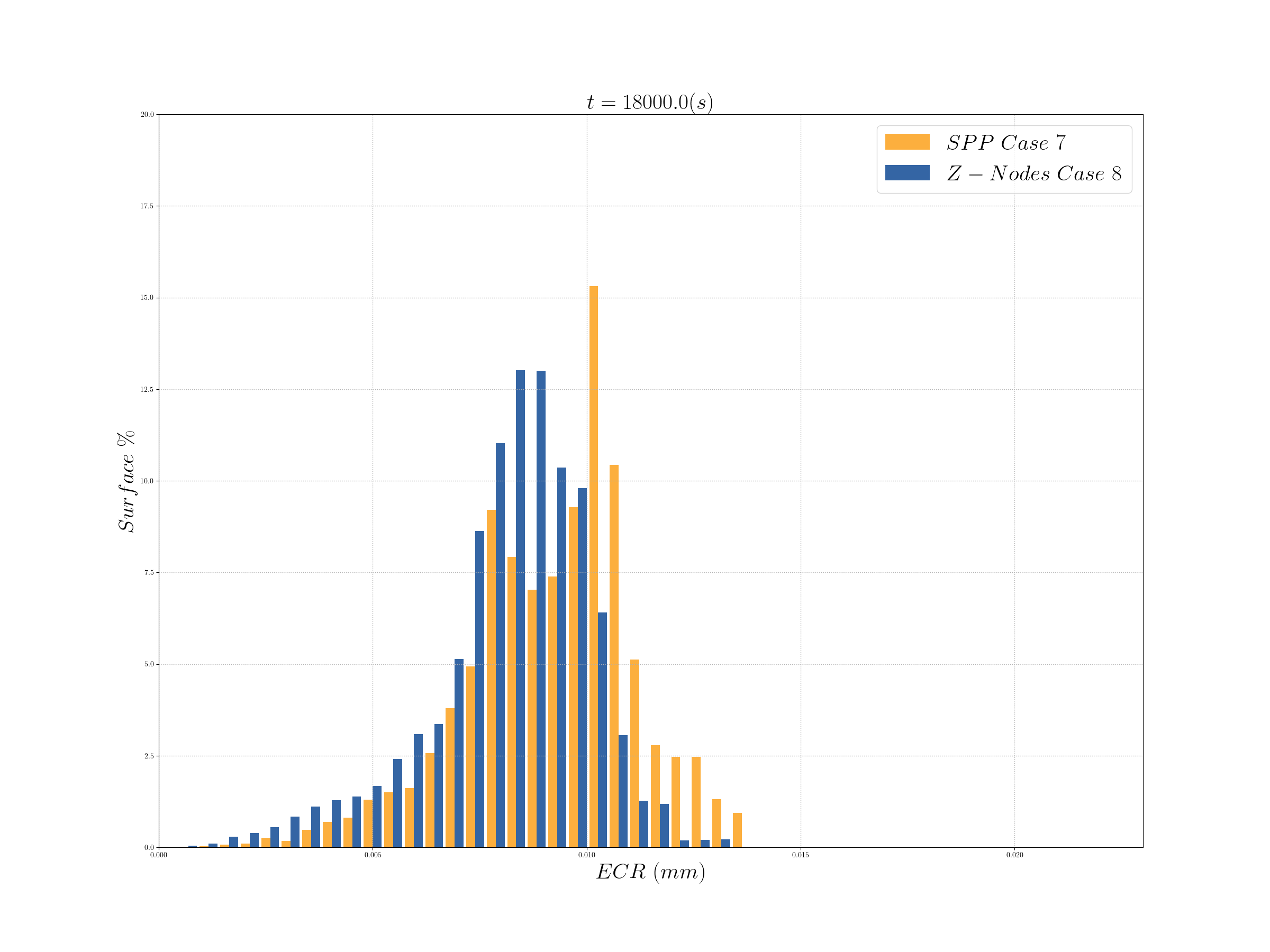}
    \label{fig:D4}
  \end{subfigure}
  \begin{subfigure}{0.49\textwidth}
    \centering
    \includegraphics[scale=0.1]{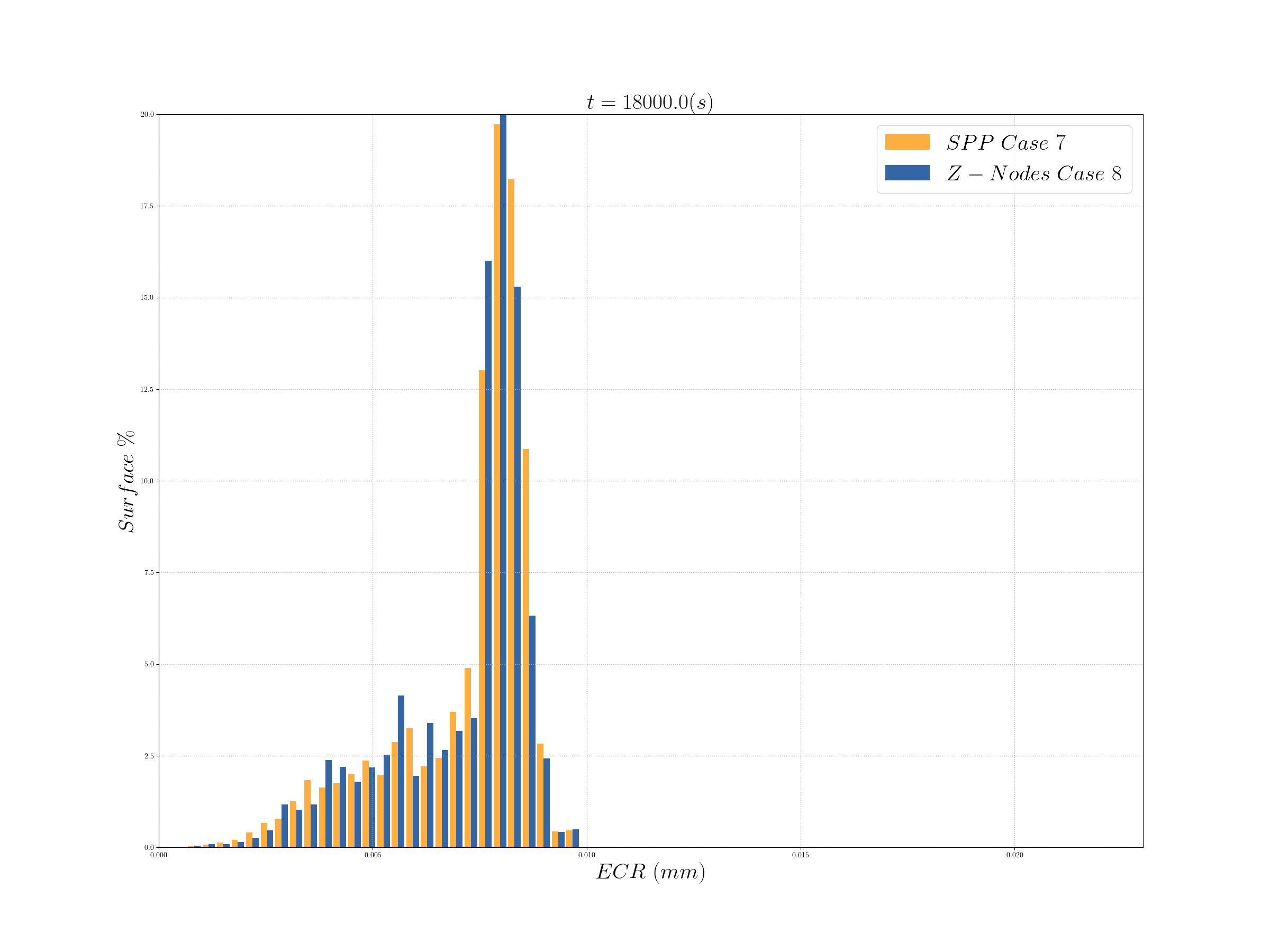}
    \label{fig:D5}
  \end{subfigure}
  \caption{For $p\in\llbracket 1,4\rrbracket$, comparisons at $t=\SI{5}{h}$ of ECR distribution (weighted in surface) evolutions between configurations with SPP discretization (cases $2p+1$) and with Z-Nodes (cases $2p+2$).}
  \label{fig:devolution}
\end{figure}

As illustrated in Table \ref{tab:simulation_settings}, four additional configurations of increasing complexity have been considered for both numerical strategies. For $p\in\llbracket 0,4\rrbracket$, Cases $2p+1$, resp. Cases $2p+2$, correspond to the same configuration simulated by considering the discretization of SPP, resp., the Z-Nodes strategy. This table summarizes, for each test case, the domain dimensions $\Omega$, the initial ($\# G_0$) and final number of grains at $t=\SI{5}{\hour}$ ($\# G_f$), the used FE mesh size $h$, the size and surface fraction of static circular SPP $\bar{r}$ and $f\%$ when they are discretized, the initial number of SPP (even cases) or Z-Nodes (odd cases) $\# SPP$,  the final arithmetic mean grain size at $t=$\SI{5}{\hour} $\overline{ECR}_f$, and the total calculation time necessary to simulate the test cases until $t=$\SI{5}{\hour} with output data at each minute (CPU Time).
The Fig.\ref{fig:cases3to10evolution} illustrates for cases 3 to 10 the final microstructures obtained at the final time whereas the Fig.\ref{fig:ecrevolution}, resp. the Fig.\ref{fig:devolution}, illustrates the comparisons, in terms of $\overline{ECR}$, resp. in terms of $ECR$ distribution, between the Case $2p+1$ and Case $2p+2$ for $p\in\llbracket 0,4\rrbracket$. Different comments can be done concerning the chosen parameters and the results obtained:

\begin{itemize}
\item For odd cases with discretized SPPs, the mesh size was fixed by the SPP size. More precisely, this mesh size was fixed to ensure the correct description of SPPs but also their conservation during the $t=\SI{5}{\hour}$ thermal treatment. Indeed the strategy described in Figures \ref{fig:unpinning} and \ref{fig:Figure_Example_ParticuleEvolution} can be responsible of SPP disappearance in case of a poor description of the SPP comparatively to the mesh size. Typically, a ratio of 4 between the SPP radius and the mesh size, systematically ensures a good conservation of SPPs (less than 0.5\% of SPP disappearance in number). The decreasing in SPP size between odd simulations is then correlated to a  decreasing in mesh size and then an important increasing in CPU time.
\item For even cases, this consideration is not necessary, and the same mesh size can therefore be adopted; only the number of Z-Nodes is considered as increasing.
\item To limit computation times, but also due to a grain size limit being reached more quickly, the last two configurations (simulations 7 to 10) feature smaller simulation domains. 
\item As illustrated in Figures \ref{fig:cases3to10evolution}, \ref{fig:ecrevolution}, and \ref{fig:devolution}, the results between both strategies remain consistent and even similar for all configurations both concerning the arithmetic mean grain size and the grain size distribution. Concerning the grain size limit, the maximal difference is around $\SI{1}{\micro\meter}$ (case 1 versus case 2 in the bottom left position in Fig.\ref{fig:ZNode1} and case 5 versus case 6 in the top right position in Fig.\ref{fig:ecrevolution}). As expected, the cases with Z-Nodes description tends to slightly underestimates the pinning pressure due to the SPP as the real interaction between the SPP and GB are not taken into account. This effect tends to disappear with the decreasing of SPP size where the predictions with Z-nodes are similar in accuracy to those with discretization as illustrated in Figs.\ref{fig:ecrevolution} and \ref{fig:devolution} in the bottom parts. When the CPU times are compared in the last column of Tab.\ref{tab:simulation_settings} between case 7 versus case 8 and case 9 versus case 10, this tends to definitively validate the approach proposed by Z-Nodes when the particle size is well below the grain size. The accuracy/computation time ratio clearly favors this strategy. When the particle size is relatively large, the computation times for the discretization approach are quite reasonable (and comparable to the Z-Nodes approach, as observed in case 1 compared to case 2 or case 3 compared to case 4 in Tab.\ref{tab:simulation_settings}). In this scenario, it seems entirely appropriate to prioritize the discretization approach for SPPs.
\item A consistency can be noted, in Tab.\ref{tab:simulation_settings} across all the simulations considered regarding the $f/\bar{r}$ ratio and, as illustrated, the prediction obtained are far from being equivalent in terms of grain size limits as well known in the state of the art \cite{Bernacki2024,manohar1998,Scholtes2016b,Bignon2024}. The proposed numerical framework would enable a more detailed exploration of this discussion in the case of SPP with sizes ranging from tens of nanometers to hundreds of nanometers, a domain currently overlooked in full-field approaches at the polycrystal scale due to prohibitive computation times \cite{Villaret2020}. This aspect represents a potential avenue for future research of the proposed methodology.
\end{itemize}

\begin{figure}[!h]
  \centering
  \begin{subfigure}{0.49\textwidth}
    \centering
    \includegraphics[scale=0.085]{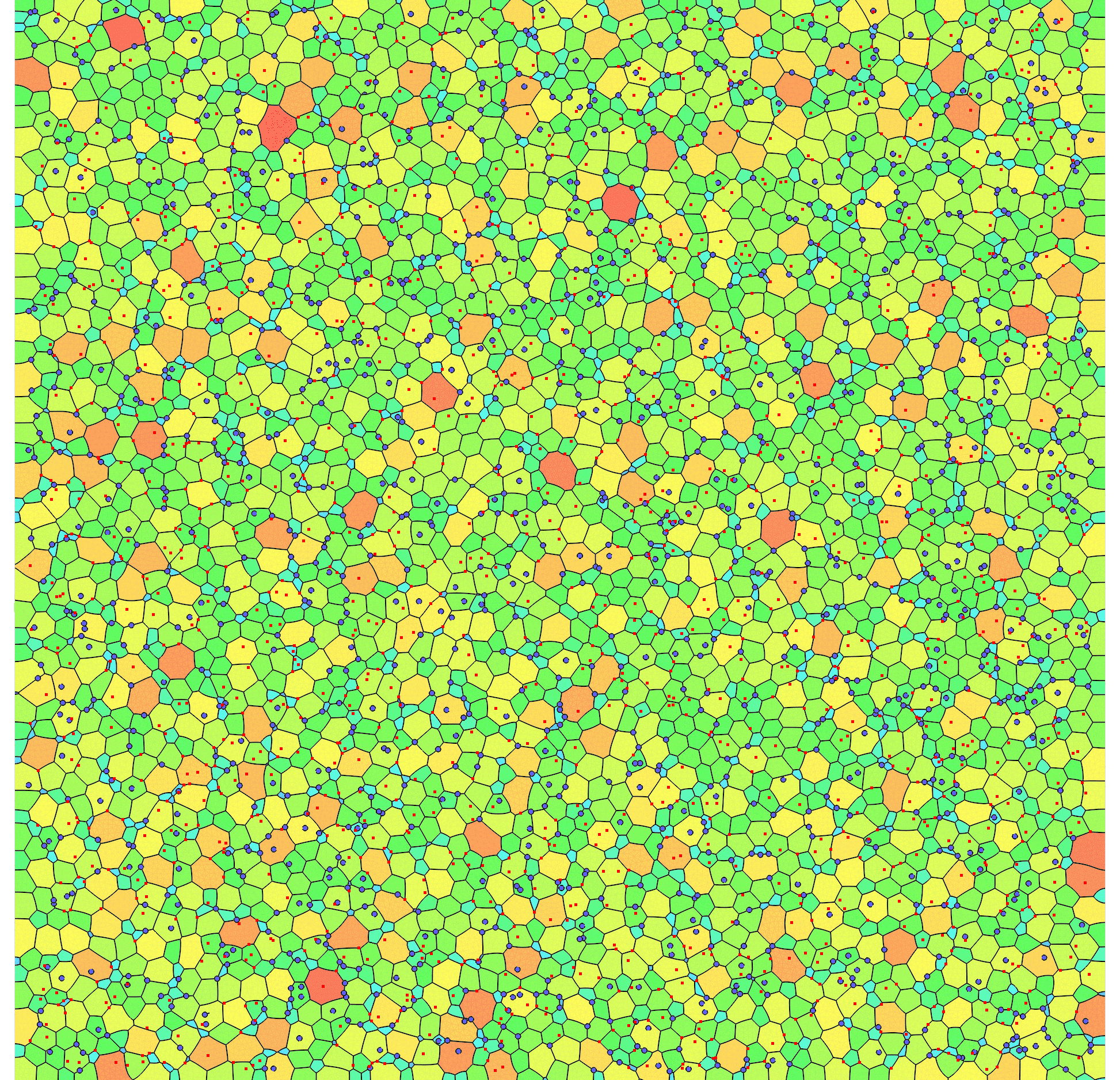}
    \label{fig:mixte1}
  \end{subfigure}
  \begin{subfigure}{0.49\textwidth}
    \centering
    \includegraphics[scale=0.085]{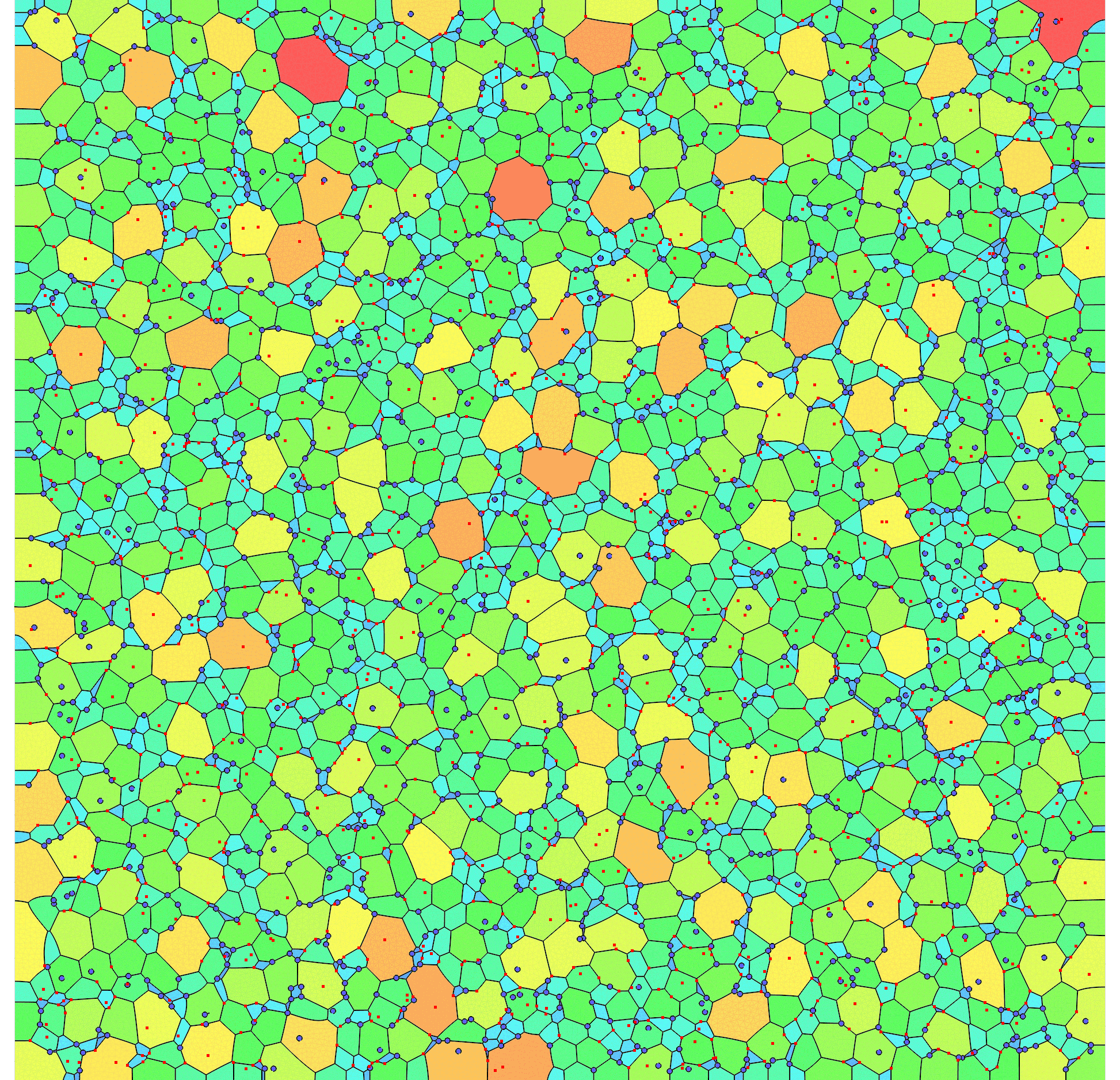}
    \label{fig:mixte2}
  \end{subfigure}
   \begin{subfigure}{0.95\textwidth}
    \centering
    \includegraphics[scale=0.12]{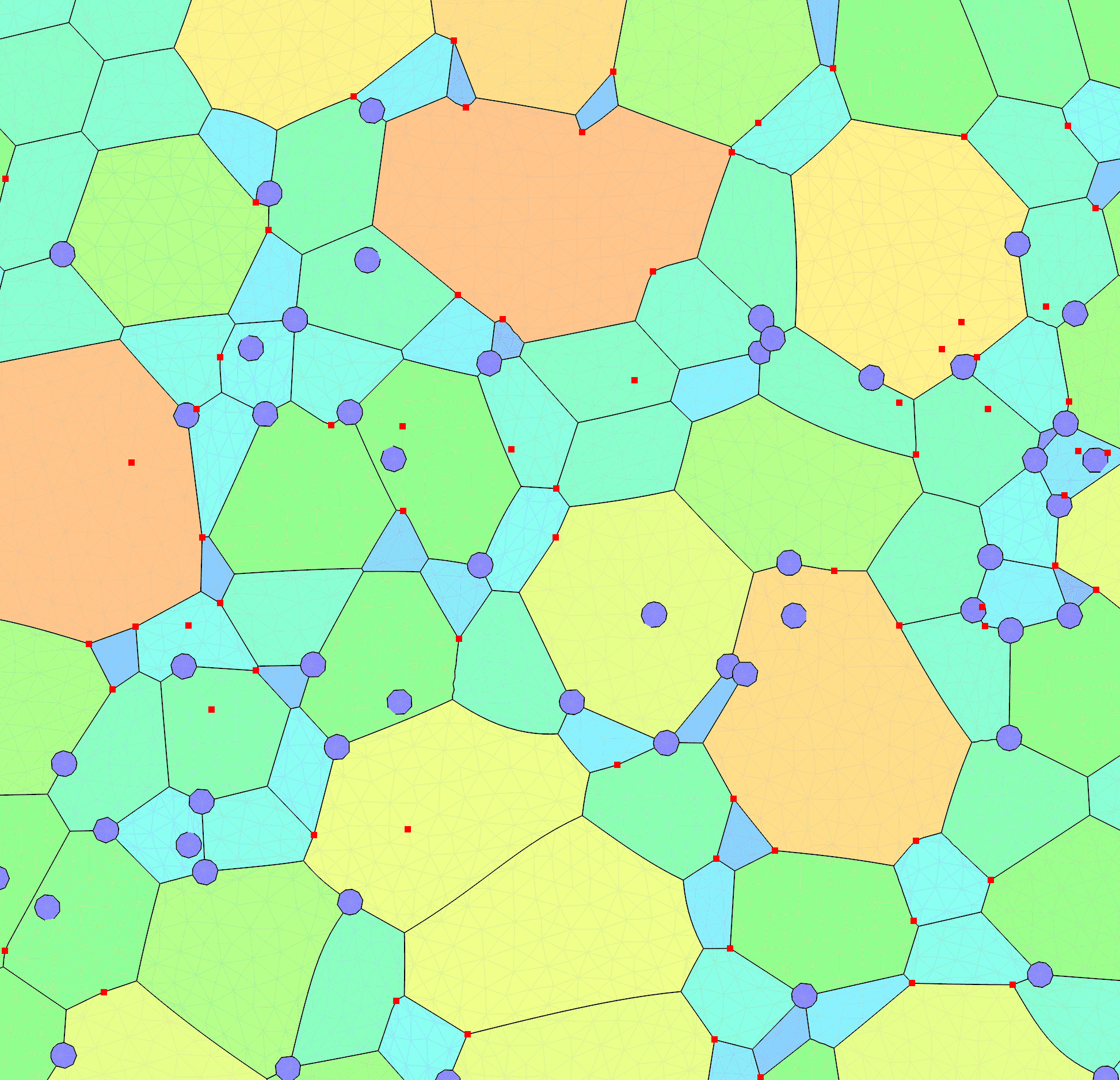}
    \label{fig:mixte3}
  \end{subfigure}
\caption{$\SI{5}{\hour}$ annealing at $\SI{1060}{\celsius}$ of a bimodal configuration concerning the SPP where the biggest ($r=\SI{2}{\micro\meter}$) are discretized whereas the smallest ones ($r=\SI{200}{\nano\meter}$) are considered through Z-Nodes. Top left: the initial microtructure. Top right: the final one. Bottom: zoom on the central zone at the final time with the large SPP in blue and the Z-Nodes in the the red points.}
  \label{fig:mixte}
\end{figure}

\begin{figure}[!h]
  \centering
  \begin{subfigure}{0.49\textwidth}
    \centering
    \includegraphics[scale=0.12]{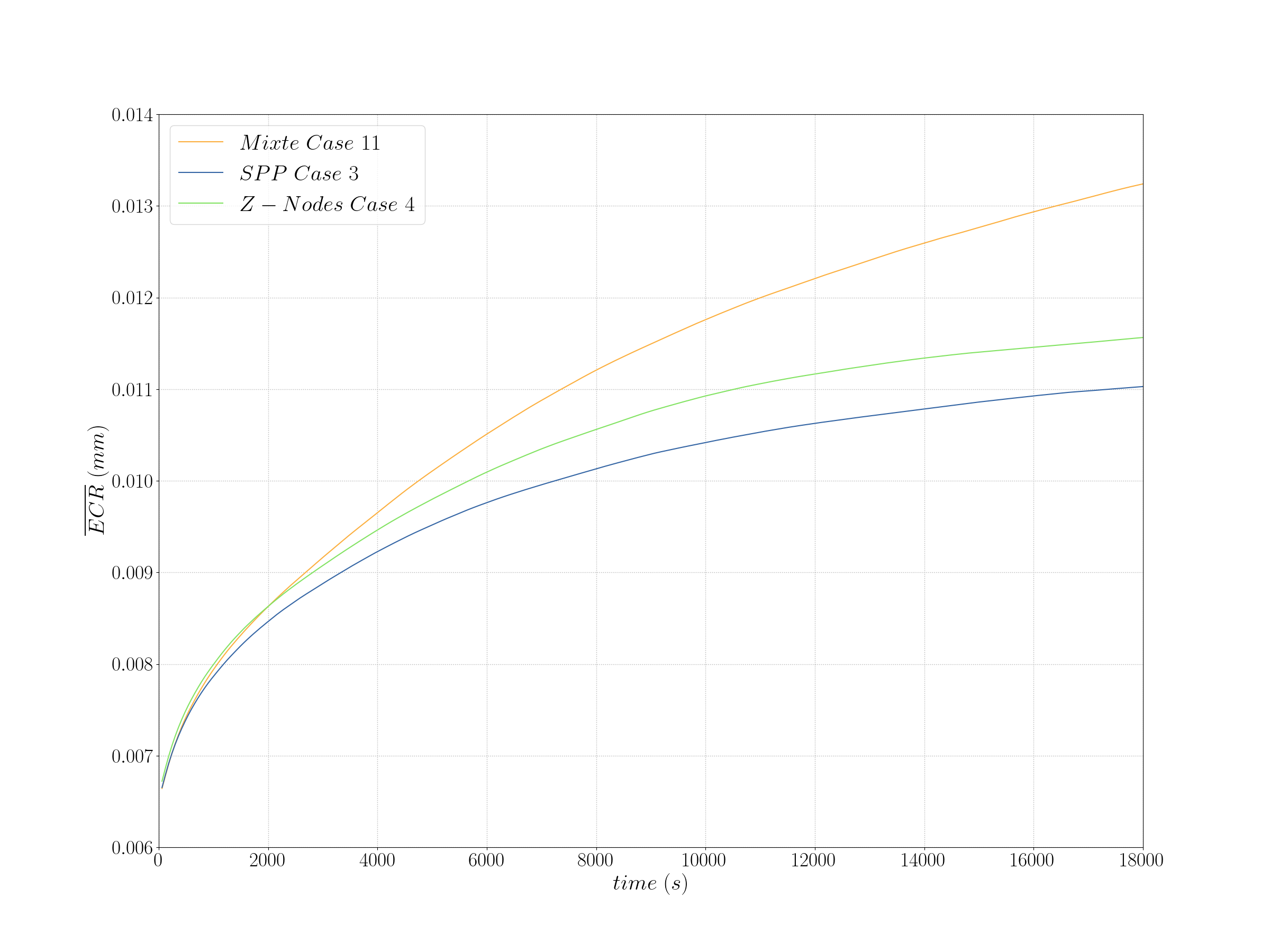}
    \label{fig:mixte4}
  \end{subfigure}
  \begin{subfigure}{0.49\textwidth}
    \centering
    \includegraphics[scale=0.12]{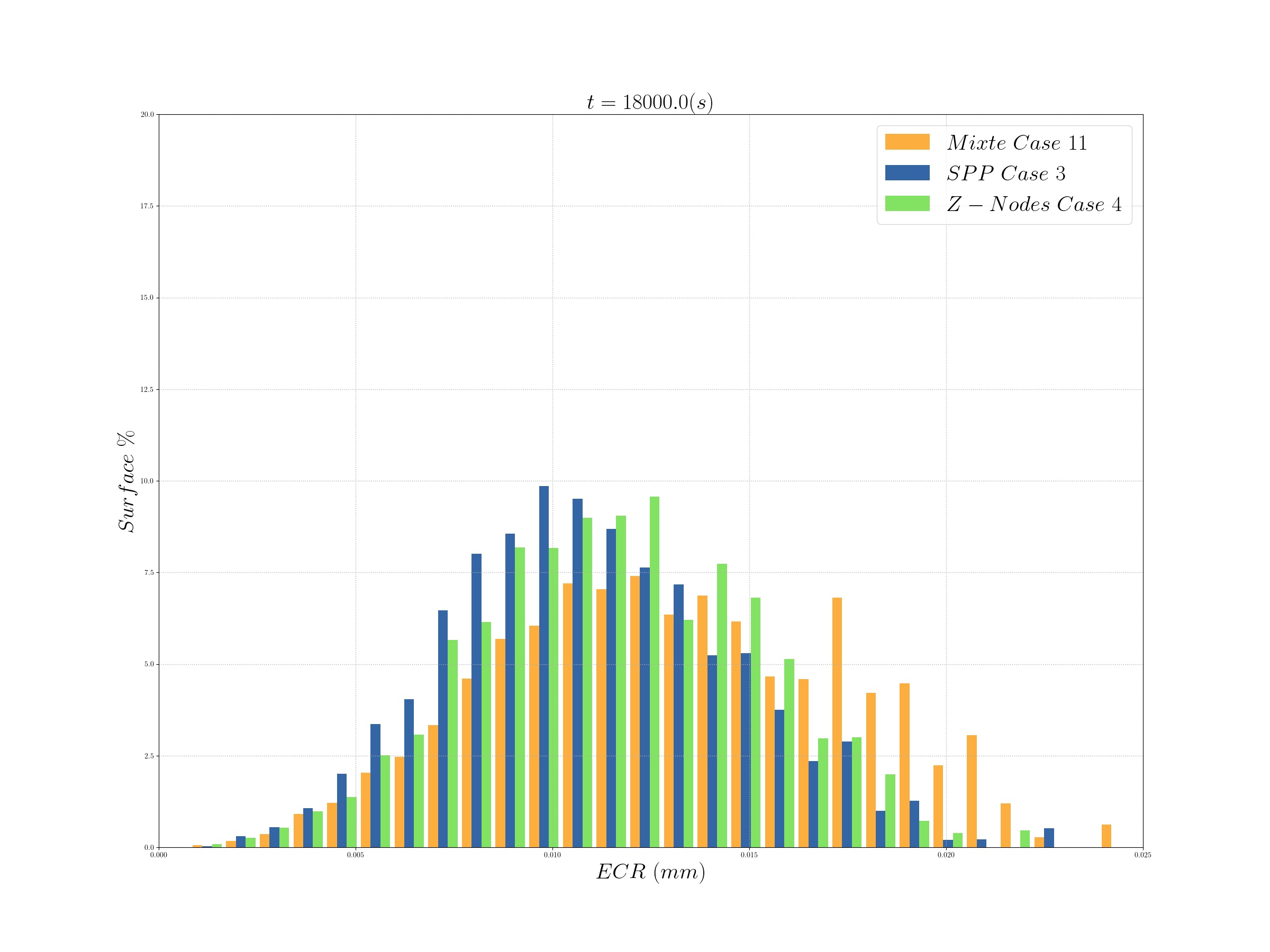}
    \label{fig:mixte5}
  \end{subfigure}
\caption{Comparison between the bimodal case with the cases 3 and 4 (which presents the same arithmetic mean particle size and surface fraction of SPP).}
  \label{fig:mixteradius}
\end{figure}

Finally, it is important to note that nothing prevents combining the advantages of both proposed methods in the case of microstructures with complex populations of second-phase particles that may be highly heterogeneous, such as bi-modal or tri-modal distributions. Indeed, it is not uncommon to encounter industrial alloys exhibiting large primary particles associated with finer populations of secondary or even tertiary particles \cite{alvarado2021dissolution}. For this type of microstructure, it is perfectly conceivable to discretize particles with sizes comparable to the grain size while implementing a Z-Nodes strategy for particles significantly smaller than the grain size. This idea is illustrated here with the last test case considered in this work. A bimodal SPP population was considered on a $0.8\SI{}{\milli\meter}\times 0.8\SI{}{\milli\meter}\ \Omega$ domain with the same polycrystalline structure considered in cases 1 to 6 from the Tab.\ref{tab:simulation_settings}. This bimodal SPP distribution consists to as 2.47\% surface fraction of large SPP with a radius of $\SI{2}{\micro\meter}$
(1260 particles) and a 0.031\% surface fraction of small SPP with a radius of $\SI{200}{\nano\meter}$ (1574 particles). The figure \ref{fig:mixte} illustrates the initial state and the final one after the $\SI{5}{\hour}$ thermal treatment at $\SI{1060}{\celsius}$ by discretizing the large particles and introduced the small ones through Z-Nodes. Interestingly, if an arithmetic mean radius of the second phase particle  population is considered along with an overall fraction, this population would be seen, according to a monomodal filter, as entirely equivalent to those in simulations 3 and 4 ($\bar{r}=\SI{1}{\micro\meter}$ and $f=$2.5\%). However, the comparisons exhibit in Fig.\ref{fig:mixteradius} illustrates that this kind of microstructures can not be correctly depicted through a such monomodal filter when the grain size limit and the Smith-Zener pinning pressure are discussed.

\section{Conclusion and perspectives}\label{sec:conclusions}

In the state of the art, there are numerous front-tracking approaches to describe grain boundary migration at the polycrystalline scale. However, when second-phase particles need to be considered, current approaches are highly restrictive: either only a few grain boundaries interacting with discretized particles are accounted for, or the second-phase particles are reduced to pinning nodes, regardless of the size ratio between particles and grains. This work introduces a new approach to address this issue.
First, a novel front-tracking approach was developed, enabling the discretization of particles for a large number of grains. This new method was validated against an existing front-capturing approach. It was shown to be just as accurate while being faster.
Furthermore, an enhancement was proposed to couple this new approach for large particles, while considering smaller particles through pinning points called Z-Nodes. A series of numerical experiments was conducted to validate the approach and demonstrate its ability to maintain reasonable computation times (a few hours on a single CPU for simulating a $\SI{5}{\hour}$ heat treatment) for millimeter-scale computational domains and second-phase particles as small as a few tens of nanometers.
To our knowledge, such calculations—essential for alloys with complex populations of second-phase particles—had never been proposed before, even in 2D. Many perspectives accompany this work. First, an extension to more complex particle shapes should be considered, along with the development of a 3D version. This model will also soon be compared with experimental data for a material featuring multimodal populations of second-phase particles (e.g. AD730 nickel-based superalloy).

\section*{Acknowledgments}
The authors thank ArcelorMittal, Aperam, Aubert \&
Duval, CEA, Constellium, Framatome, and Safran companies and the
ANR for their financial support through the DIGIMU consortium and
RealIMotion ANR industrial Chair (Grant No. ANR-22-CHIN-0003).

\bibliography{main} 

\end{document}